\documentclass[longauth]{aa}  
\usepackage{graphicx}
\usepackage{txfonts}
\usepackage[colorlinks=true, 
urlcolor=blue, 
citecolor=blue, 
linkcolor=blue]{hyperref}
\usepackage{natbib}
\usepackage{booktabs}
\usepackage{xcolor}
\usepackage[switch]{lineno}
\linenumbers
\usepackage{orcidlink}
\usepackage{longtable}     
\usepackage{multicol}
\begin{document}

   \title{Bridging the gap between SLSNe and SE-SNe}
   \subtitle{Multiwavelength analysis of the SLSN-Ib SN~2024jlc}
    
   \titlerunning{The SLSN 2024jlc}

\author{
    A. Simongini$^{\orcidlink{0009-0000-3416-9865}}$\inst{1,2}\thanks{E-mail: andrea.simongini@inaf.it}
    \and
    F. Acero$^{\orcidlink{0000-0002-6606-2816}}$\inst{3,4}
    \and 
    M. Imbrogno$^{\orcidlink{0000-0001-8688-9784}}$\inst{5,6,1}
    \and
    J. L. Wise$^{\orcidlink{0000-0003-0733-2916}}$\inst{7} 
    \and 
    J. Sollerman$^{\orcidlink{0000-0003-1546-6615}}$\inst{8} 
    \and 
    S. Schulze$^{\orcidlink{0000-0001-6797-1889}}$\inst{9}  
    \and 
    N. Paul M. Kuin$^{\orcidlink{0000-0003-4650-4186}}$\inst{10}
    \and    
    T. X. Chen$^{\orcidlink{0000-0001-8472-1996}}$\inst{11}
    \and 
    C. Fremling$^{\orcidlink{0000-0002-4223-103X}}$\inst{12, 13}
    \and 
    A. Gangopadhyay$^{\orcidlink{0000-0002-3884-5637}}$\inst{8}
    \and 
    M. M. Kasliwal$^{\orcidlink{0000-0001-9152-6224}}$\inst{11}
    \and 
    R. R. Laher$^{\orcidlink{0000-0003-2451-5482}}$\inst{11}
    \and 
    Z. McGrath$^{\orcidlink{0009-0006-0726-1328}}$\inst{7} 
    \and 
    D. A. Perley$^{\orcidlink{0000-0001-8472-1996}}$\inst{7, 14} 
    \and 
    J. N. Purdum$^{\orcidlink{0000-0003-1227-3738}}$\inst{12}
    \and 
    N. Rehemtulla$^{\orcidlink{0000-0002-5683-2389}}$\inst{15, 9, 16}
    \and 
    R. M. Rich\inst{17}
    \and 
    R. Riddle$^{\orcidlink{0000-0002-0387-370X}}$\inst{12}
    }
    \authorrunning{A. Simongini et al.}

   \date{Received 3 March, 2026; accepted 28 July, 2026}

    % List of institutions
    \institute{
    INAF - Osservatorio Astronomico di Roma, Via di Frascati 33, I-00078 Monteporzio Catone, Italy  %1
    \and 
    Università Tor Vergata, Dipartimento di Fisica, Via della Ricerca Scientifica 1, I-00133 Rome, Italy %2
    \and 
    FSLAC IRL 2009, CNRS/IAC, La Laguna, Tenerife, Spain %3
    \and
    Universit\'e Paris-Saclay, Universit\'e Paris Cit\'e, CEA, CNRS, AIM, 91191 Gif-sur-Yvette, France %4
    \and 
    Institute of Space Sciences (ICE, CSIC), Campus UAB, Carrer de Can Magrans s/n, E-08193 Barcelona, Spain %5
    \and 
    Institut d’Estudis Espacials de Catalunya (IEEC), E-08860 Castelldefels (Barcelona), Spain %6
    \and 
    Astrophysics Research Institute, Liverpool John Moores University, 146 Brownlow Hill, Liverpool L3 5RF, UK % 7
    \and
    The Oskar Klein Centre, Department of Astronomy, AlbaNova, SE-106 91 Stockholm, Sweden % 8
    \and
    Center for Interdisciplinary Exploration and Research in Astrophysics (CIERA), 1800 Sherman Avenue, Evanston, IL 60201, USA %9
    \and
    University College London, Mullard Space Science Laboratory, Holmbury St. Mary, Dorking, RH5 6NT,  U.K. %10
    \and 
    IPAC, California Institute of Technology, 1200 E. California Blvd, Pasadena, CA 91125, USA %11 
    \and
    Caltech Optical Observatories, California Institute of Technology, Pasadena, CA 91125, USA %12
    \and
    Division of Physics, Mathematics and Astronomy, California Institute of Technology, 1200 E. California Blvd, Pasadena, CA 91125, USA % 13
    \and
    Nordic Optical Telescope, Rambla José Ana Fernández Pérez 7, ES-38711 Breña Baja, Spain %14
    \and
    Department of Physics and Astronomy, Northwestern University, 2145 Sheridan Road, Evanston, IL 60208, USA %15 
    \and
    NSF-Simons AI Institute for the Sky (SkAI), 172 E. Chestnut St., Chicago, IL 60611, USA %16 -
    \and 
    Department of Physics $\&$ Astronomy, University of California Los Angeles, 430 Portola Plaza, Los Angeles, CA 90095-1547, US %17 
    }

  \abstract
  {The Type I super-luminous supernova SN~2024jlc (ZTF24aapadbb) exploded on 25 May 2024 at $z = 0.039$. 
  This was the closest supernova of this class discovered in recent years and one of the closest ever. 
  It therefore represented a rare opportunity to study this type of object in detail.
  We performed a multiwavelength analysis that spanned ten orders of magnitude in frequency and included optical and UV photometry and spectroscopy, soft and hard X-rays, and high-energy $\gamma$-rays.
  We characterized the event as a slow-evolving and He-rich supernova, with one of the lowest peak luminosities reported for a super-luminous event $M_g\sim-19.37$ mag, and a light-curve evolution compatible with either a circumstellar interaction and a magnetar spin-down model, with noticeable contribution from $^{56}$Ni decay. 
  No significant excess was found in the soft and hard X-ray bands, for which we provide upper limits on the flux. 
  Additionally, we analyzed two years of \textit{Fermi}-LAT data, from which we report an intriguing hint of a $\gamma$-ray signal at the $\sim 3.6 \sigma$ level, although no firm detection can be claimed. 
  The ratio of gamma to optical efficiency, $\eta = 0.38$, suggests a central-engine scenario, similar to SN~2017egm. 
  Our analysis suggests that SN~2024jlc might bridge the gap between super-luminous supernovae and classical stripped-envelope supernovae.
  While still poorly populated, this bridge might consist of all super-luminous Ib supernovae, with the key difference residing in the powering mechanism.
  } 


   \maketitle
    \nolinenumbers 

%-------------------------------------------------------------------

\section{Introduction}

    Super-luminous supernovae (SLSNe) defy the standard models of stellar explosions. 
    They significantly outshine normal thermonuclear and core-collapse supernovae (CCSNe).
    Mimicking the standard spectral classification scheme of CCSNe, all SLSNe can be classified based on the presence (SLSN-II) or absence (SLSN-I) of clear hydrogen features. 
    Recently, helium has also been identified in the spectra of an increasing number of SLSNe-I (e.g., \citealt{yan2020helium}), suggesting a further subclassification into SLSNe-Ib and SLSNe-Ic, based on the presence and absence of helium, respectively. 
    Although the broad and often diverse properties exhibited by SLSNe-I challenge the definition of a unified model, luminosities exceeding $M \sim -20$ mag and the presence of spectral \ion{O}{ii} features have been identified as key diagnostics for recognizing a super-luminous event (e.g., \citealt{de2018light, chen2023hydrogen, gomez2024type}).
    However, the discovery of several less luminous objects ($-20\le M \le -19$) raises the question whether the canonical definition should be revised. 
    These objects, also referred to as luminous supernovae (LSNe; \citealt{gomez2022luminous}), exhibit a broad diversity of features, with some resembling normal SLSNe in terms of spectral sequence and diffusion time, and others being more similar to normal stripped-envelope supernovae (SE-SNe), namely types Ib, Ic, and Ic-BL.
    The key difference appears to be related to the powering mechanism. 
    Normal SE-SNe are mainly powered by radioactive decay of \ion{Fe}-group elements, and in some cases, they can also receive a contribution from the shock interaction between the ejecta and the circumstellar medium (CSM). 
    On the other hand, while CSM interaction can still account for the properties of some events (e.g., \citealt{wheeler2017circumstellar, zhu2023sn, chen2023hydrogenb}), the preferred powering mechanism for SLSNe-I is the spin-down of a rapidly rotating young magnetar. 
    In some cases, both mechanisms might contribute (see for a review, e.g., \citealt{howell2017superluminous, gal2019most, inserra2019observational, moriya2024superluminous}). 

    Helium is typically absent from SLSN-I ejecta or appears in a too limited quantity to form detectable spectral features (i.e., \citealt{kumar2025near}). 
    Only a few events have been claimed to show distinguishable helium features\footnote{To date: SN~1991D, SN~2003L, PTF10hgi, SN2017egm, SN~2018beh, SN~2018fcg, SN~2018kyt, SN~2019hge, SN~2019gam, SN~2019kws, SN~2019obk, SN~2019unb, SN~2020qef, SN~2021bnw, SN~2024afav, SN~2024rmj.} \citep{yan2020helium, 2020ATel13970....1T, gomez2022luminous, zhu2023sn, kumar2025detection, kumar2026sn, fiore2026nearby}. 
    The majority of these events (with the only exceptions of SN~2017egm, SN~2021bnw, SN~2024afav, and SN~2024rmj) exhibits a relatively low peak luminosity, with a cut at $M_\text{g} \le -20.2$. 
    Except for a few cases, they exhibit broad light curves, with long diffusion times and high ejecta masses (i.e., $\ge 10\,M_\odot$), although the velocity of the ejecta remains relatively low. 
    Moreover, some of them show noticeable contribution from $^{56}$Ni decay \citep{gomez2024type}, with some of them having $M_{Ni} \ge 1\, M_\odot$.
    Within the magnetar powering scenario, they tend to prefer shorter spin periods than normal SLSNe, with average magnetic field intensities. This difference can account for the low observed luminosity \citep{gomez2022luminous}.
    However, as pointed out by \citet{chen2023hydrogenb}, some of these objects clearly prefer a CSM+$^{56}$Ni powering model and their light curve undulate. 
    Although the true origin of these undulations is still debated, the authors proposed ejecta-CSM interaction as a possible solution, as well as ejecta clumpiness and changes in opacity. 
    Interestingly, all these events show an early-time blue continuum, and except for SN~2017egm and SN~2024rmj, they did not exhibit obvious spectral W-shaped features. 
    These \ion{O}{ii} features around 4200 and 4450 \AA\, are considered a key spectral hallmark of SLSNe, distinguishing them from SNe Ic \citep{quimby2018spectra}. 
    \citet{gomez2022luminous} suggested that the absence of these lines might result from low temperatures that prevent sufficient ion excitation at early times. 
    However, as also noted by \citet{chen2023hydrogenb}, the early-phase temperatures are quite high, indicating that the lack of these features must have a different origin. 
    \citet{mazzali2016spectrum} proposed that the W-shaped features arise from excitation powered by magnetar spin-down, which would further support CSM interaction as the dominant powering mechanism in some cases.
    Last, SLSNe tend to prefer low-mass host galaxies ($M\le 10^{9} M_\odot$), high star formation rates, and low local metallicites ($\le 0.5$ solar), with some exceptions \citep{gal2019most}. 

    Owing to their extraordinary optical brightness, SLSNe have been the focus of extensive multiwavelength observational campaigns. 
    Previous observations indicated that these events tend to be faint in the other regions of the electromagnetic spectrum, and only a few have been detected beyond the optical. 

    X-ray observations were performed for over 30 SLSNe \citep{ofek2013x, levan2013superluminous, inserra2017complexity, bhirombhakdi2018engine, margutti2018results}, but only 3 have been tentatively detected.
    X-ray emission occurs when the optical photons are up-scattered by the electrons accelerated in the shock front. 
    The inverse-Compton (IC) scattering mechanism is considered the primary driver of X-ray emission in young SNe and SLSNe, although thermal Bremsstrahlung can dominate at early times in cases of dense CSM environments such as those seen in Type IIn SNe. 
    Previous observations tended to favor the first scenario as dominant at around peak luminosity, suggesting the presence of a central engine rather than strong CSM interaction (e.g., \citealt{margutti2018results}), although a very thin layer of dense CSM around the exploded star could not be ruled out entirely. 
    This highlights how significantly the powering mechanism of optical light curves can shape the emission at different wavelengths. 

    Another possible signature for CSM-interaction or magnetar spin-down is the emission in $\gamma$-rays. 
    Potential $\gamma$-ray emission was searched for in a population of more than 200 SLSNe in the energy range between 500 MeV and tens to hundreds of GeV \citep{renault2018search, crnogorvcevic2026gamma}. 
    Two of these events were also observed at higher than TeV energies \citep{acharyya2023veritas}. 
    No significant excess was found for any of these SNe, except for SN~2017egm, which independent analyses supported as a strong candidate with $\gamma$-ray emission powered by magnetar spin-down \citep{li2026evidence, acero2026gamma, crnogorvcevic2026gamma, mukhopadhyay2026neutrino}. 
    Bright $\gamma$-ray emission has been predicted to arise in the CSM-interaction and in the magnetar spin-down model (e.g., \citealt{murase2011new, murase2015gamma}).
    In the CSM-interaction scenario, collisionless shocks following the interaction between the fast-moving ejecta and the slow CSM lead to efficient cosmic-ray acceleration and nonthermal radiation from radio to $\gamma$-rays.
    This scenario likely causes the evidence of radio emission from some SLSNe \citep{eftekhari2019radio, margutti2023luminous}. 
    Alternatively, the energy injection from a fast-rotating magnetar into the SN ejecta can induce strong synchrotron and IC radiation that appears a few months to years after the explosion \citep{kasen2010supernova, vurm2021gamma}. 
    The two models predict strong attenuation at high energies in the first days to weeks through pair-production processes established between $\gamma$-rays and the optical photons of the photosphere, namely $\gamma$-$\gamma$ absorption \citep{Brose2022}, and between $\gamma$-rays and the dense field of nuclei and thermal electrons within the ejecta, namely Bethe-Heitler (BH) absorption \citep{crnogorvcevic2026gamma}. 
    The key difference between the two models is that in the CSM-interaction scenario, the $\gamma$-ray flux is confined to the brief period when the ejecta encounter the CSM. 
    If this material is restricted to small radii around the exploding star, the resulting $\gamma$-rays can be fully suppressed by pair production, leaving no detectable signal. 
    In contrast, the magnetar model is not limited by external conditions, so that particle acceleration, and thus $\gamma$-ray production, can occur continuously without significant absorption.

    We present a multiwavelength analysis of SN~2024jlc (ZTF24aapadbb), which is one of the closest and least luminous SLSNe-I discovered to date. 
    The paper is laid out as follows: we present the event and its host galaxy in Sect.~\ref{sec:discovery} and report our observations in Sect.~\ref{sec:obs}. 
    Section~\ref{sec:lcs} and Sect.~\ref{sec:Spectra} are entirely dedicated to the characterization of the optical light curves and spectra.
    In Section~\ref{sec:mwl} we model our X-rays and $\gamma$-rays data. 
    Finally, we discuss our results in Sect.~\ref{sec:discussion}, and our conclusions are reported in Sect.~\ref{sec:conclusions}.

    \section{Discovery and host galaxy} \label{sec:discovery}

    \begin{figure*}[t]
    \centering
    \includegraphics[width=0.8\textwidth]{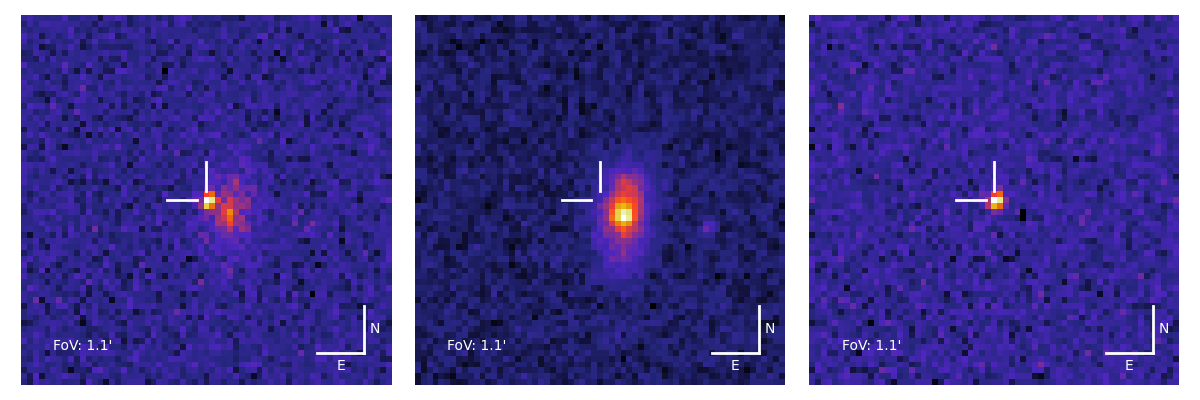}
    \caption{Detection of SN~2024jlc in ZTF g-band images. (Left): Target image from 28 May 2024. (Middle): Deep pre-explosion reference image. (Right): Difference image post explosion.}
    \label{fig:host}
    \end{figure*}

    SN~2024jlc was discovered on 28 May 2024 at 06:14:51 UTC by the Zwicky Transient Facility (ZTF; \citealt{bellm2018zwicky, graham2019zwicky, masci2018zwicky, dekany2020zwicky}) as ZTF24aapadbb with a discovery magnitude of $g = 19.39$, at coordinates $\rm RA(J2000) = 15^\text{h}21^\text{m}58.812^\text{s}$, $\rm DEC(J2000) = +62^\text{d}48^\text{m}22.69^\text{s}$ \citep{Sollerman2024}.
    SN~2024jlc was initially classified as an SN Ib at redshift $z = 0.049$ \citep{Perez2024} and later as an SLSN-I at redshift $z = 0.039$ \citep{Wise2024} \footnote{Note that the two estimates were obtained using two different techniques: \protect\cite{Perez2024} used a template-matching software, while \cite{Wise2024} estimated it directly from the spectrum using host emission lines, as we did in this work.}.
    The last non-detection was reported on 25 May 2024 at 09:14:58\,UTC. 
    Searching deeper into the ZTF forced photometry, we found that the first detection was on 25 May 2024 at 04:56:30\,UTC (MJD\,60455.21) with a discovery magnitude of $g = 20.15 \pm 0.17$. 
    We adopt this time, hereafter $t_0$, as the reference epoch of explosion throughout the paper. 
    
    The SN is associated with the Panoramic Survey Telescope and Rapid Response System (Pan-STARSS, PS1; \citealt{flewelling2020pan}) galaxy with $\rm id = 183362304927039129$, centered at $\rm RA(J2000)\,=\,15^\text{h}21^\text{m}58.178^\text{s}$, $\rm DEC(J2000) = +62^\text{d}48^\text{m}25.16^\text{s}$ and with a reported luminosity of $m_\text{r} \sim 18.47$. 
    If the association is confirmed, SN~2024jlc would be offset by $4.9'' \approx 4.3$\,kpc from the host galaxy center. 
    To analyze the integrated host properties, we retrieved science-ready co-added images from the PS1-DR1 \citep{Chambers2016a} and the reprocessed images from the \textit{Wide-field Infrared Survey Explorer} (WISE; \citealt{Wright2010a}) from the unWISE archive \citep{Lang2014a}.
    The unWISE images include data from the NEOWISE reactivation mission \citep{mainzer2014initial, meisner2017full} through year 6/7. We measured the brightness with the \texttt{LAMBDAR}~\citep[Lambda Adaptive Multi-Band Deblending Algorithm in R;][]{Wright2016a} and used tools developed by \citet{Schulze2021a} to measure the input parameters for \texttt{LAMBDAR}. 
    See Table~\ref{tab:host} for the measurements. 
    We model the host spectral energy distribution (SED) with the software package \texttt{Prospector} version 1.4 \citep{Johnson2021a}. 
    \texttt{Prospector} uses the \texttt{Flexible Stellar Population Synthesis} (\texttt{FSPS}) code \citep{Conroy2009a} to generate the underlying physical model and \texttt{python-fsps} \citep{ForemanMackey2014a} to interface with \texttt{python}. 
    We assume a parametric star formation history of the form $t \times \exp\left(-t/\tau\right)$, where $t$ is the age of the SFH episode and $\tau$ is the $e$-folding timescale, the \citet{Chabrier2003a} initial mass function, the \citet{Calzetti2000a} attenuation model, and the \citet{Byler2017a} model for the ionized gas contribution. 
    The priors were set similar to those in \citet{Schulze2021a}. 
    All parameters were inferred in a Bayesian way by sampling the posterior probability functions with the dynamic nested sampling package \texttt{dynesty} \citep{Speagle2020a} version 2.1.4. 
    We infer a galaxy mass of $\log(M_*/M_\odot) = 9.25^{+0.15}_{-0.18}$ and a star formation rate of $0.05^{+0.13}_{-0.05}~M_\odot\,{\rm yr}^{-1}$, albeit the lack of data blueward of the $g$ band limits the constraints on the star formation rate.

    We collected one spectrum of the host galaxy with the MMT-Binospec \citep{fabricant2019binospec} at 335 days after the explosion. 
    The MMT spectrum shows emission lines from the ionized gas in \ion{H}{ii} regions along the line of sight. 
    Their luminosities and flux ratios allow to determine the metallicity of the gas in the star-forming regions, the metal enrichment and the level of attenuation. 
    The MW-extinction corrected H$\alpha$-H$\beta$ flux ratios is $\approx3.02$, marginally larger than the theoretically expected value of 2.86 for Case B recombination \citep{Osterbrock2006a}. 
    The nominal excess in the flux ratio translates to $E_{\rm host}(B-V) = 0.05\pm0.07$~mag, assuming the Calzetti attenuation model with $R_V=4.05$. 
    The attenuation-corrected H$\alpha$ luminosity translates into a star formation rate of $0.010 \pm 0.002~M_\odot\,{\rm yr}^{-1}$ using \citet{Kennicutt1998a} and \citet{Madau2014a} to convert from the Salpeter IMF (assumed in \citealt{Kennicutt1998a}) to the Chabrier IMF (assumed in our galaxy SED modeling). 
    We determine the metallicity of the gas in the star-forming region using the O3N2 and N2 diagnostics \citep{Pettini2004a} and the parameterizations from \citet{Curti2017a}. 
    We measure a metallicity of $0.50\pm0.03$ solar. 
    The local metallicity is a bit lower than the expected galaxy metallicity based on the mass-metallicity relation \citep{Andrews2013a}, but consistent within errors.

\section{Observations and data}\label{sec:obs}

    We report observations of SN~2024jlc, spanning ten orders of magnitude in frequency, including UV, optical, soft and hard X-rays, and high-energy $\gamma$-rays. 
    All UV and optical data are reported in Table~\ref{tab:photometry} and Table~\ref{tab:spectra}, while the X-rays data are given in Table~\ref{tab:xrt}. 
    The $\gamma$-ray data are publicly available on the \textit{Fermi}-LAT database. 
    
\subsection{Optical photometry}

    We obtained ZTF $gri$ photometry using the ZTF forced-photometry service \citep{masci2018zwicky} via the ZTF Fritz marshal, an instance of SkyPortal \citep{van2019skyportal, coughlin2023data}.
    We also obtained photometry in the Sloan Digital Sky Survey (SDSS ; \citealt{albareti201713th}) $g$, $r$, $i$ and $z$ filters with the IO:O photometer mounted on the Liverpool Telescope (LT; \citealt{Steele2004}). 
    Images were processed using the \texttt{subphot\_pipe}\footnote{\url{https://github.com/kryanhinds/subphot_pipe}} image subtraction and PSF photometry pipeline, with PS1 reference images used for subtraction, and photometry measured using PSF fitting methodology relative to PS1 standards based on the technique outlined in \citet{fremling2016ptf12os}.    
    We retrieved data from the Asteroid Terrestrial-impact Last Alert System (ATLAS; \citealt{tonry2018atlas, smith2020design, Shingles2021}) in the $c$ and $o$ filters. 
    These light curves range from $t_0$ to $t_0 + 452\, \rm d$ and cover the rise, the peak and the decline phase in all filters. 
    The $g$-band light curve peaks at $t_{\text{max}} = 60504.43 \pm 2.64$ MJD (2024-07-13 10:23:09.442 UTC), indicating a rest-frame rise time of $t_{\text{rise}} = 49.23 \pm 2.64$\,d. 
    This is similar to the rise times in the $r$ band ($50.02 \pm 2.42$\,d), in the $i$ band ($49.64 \pm 3.31$\,d) and in the $o$-band ($51.49 \pm 0.44$\,d).   
    The rise time of SN~2024jlc aligns with the average of typical slow evolving SLSNe-I (i.e., 52 days; \citealt{inserra2019observational}). 

\subsection{Optical spectroscopy}

    We collected 21 spectra of SN~2024jlc and one of its host galaxy (see Table~\ref{tab:spectra} for details).
    In particular, we collected 14 spectra with the Spectral Energy Distribution Machine (SEDM; \citealt{blagorodnova2018sed, rigault2019fully, kim2022new}).

    We also obtained two spectra with the SPectrograph for the Rapid Acquisition of Transients (SPRAT; \citealt{piascik2014sprat}) at the LT using a blue-optimized setup, which were reduced using a custom \texttt{python} pipeline. 
    This pipeline is based on the packages \texttt{lacosmic} \citep{van2001cosmic}, \texttt{NumPy} \citep{van2011numpy}, \texttt{SciPy} \citep{virtanen2020scipy}, and \texttt{Astropy} \citep{astropy2022astropy}.  
    Polynomial fits for sky subtraction and trace fitting, as well as the setting of the aperture sizes, were performed manually to optimize the signal to noise ratio. 
    We corrected for airmass differences between the science and standard star exposures using Table 1 from La Palma Technical Note No.~31\footnote{\url{https://www.ing.iac.es/Astronomy/observing/manuals/ps/tech_notes/tn031.pdf}}.
    One spectrum was obtained with the KAST spectrograph at the Lick Observatory \citep{MillerStone1994}. 
    The 300/7500 grating (2.55 \AA/pixel dispersion) was used for the red side, and the 600/4310 grism (1.02 \AA/pixel dispersion) was used for the blue side. 
    A slit width of $2.0^{''}$ was used for all observations. 
    The spectrum was reduced using the UCSC spectral pipeline\footnote{\url{https://github.com/msiebert1/UCSC\_spectral\_pipeline}} \citep{Siebert20}, a custom data-reduction pipeline based on procedures outlined by \citet{Foley03} and \citet{Silverman2012}, utilizing the optimal one-dimensional spectral extraction algorithm from \citet{Horne86}.
    We also observed SN~2024jlc with the Binospec instrument \citep{fabricant2019binospec} installed on the 6.5\,m MMT telescope at MJD 60791.39. 
    Binospec was configured to use the 270 lines/mm grating, achieving wavelength coverage between 3820-9210~\AA\ with $R\sim1340$. 
    The data are reduced using standard procedures implemented in the \texttt{pypeit} software package \citep{prochaska2020pypeit}. 
    Additionally, we collected two spectra with the Alhambra Faint Object Spectrograph and Camera (ALFOSC; \citealt{djupvik2010nordic}), and one with the Double Spectrograph (DBSP, \citealt{oke1982efficient}). 
    Our spectra cover the time range from $13$ to $337$ days after the explosion and the wavelength range $3500 - 10500\,\AA$. 
    We estimate the redshift using the narrow H$\alpha$ emission line in the host galaxy, obtaining $z = 0.0392 \pm 0.0058$, similar to the previous estimation by \citet{Wise2024}.
    Assuming a flat $\Lambda$CDM Universe with $H_0 = 67.4\,\rm km\,s^{-1}\,Mpc^{-1}$, $\Omega_m = 0.315$ and $\Omega_\Lambda = 0.685$ \citep{2020A&A...641A...6P}, we obtain a luminosity distance of $D_L = 179.38 \pm 17.94$\,Mpc, making SN~2024jlc one of the closest classified SLSNe-I to date, after SN~2018bsz \citep{anderson2018nearby}, SN~2017egm \citep{zhu2023sn}, SN~2020wnt \citep{tinyanont2023supernova} and SN~2019ieh \citep{Dahiwale2019}. 
    Consequently, we obtain a distance modulus of $\mu = 36.27 \pm 0.22$ mag.

\subsection{\textit{Swift}-UVOT photometry}

    The observations of SN~2024jlc with the Neil Gehrels \textit{Swift} space telescope, hereafter \textit{Swift} \citep{gehrels2004swift}, were triggered as a target of opportunity (P.I. T. Moore; target ID 16706) three nonconsecutive nights in July 2024, before and after the peak of optical luminosity.
    The Ultra-Violet Optical Telescope (UVOT; \citealt{roming2005swift}) took data with all six filters (w2, m2, w1, u, b, v), resulting in a total exposure time of 4885 seconds. 
    UVOT photometry was processed using the standard \texttt{uvotsource} module \citep{uvot-poole2008, uvot-breeveld2011} within the \texttt{HEASOFT} package for \textit{Swift}-UVOT analysis, using CALDB 2023-12-08\footnote{\url{https://swift.gsfc.nasa.gov/caldb/}}. 
    The complete photometry is reported in Table~\ref{tab:photometry}.

\subsection{\textit{Swift} X-ray observations}

    The X-ray Telescope (XRT; \citealt{burrows2005swift}) on board the \textit{Swift} satellite took data in PC mode simultaneously to UVOT. The observations were conducted in an energy range of $0.3$ -- $10$ keV for a total exposure of 4880 s. 
    Data were reduced using the \texttt{xrtpipeline} and count rates were estimated with the \texttt{sosta} tool from the \texttt{HEASOFT} package. 
    Fluxes were derived in the \texttt{WebPIMMS} portal. 
    Assuming a photon spectral index of 2 and a Galactic hydrogen column density of $N_\text{H} = 1.38\times 10^{20}$ cm$^{-2}$ \citep{bekhti2016hi4pi}, we derived $3\sigma$ flux upper limits (Table~\ref{tab:xrt}) in the full energy band. 
    Integrating over the full time range, we obtain a total absorbed flux upper limit of $F_x \le 0.838 \times 10^{-13}$ erg cm$^{-2}$ s$^{-1}$, corresponding to a luminosity of $L_x \le  3.36 \times 10^{41}$ erg s$^{-1}$. 
    Owing to the vicinity of the event, despite the relatively low exposure time, these upper limits are among the deepest for SLSNe-I in the range $0.3$ -- $10$ keV, slightly above the possible detection of PTF12dam ($L_x \sim 2\times 10^{40}$ erg s$^{-1}$; \citealt{margutti2018results}) and the upper limits of SN~2017egm ($L_x \sim  1.9\times 10^{39}$ erg s$^{-1}$; \citealt{zhu2023sn}). 

    During the \textit{Swift} follow-up, the Burst Alert Monitor (BAT; \citealt{barthelmy2005burst}) took data in survey mode. 
    This mode allows the instrument to monitor the entire field of view for hard X-ray transients in the 15 -- 150 keV band, collecting count-rate data in five-minute time bins. 
    In case of a significant burst, the instrument switches to a photon-by-photon mode, specifically designed to localize and follow fast transients. 
    BAT data are processed using the \texttt{BATIMAGER} software \citep{segreto2010palermo}. This software is specifically designed to perform image reconstruction and generate spectra and light curves of a source. 
    Count rates on the position of the SN are estimated for one day of observation (60517 MJD) and for the full month of July 2024. 
    We obtain $(-2.63 \pm 1.48)\times 10^{-4}$ cts s$^{-1}$ and $(-6.71 \pm 2.55)\times 10^{-5}$ cts s$^{-1}$ respectively. 
    The statistical errors follow a Gaussian distribution at 1$\sigma$. 
    Subsequently, fluxes are derived from the \texttt{WebPIMMS} portal assuming the same properties as those used for the XRT observations and a power-law fit. 
    We obtain absorbed fluxes upper limits of $F_x \le 9.06\times10^{-11}$ erg cm$^{-2}$ s$^{-1}$ and $F_x \le 1.56\times10^{-11}$ erg cm$^{-2}$ s$^{-1}$ corresponding to a luminosity upper limit of $L_x \le 3.49\times10^{44}$ erg s$^{-1}$ and $L_x \le 6.01\times10^{43}$ erg s$^{-1}$ respectively.

\subsection{\textit{Fermi}-LAT monitoring}\label{sect:fermi-lat}

    High-energy (MeV -- GeV) $\gamma$-ray data were collected with the \textit{Fermi}-Large Area Telescope (LAT) space telescope \citep{atwood2009large}.
    Because \textit{Fermi}-LAT scans the entire sky approximately every three hours, the position of SN~2024jlc was within its field of view for almost 50\% of the time. 
    Data analysis was performed using the \textit{Fermitools} version 2.4.0\footnote{\url{https://Fermi.gsfc.nasa.gov/ssc/data/analysis/documentation/}} and \texttt{fermipy v1.4} packages \citep{2018ascl.soft12006W}, with standard models for Galactic (\texttt{gll\_iem\_v07.fits}) and isotropic diffuse emission (\texttt{iso\_P8R3\_SOURCE\_V3\_v1.txt}). 
    To optimally handle the LAT point-spread function (PSF) strong energy dependence, we performed a  summed PSF likelihood method.
    Specifically, events were simultaneously fit using the four PSF event types\footnote{\url{https://Fermi.gsfc.nasa.gov/ssc/data/analysis/documentation/Cicerone/Cicerone_Data/LAT_DP.html}}, each corresponding to a different level of angular reconstruction quality, and the total likelihood was computed as the sum over these components.  
    We used 4 components, of which three (PSF1, PSF2, and PSF3) in the 100 MeV -- 1 GeV energy range with a zenith angle cut of $< 90^{\circ}$, and a single component above 1 GeV with all event types and a broader maximum zenith angle cut of 105$^{\circ}$.
    The analysis was performed over a $10^\circ\times10^\circ$ region centered on the SN coordinates, using the latest release of the 4FGL-DR4 catalog as the baseline source model \citep{ballet2023fermi}. 
    The details of the fitting procedure are similar to the one described in Sect. 2.3 of \citet{acero2026gamma}.

    To evaluate the significance of the $\gamma$-ray signal from the SLSN, we modeled a point source at the optical coordinates using a power-law spectrum, $E^{-\Gamma}$, with a fixed spectral index of $\Gamma = 2$. 
    We assessed the statistical significance of this signal using the likelihood ratio test statistic, defined as ${\rm TS}=2(\ln \mathcal{L}_1 - \ln \mathcal{L}_0)$, where $\mathcal{L}_0$ and $\mathcal{L}_1$ represent the likelihoods of the background-only and source-plus-background hypotheses, respectively.

\begin{figure}
        \centering
        \includegraphics[width=1\linewidth]{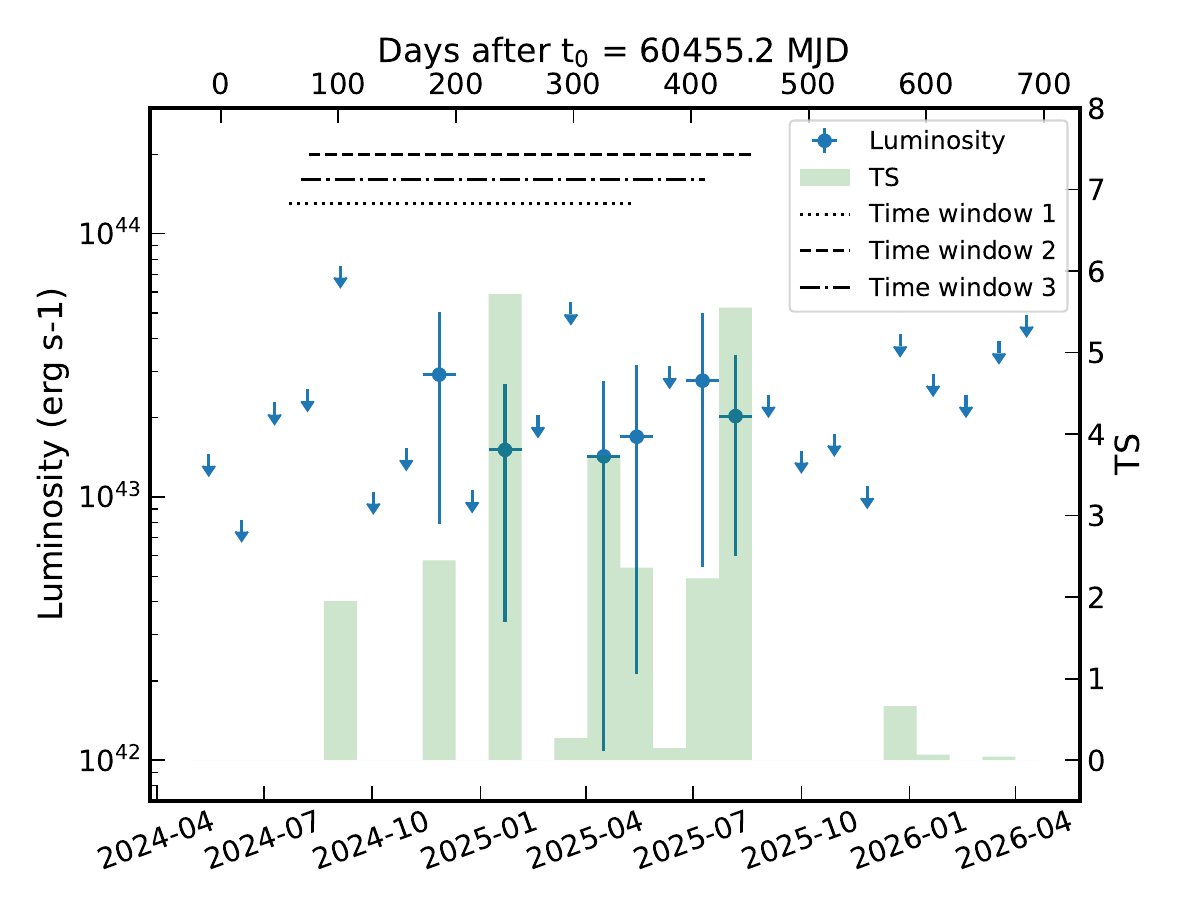}
        \caption{\textit{Fermi}-LAT $\gamma$-ray light curve of SN~2024jlc between April 2024 and April 2026 showing the $\gamma$-ray energy flux and the TS in every monthly time bin. 
        Upper limits at a $95\%$ confidence level are given for time bins with significance TS$<$2. 
        The time integration windows discussed in Sect.~\ref{sec:gamma} are also represented, each representing a different set of parameters: model-independent (time window 1), CSM+$^{56}$Ni model (2), and magnetar spin-down model (3).  
        }
        \label{fig:fermi-lat}
    \end{figure}

    The results are presented and discussed in Sect.~\ref{sec:gamma}. 
    To put these results in context of the global time evolution of the $\gamma$-ray signal, we extracted a light curve with monthly time bin keeping the diffuse background normalizations fixed to the best-fit value of the full time period. 
    In order to account for source variability other than our target in the region, the normalization was set free for any source reaching TS$>$16 in any time bin. 
    The resulting $\gamma$-ray light curve is presented in Fig.~ \ref{fig:fermi-lat}.

\section{Light-curve analysis}\label{sec:lcs}

\subsection{Light-curve evolution}

    The optical photometric evolution of SN~2024jlc is shown in Fig.~\ref{fig:lcs}. 
    Every magnitude is expressed in the AB standard photometric system.
    The light curves were analyzed with \texttt{CASTOR} \citep{Simongini2024, simongini2025a}, which applies Gaussian processes to smoothly interpolate data with no prior information.
    We adopt a foreground Galactic extinction of $E(B-V) = 0.0182$ mag \citep{schlegel1998maps, schlafly2011measuring}, assuming the $R_\lambda$ values from \citet{McCall2004}.
    Absolute magnitudes are obtained after correcting for foreground extinction, host extinction (see Sect.~\ref{sec:discovery}) and K-correction. 

    We obtained a maximum absolute magnitude of $M_{\rm g, peak} = -19.37 \pm 0.22$ and $M_{\rm r, peak} = -19.39 \pm 0.22$, similar to the first rough estimates from \citet{Wise2024} who reported $M_{\rm g} \sim M_{\rm r} \sim -19.2$ at 49 days after discovery.
    Similarly, we obtained $M_{\rm i} = -19.28\, \pm\,0.23$, $M_{\rm z} = -19.32\,\pm\,0.22$ and $M_{\rm o} = -19.33\,\pm\,0.22$, where the overall uncertainty is dominated by the uncertainty in the distance estimate.
    The low cadence of \textit{Swift}-UVOT observations prevent a clear estimate of the peak of luminosity in any of its bands, although the observations were performed very close to the $g$-band maximum. 
    Therefore, we obtain maximum absolute magnitude lower limits of $M_{\rm u} \le -18.73$, $M_{\rm b} \le -19.53$ and $M_{\rm v} \le -19.63$ in the near-UV bands and $M_{\rm w2} \le -17.32$, $M_{\rm m2} \le -17.45$ and $M_{\rm w1} \le -17.95$ in the UV bands.
    These values are generally under-luminous when compared to other SLSNe and seem more similar to the higher end of Type Ic-BL events, suggesting that SN~2024jlc could be a transitional event between the two classes. 

    Using the $griz$ interpolated light curves we estimated an average rise rate between $t_0$ and $t_{\rm max}$ of $0.058 \pm 0.004\,\text{mag}\,\text{d}^{-1}$, with $g$ being the fastest to rise ($49.23\pm2.64\,\text{d}$) and $z$ the slowest ($52.84\pm4.12\,\text{d}$).  
    After, between the peak and day 80, the $g$ band exhibits a steep decline of $0.049 \pm 0.003\,\text{mag}\,\text{d}^{-1}$, although the $rzi$ light curves decrease more slowly, with an average of $0.018 \pm 0.07\,\text{mag}\,\text{d}^{-1}$ 
    Then, between 80 and 150 days after the explosion, the decline is halted. 
    During this time window, especially in the $g$, $r$ and $o$ bands, the light curve is almost flattened, with a global rate of $0.004 \pm 0.002$ mag d$^{-1}$. 
    This behavior can be observed at the same time with different instruments and in different filters.  
    Afterwards, the luminosity starts dropping again, with a rate of $0.013 \pm 0.002$ mag d$^{-1}$, measured between 200 and 400 days post-explosion. 
    This late-time evolution is consistent with a $^{56}$Co nuclear decay-powered model, and it does not show the typical magnetar-powered tail as discussed in \citet{Inserra2013}.

\subsection{Color evolution}

    Figure~\ref{fig:lcs} presents the $(g-r)$ color evolution of SN~2024jlc.
    Whereas the \textit{Swift}-UVOT data are sparse and do not allow for a clear reconstruction of the color evolution, the ZTF dataset has a dense cadence, enabling a detailed tracking of the color trend over the full observation time. 
    The $(g-r)$ color displays an almost linear reddening from the explosion up to the peak of luminosity, reaching a value of $(g-r)_{\text{MW + HG}} = 0.02 \pm 0.03$ mag when accounting for host-galaxy and Milky Way extinction.
    This value is consistent within the uncertainties of peak $(g-r)$ colors in the SLSN-I sample of \citet{chen2023hydrogen}, who found $(g-r)_{\text{median}} = -0.03^{+0.12}_{-0.11}$ mag after host-galaxy correction and without correcting for K.   
    Afterwards, the color shifts towards the blue.
    Finally, after $\sim 90$\,days, the color exhibits an almost linear reddening trend. 
    The early-time UV excess also indicates a strong ultraviolet contribution at early phases, which rapidly diminishes after the peak. 

    \subsection{Photospheric evolution}

\begin{figure}
    \begin{center}
    \includegraphics[width=1\columnwidth]{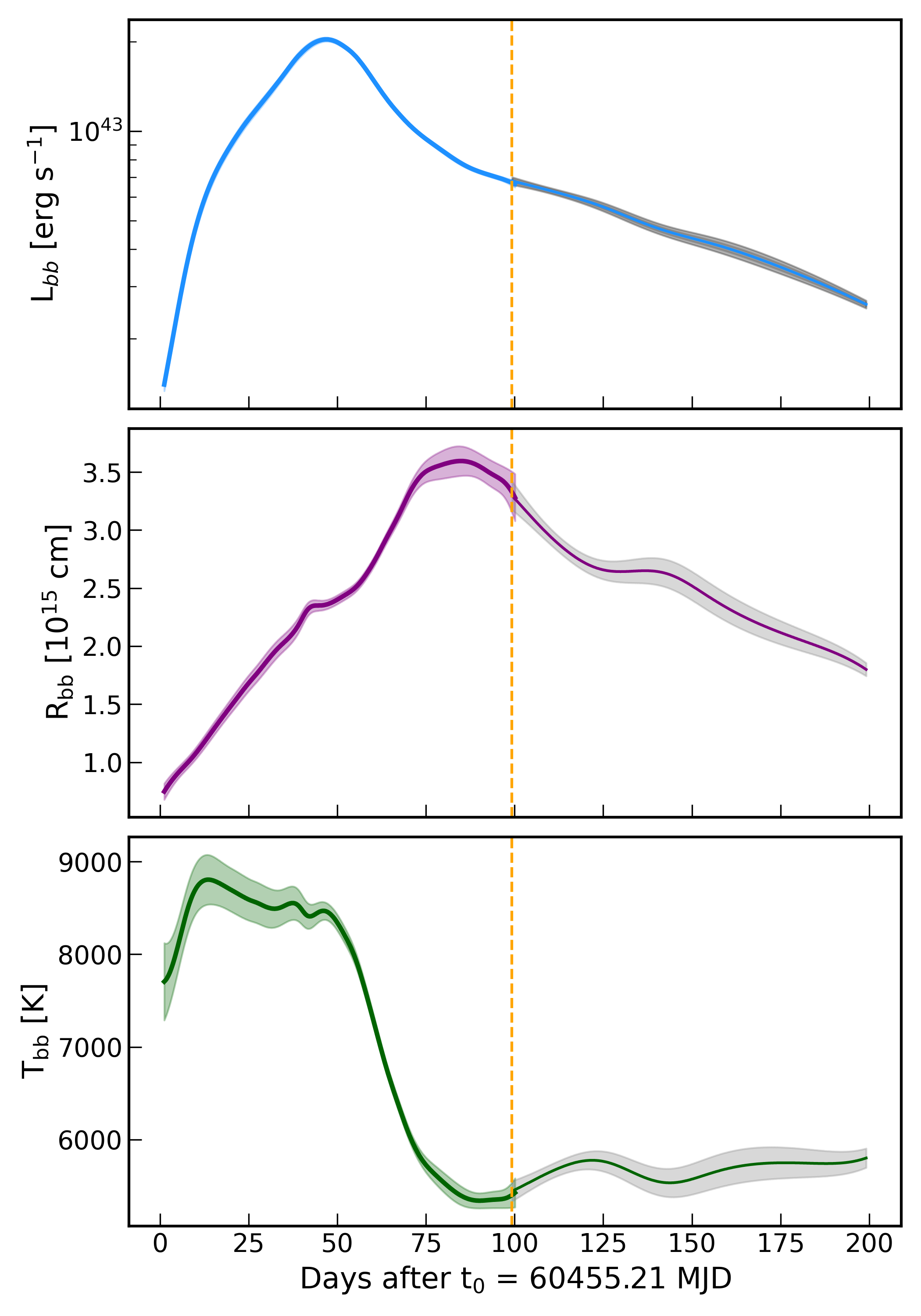}
    \caption{(Top panel): Bolometric luminosity. (Central panel): Photospheric radius. (Bottom panel): Photospheric temperature. 
    The vertical dashed line at $t_0 + 100$ identifies the separation between the first fit (all filters) and the second fit (only $griz$ filters). 
    }
    \label{fig:phot}
    \end{center}
    \end{figure}

    The evolution of the bolometric luminosity and the photospheric temperature and radius during the first 200 days after the explosion is shown in Fig.~\ref{fig:phot}. 
    Calculations are performed with the software \texttt{extrabol} \citep{extrabol_zenodo}, which derives the relevant parameters fitting a black-body to the available data. 
    Notably, \texttt{extrabol} interpolates the available light curves (including UV data) with Gaussian process techniques. 
    However, the available time coverage of UV and near-UV filters does not allow a complete reconstruction of the bolometric luminosity, although the black-body is often suppressed in the UV due to line blanketing.
    Thus, we perform two separate fits: the first, between $t_0$ and $t_0+100$ d including all interpolated filters, and the second, after $t_0 + 100$d using only $griz$ filters. 

    We obtain a peak luminosity of $\rm L_{BB} = (2.03 \pm 0.03) \times 10^{43}\,\rm erg\,s^{-1}$, corresponding to $\rm M_{BB} = -19.57 \pm 0.02$. 
    These values indicate that SN~2024jlc is rather under-luminous if compared to the general definition of SLSNe-I, which corresponds to a peak magnitude lower than $-20$ and a peak luminosity higher than $3 \times 10^{43}$ erg s$^{-1}$ \citep{moriya2018superluminous, moriya2024superluminous}. 
    Finally, the total radiated energy during the entire duration of the light curve is $E = 1.34\times 10^{50}$ erg.

    The temperature rises to its peak of $\rm T_{\text{BB}} = 8806 \pm 269\,\text{K}$ during the first 15 days after the explosion due to the high UV contribution, and then decreases significantly to a minimum of $T_{\text{BB}} = 5339 \pm 82$ K at day $\sim 90$. 
    Afterwards, it remains almost constant, to then rise again at about 200 days, where the black-body assumption is no longer valid due to the onset of the nebular phase. 
    The photospheric radius grows almost linearly from $\rm R_{\text{BB}} = 1\times10^{15}\,\text{cm}$ during the first 90 days, reaching a peak of $R_{\text{BB}} = 3\times10^{15}$ cm, then recedes back to the starting value in the next hundred of days. 
    The corresponding value of photospheric velocity at $t_{\text{max}}$ is $v_\text{phot} = 5600\,\text{km}\,\text{s}^{-1}$. 
    This evolution is in line with the average for SLSNe-I reported by \citet{chen2023hydrogen} and \citet{gomez2024type}.

    \subsection{Diffusion time and mass of the ejecta}\label{sec:diffusion}

    The mass of the ejecta has been identified as a good diagnostic for distinguishing SLSNe from SE-SNe explosions. 
    On average, assuming similar opacities, SLSNe tend to eject masses that are 2 --3 times larger than those of SE-SNe \citep{Nicholl2015, chen2023hydrogen}, although there are a few exceptions \citep{karamehmetoglu2023population}. 
    Regardless of the powering mechanism behind the explosion, it is possible to evaluate the mass of the ejecta based on the diffusion time. 
    We define the diffusion time as $\tau_{\text{m}} \sim (\tau_{\text{rise}} + \tau_{\text{dec}})/2$, where $\tau_{\text{rise}} = t(L_{\text{peak}}/e)$ for $t<t_{\text{max}}$ and $\tau_{\text{dec}} = t(L_{\text{max}}/e)$ for $t>t_{\text{max}}$. 
    Subsequently, the mass of the ejecta can be expressed as a function of the width of the light curves, adapting the equations from \citet{kasen2010supernova} and \citet{Inserra2013}, by 
    \begin{equation}
        M_{\text{ej}} = 7.7 \times 10^{-7} \left(\frac{\kappa}{0.1 \text{cm}^2 \text{g}^{-1}}\right)^{-1} \frac{v}{\text{km s}^{-1}} \left(\frac{\tau_{\text{m}}}{\text{days}}\right)^2\, M_\odot 
    \label{eq:mej},
    \end{equation}
    where $\kappa$ is the opacity and $v$ is the "scale velocity", which is assumed as the expansion velocity of the ejecta. 
    Here, we use $v = 9800\,\text{km}\,\text{s}^{-1}$ (see Sect.~\ref{sec:Spectra} for the derivation of this parameter).  
    For SN~2024jlc we find $\tau_{\text{rise}} = 16.2$ days, $\tau_{\text{dec}} = 87.8$ days and consequently $\tau_{\text{m}} = 52.02$ days. 
    Alternatively, following \citet{chen2023hydrogen}, we define the rise time as the time interval between the peak luminosity and the time at which the luminosity reaches 10$\%$ of its peak value.
    In this case, we obtain $\tau_\text{rise, 10\%}$ = 42.9 days. 
    When compared to other events (e.g., \citealt{Nicholl2015, chen2023hydrogen}), SN~2024jlc rise time is well within the mean distribution found by \citet{chen2023hydrogen} for SLSNe-I, $41.9\pm17.8$ days, and faster than SE-SNe. 
    The decline time is larger than for most SLSNe, but still within the distribution, and significantly higher than any SE-SNe. 
    For comparison, as of \citet{Nicholl2015}, the SN Ic SN~2011bm has the longest decline, of 57.2 days. 
    Nevertheless, the diffusion time is consistent with other SLSN-I slow-evolving events, such as SN~2007bi (58.3 days), PTF12dam (55.2 days) and PS1-11ap (61.6 days).  
  
    Assuming $\kappa = 0.1$ cm$^{2}$ g$^{-1}$ we found $M_{\text{ej}} = 20.4^{+37.8}_{-18.4} \, M_\odot$. 
    Note that the uncertainty is obtained by applying Eq.~\ref{eq:mej} with $\tau_{\text{m}} = \tau_\text{dec}$ and $\tau_{\text{m}} = \tau_\text{rise}$ for the high and low error respectively.
    Finally, assuming the model by \citet{arnett1982}, we get the kinetic energy of the ejecta of $E_\text{k} = 1.17^{+3.0}_{-0.1} \times 10^{52}$ erg. 
    Note that additionally to the case with $\kappa = 0.1$ cm$^{2}$ g$^{-1}$, \citet{Nicholl2015} investigate the scenario with $\kappa = 0.2$ cm$^{2}$ g$^{-1}$. 
    In this case, the values of the mass of the ejecta and the kinetic energy are all halved and can be used as lower limits.
    
    Although accompanied with high uncertainty, this result suggests the massive nature of the progenitor of SN~2024jlc, yielding an ejected mass at the high end of the distribution shown by \citet{Nicholl2015}. 
    For instance, assuming a neutron star with $M_{\text{NS}}\ge 1.2\, M_\odot$ as the most probable outcome of the explosion, by assuming perfect conservation of mass and using the lower limit of the mass of the ejecta obtained with $\kappa = 0.2$ cm$^{2}$ g$^{-1}$, we set a conservative lower limit to the mass of the progenitor of $M_{\text{pro}} \ge 11\,M_\odot$.

    \subsection{Powering mechanism}\label{sec:model}

    \begin{figure}
    \begin{center}
    \includegraphics[width=1\columnwidth]{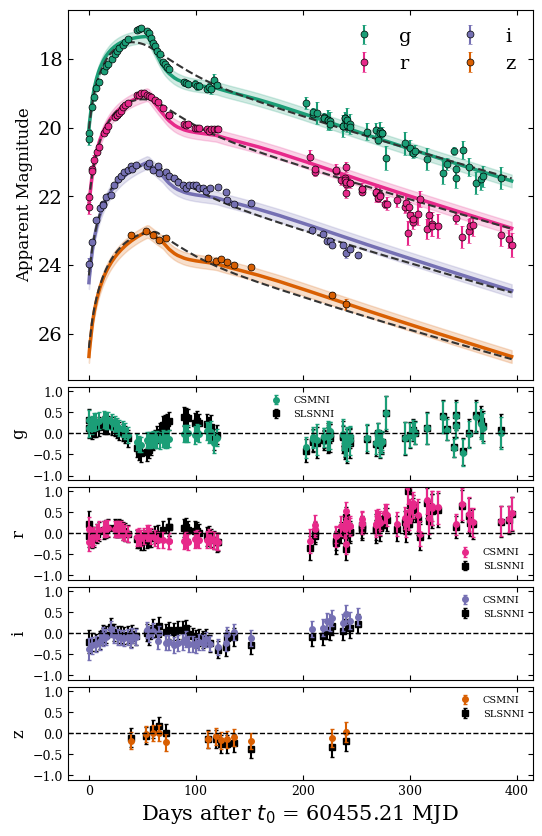}
    \caption{Main panel: Best fit of SN~2024jlc light curves (filled points, offset for clarity), using the \texttt{csmni} (solid color lines) and \texttt{slsnni} (dashed black lines) models from \texttt{MOSFiT}. 
    Lower panels: Normalized residuals for the two models. 
}
    \label{fig:mosfit}
    \end{center}
    \end{figure}

    To investigate the powering mechanism behind the evolution of SN~2024jlc, we used the \texttt{MOSFiT} software \citep{guillochon2018mosfit}, as it is largely used in the literature for SLSNe.     
    In particular, we tested the \texttt{slsnni} and the \texttt{csmni} models, as CSM-interaction and magnetar spin-down are the most common scenarios for SLSNe, and the shape of the light curves suggest the tails to be powered by $^{56}$Ni decay. 
    Results and priors for the parameters are reported in Table~\ref{table:mosfit_slsnni} and Table~\ref{tab:mosfit_csmni}. 
    In addition to the \texttt{MOSFiT} fit, we cross-checked our results for the magnetar model with \texttt{REDBACK} \citep{sarin2024redback}. 
    Results are reported in Table~\ref{tab:mosfit_csmni} and Table~\ref{table:mosfit_slsnni}, respectively.  
       
    Figure~\ref{fig:mosfit} shows the final results of the fit. 
    Note that, given the low cadence of UVOT data, we used only our $griz$ light curves.
    In both cases, we fixed the redshift, Galactic extinction and luminosity distance of the source, employing 5000 walkers to ensure a wide exploration of the parameter space. 
    In general, the two models reproduce the shape of the light curves well, with some caveats. 
    The \texttt{csmni} model reproduces the rise and steep decline, as well as the following flattening. 
    However, it fails in reproducing the $g$-band peak and the $r$- and $i$-bands late time decay. 
    On the other hand, the \texttt{slsnni} model does not reproduce the undulation at around $80$ days and the $g$-band rise, but yields a better fit of the $r$- and $i$- bands at late times.  
    A similar underestimation of the $g$-band peak reconstruction was also noted by \cite{chen2023hydrogenb}. 
    To assess which model yielded the best fit, we evaluated the reduced $\chi^2$ for each pairs of filters and for the global fit. 
    The \texttt{csmni} model yielded $\chi^2/\text{d.o.f.} = (0.71, 1.24, 1.27, 0.67)$ for the $g$, $r$, $i$, and $z$ bands respectively, while the \texttt{slsnni} model yielded $\chi^2/\text{d.o.f.} = (1.12, 0.90, 0.64, 11.78)$. 
    When all filters are combined, the \texttt{csmni} yielded $\chi^2_{\rm tot}/\text{d.o.f.} = 1.03$ and the \texttt{slsnni} $\chi^2_{\rm tot}/\text{d.o.f.} = 1.00$. 
    This shows that the two models describe the SN~2024jlc light curves, and no mechanism can be confidently ruled out. 

    Notably, we performed additional tests changing the prior parameters for the two models, and the presented realizations are those that yielded the lowest $\chi^2$ in both cases. 
    In particular, we mostly focused on the mass of the ejecta, CSM mass, the nickel fraction, the magnetic field, and the magnetar spin period. 
    We note that in both cases, better fits were achieved by minimizing the secondary power sources: \texttt{slsnni} preferring a weak magnetar field, while \texttt{csmni} preferring a low CSM envelope, leaning heavier on the nickel decay component. 
    In both cases, intermediate values of $M_{\rm ej}$ are preferred, with too high or too low values degrading the first 100 days; similarly, higher nickel fractions are necessary to reproduce the tails. 
    Specifically, in the \texttt{slsnni} model, no attempt resulted in a good reconstruction of the $g$-band peak and the undulation at $80$ days, regardless of the prior exploration.
    Conversely, in the \texttt{csmni} model, increasing the CSM mass resulted in an artificial plateau at early phases. 
    
    The overall behavior of the luminosity tail is reasonably well reproduced by the two models, although they deviate slightly from the observed decay rate. 
    In particular, the light curves exhibit a mean decline of about $0.013\,\text{mag}\,\text{d}^{-1}$ (with some variation expected due to outliers and scatter in the data), whereas the \texttt{csmni} and \texttt{slsnni} fits yield decay rates of 0.0097 and 0.0100  mag d$^{-1}$, respectively.
    This deviation may point to partial $\gamma$-ray escape from the ejecta, a scenario that is plausible in the context of high expansion velocities and large $^{56}$Ni fractions, as proposed by \citet{de2018light}.
    Assuming that the tail is completely powered by radioactive decay, we can use Arnett's model \citep{arnett1982} to estimate the synthesized nickel mass. 
    By fitting the luminosity tail, we obtained $M_{\text{Ni}} = 0.60 \pm 0.02\,M_\odot$, which corresponds to $f_{\text{Ni}} \sim 0.06 - 0.03$ depending on the opacity value. 
    This result is in the high end of the distribution by \citet{gomez2024type}, and indicates the relevant contribution from radioactive decay in powering the light curves.

\section{Spectral analysis}\label{sec:Spectra}

    We report spectroscopic observations of SN~2024jlc between 13 and 336 days after the explosion (see Fig.~\ref{fig:spectra}). 
    The early time spectra are characterized by a blue continuum and black-body temperatures of the order of $8000$ K, in good agreement with the photospheric behavior shown in Fig.~\ref{fig:phot}.
    This condition remains stable until $\sim$ day 57 (i.e., 8 days after maximum), when the black body temperature starts dropping rapidly, reaching $5000$ K in only ten days. 
    In this short time window, the blue continuum fades almost entirely, marking the transition from photospheric to nebular phase. 

    In the following, we discuss the discernible features following the work by \citet{quimby2018spectra}. 
    Then, we tentatively match these features with known chemical features and, in particular, we discuss possible helium detections, which would make SN~2024jlc one of the few SLSNe-Ib known to date.

    \subsection{Spectral features} 

    We investigate the spectral features of SN~2024jlc using the near-maximum spectrum taken at +48.71~d. 
    In the work by \citet{quimby2018spectra}, they divide SLSN-I spectra into two groups, based on the similarity with either SN~2011ke or PTF12dam. 
    In agreement with the definition of PTF12dam-like events, when centering the spectrum on the emission feature at around 4600~\AA, the emission feature at 5400~\AA\, is blue-shifted. 
    Broad emission features are clearly visible at $\sim$ 5000, 5800, 6100, 6500, 7800 and 8700~\AA\,, thanks to the good signal-to-noise ratio in the optical range. 
    The emission features at 3600 and 4100~\AA\,are less clearly discernible due to the low signal. 
    Other features that are more prominent in PTF12dam-like spectra are the absorption dips at 5700 and 7600~\AA. 
    Adding to those, we find clear features at $\sim$ 4750, 6200, 7000, 7350, 8300 and 9000~\AA. 
    The features at around 6500 \AA\, and 7350 \AA\, are telluric.
    Lastly, in general, the low signal-to-noise ratio in the bluest part of the spectra, does not allow for a clear identification of spectral features, although the smoothed spectra hint at possible features at $\sim$ 3600, 3900 and 4400 \AA. 

    We identify the ions responsible for the observed spectral features by comparing our spectra with those of other SLSNe-I, namely SN~2017egm \citep{zhu2023sn}, SN~2019hge \citep{yan2020helium} and PTF10hgi \citep{yan2020helium}. 
    Our first spectrum was taken at 12.51~d post-explosion, when the photospheric temperature was already at its peak, entering a plateau that lasted for about 40 days. 
    Consequently, the shape and the line intensity of all the early-time spectra taken during this time window remain very stable.
    The most discernible features, typical to SLSN objects, are the \ion{Ca}{ii} $\lambda\lambda$3934, 3969 H$\&$K doublet and the narrow (i.e., FWHM $\sim$ 2000-3000~km~s$^{-1}$) [\ion{Ca}{ii}] $\lambda$7300.  
    Less intense but still visible are the \ion{Fe}{ii} lines in the 4000-5000 $\AA$ range, the \ion{Na}{i} $\lambda$5890 and the near-IR \ion{Ca}{ii} $\lambda\lambda$8498, 8542, 8662 triplet. 
    Strong metal lines, broad and intense start appearing clearly from 18.76~d post maximum, when the temperature has almost reached its minimum at 5000 K. 
    The \ion{Ca}{ii} $\lambda\lambda$3934, 3969 H$\&$K  doublet remains one of the strongest features, together with the near-IR \ion{Ca}{ii} $\lambda\lambda$8498, 8542, 8662 triplet. 
    Other species are now clearly discernible due to the disappearing of the blue continuum, such us the \ion{Ca}{ii} $\lambda$3750 and \ion{Fe}{ii} $\lambda\lambda$4303, 4352, 5018, 5169, 5235, 5363. 
    The spectra exhibit also strong \ion{Mg}{i}] $\lambda$4571 broad emission lines, with velocities of the order of $10000\,\text{km}\,\text{s}^{-1}$ and [\ion{Ca}{ii}] $\lambda$7300 with similar velocities, although contaminated by telluric lines.    
    Similarly to SN~2017egm and SN~2019hge we note possible \ion{Na}{i} $\lambda$5890 and \ion{Si}{ii} $\lambda$6350 features.
    Finally, oxygen lines are clearly visible, in particular [\ion{O}{i}] $\lambda$6300 and \ion{O}{i} $\lambda$7774.
    Our last spectrum, at +286.89~d post-maximum, exhibits strong host-galaxy lines and broad metal features.

 \subsection{Helium lines}

    \begin{figure}
    \begin{center}
    \includegraphics[width=1\columnwidth]{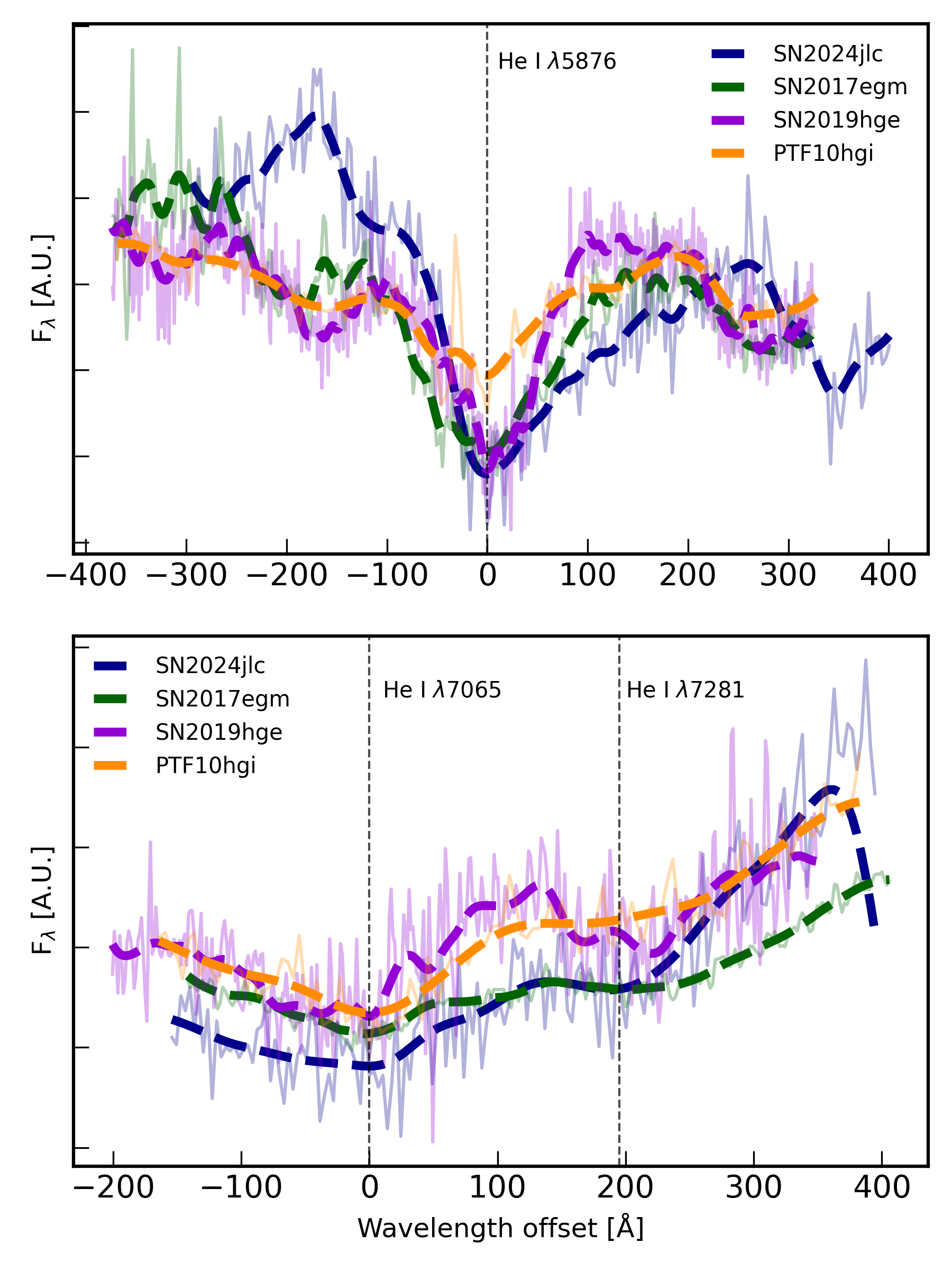}
    \caption{Zoom-in around the \ion{He}{i} $\lambda$5876 (upper panel) and \ion{He}{i} $\lambda\lambda$7065, 7281 (bottom panel) of different SLSNe-Ib. 
    All the spectra are centered on the rest-frame value of the line, normalized to the highest flux and continuum-corrected. 
    The normalization does not change between the two panels. 
    We show SN~2024jlc (71 days; this work), SN~2017egm (98 days; \protect\citealt{zhu2023sn}), SN~2019hge (67 days; \protect\citealt{yan2020helium}), PTF10hgi (105 days; \protect\citealt{yan2020helium}).
    The epochs are scaled to the epoch of maximum light. 
    } 
    \label{fig:helium} 
    \end{center}
    \end{figure} 

    Two independent analyses performed using different spectra before the peak classified SN~2024jlc first as a SN Ib \citep{Perez2024} and then as a SLSN-I similar to PTF10hgi \citep{Wise2024}. 
    This motivated a deeper search for helium features in our spectral dataset. 

    We identify \ion{He}{i} absorption features at $\lambda\lambda$5876, 6678, 7065, 7281 with a similar blue-shift velocity $\sim$ 9500\,km\,s$^{-1}$ at the time of the peak.
    The feature at $\lambda$5876 is the strongest of the four with a deep absorption dip. 
    It first appears in our spectra at 13 days post explosion, with a blueshift velocity of 15200\,km\,s$^{-1}$, and it is identified throughout the entire evolution up to 120 days after the explosion.
    We fit the evolution of the velocity of this line with a power-law, obtaining $v_\text{peak} = 9800 \pm 400\,\text{km}\,\text{s}^{-1}$ at $t_{\text{max}}$. 
    We assume this value as a reference velocity of the ejecta in our calculations. 

    The other three features are dimmer and not always above the noise level to be clearly identified, although they appear in several epochs.      
    A direct comparison between the \ion{He}{i} $\lambda\lambda$5876, 7065, 7281 features of SN~2024jlc, SN~2017egm, SN~2019hge and PTF10hgi is shown in Fig.~\ref{fig:helium}.
    Beyond a spectral template matching with prototypical events, previous SLSNe-Ib were classified based on two key factors: (i) the presence of multiple \ion{He}{i} features across different epochs (see, e.g., \citealt{yan2020helium, zhu2023sn}), and (ii) the identification of the near-IR \ion{He}{i} $\lambda$ 2.058 $\mu$m feature (see, e.g., \citealt{yan2020helium, kumar2025detection}), which is not contaminated by other elements.
    Despite the absence of near-IR data, that could reveal the \ion{He}{i} $\lambda\lambda$ 1.08, 2.05 $\mu m$ features \citep{kumar2025near, kumar2025detection}, the detection of optical helium lines in multiple epochs of SN~2024jlc is already a strong indication of the existence of helium in the ejecta, making this object a new candidate for the SLSN-Ib class. 

\subsection{Nebular phase} 

    \begin{figure}
    \begin{center}
    \includegraphics[width=1\columnwidth]{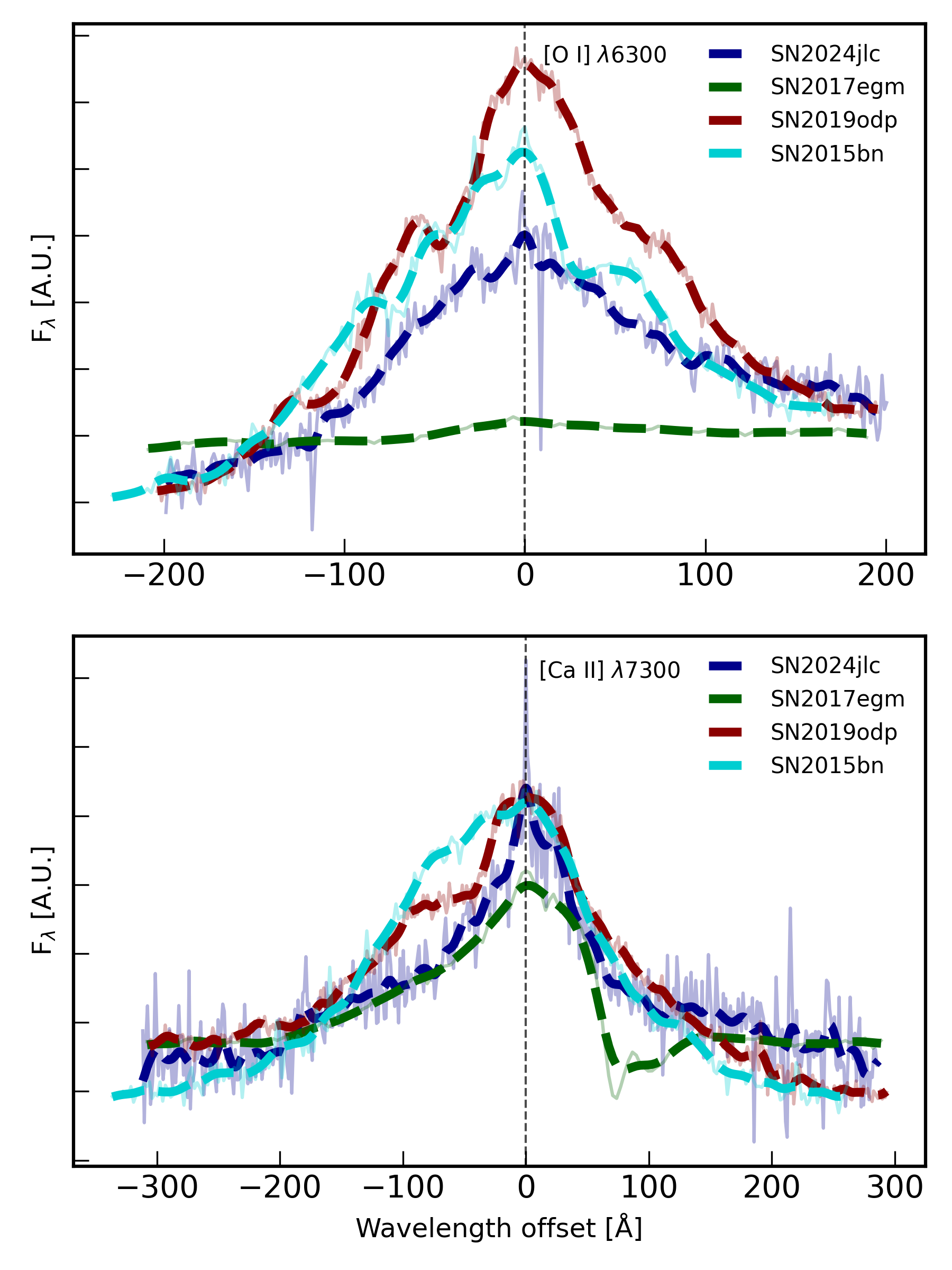}
    \caption{Zoom-in around the [\ion{O}{i}] $\lambda6300$ (upper panel) and [\ion{Ca}{ii}] $\lambda 7300$ (bottom panel) of different objects.
    All the spectra are centered on the rest-frame value of the line, normalized to the highest flux and continuum-corrected. 
    The normalization does not change between the two panels. 
    We show SN~2024jlc (286 days; this work), SN~2017egm (269 days; \protect\citealt{zhu2023sn}), SN~2019odp (138 days; \protect\citealt{schweyer2025sn}), SN~2015bn (350 days; \protect\citealt{nicholl2019nebular}). 
    The epochs are scaled to the epoch of maximum light. 
    }     
    \label{fig:nebular} 
    \end{center}
    \end{figure} 

    Nebular spectroscopy can offer a view on the explosion's core and reveal the powering source. 
    Nebular spectra can exhibit unique features as the innermost ejecta become directly observable.     
    We collected one nebular spectrum of SN~2024jlc at 286 days post maximum. 
    We found strong contribution from the [\ion{O}{i}] $\lambda 6300$ and [\ion{Ca}{ii}] $\lambda 7300$ doublets, as well as \ion{Mg}{i}] $\lambda 4571$ and \ion{O}{i} $\lambda 7774$ lines, as well as narrow host-galaxy features.
    Notably, the oxygen lines are much stronger than in the case of SN~2017egm and more similar to SLSN-I SN~2015bn \citep{nicholl2019nebular} and the Type Ic-BL SN~2019odp (\citealt{schweyer2025sn}; see Fig.~\ref{fig:nebular}). 
    We estimate a calcium to oxygen luminosity ratio of $L_{7300/6300} = 0.48 \pm 0.02$. 
    This ratio is sensitive to the core mass of the progenitor star, supporting the high-mass origin of the progenitor of SN~2024jlc \citep{karamehmetoglu2023population}. 
    Our result is compatible with a previous estimation by \citet{blanchard2025hydrogen}, who found $L_{7300/6300} = 0.5 \pm 0.01$ using a spectrum taken at 257 days post maximum. 
    For comparison, we estimate the same ratio with our second-last spectrum taken at 164 days post maximum, obtaining $L_{7300/6300} \approx 1$, although the transition to the nebular phase is not fully completed. 

    Another diagnostic useful for distinguishing between SLSNe and SE-SNe is the relative strength of the oxygen lines, quantified by the ratio $L_{L7774/6300}$.
    The oxygen line at $7774$~\AA\, is ubiquitously detected in the sample of SLSNe studied by \citet{blanchard2025hydrogen}, yielding $L_{L7774/6300}\ge 0.15$, whereas in SE-SNe this ratio is generally very small or consistent with zero. 
    We obtain $L_{L7774/6300} = 0.22 \pm 0.05$, consistent with the previous estimate of $L_{L7774/6300} = 0.24\pm0.01$ by \citet{blanchard2025hydrogen}.

\section{Multiwavelength SED}\label{sec:mwl}

\begin{figure}
    \begin{center}
    \includegraphics[width=1\columnwidth]{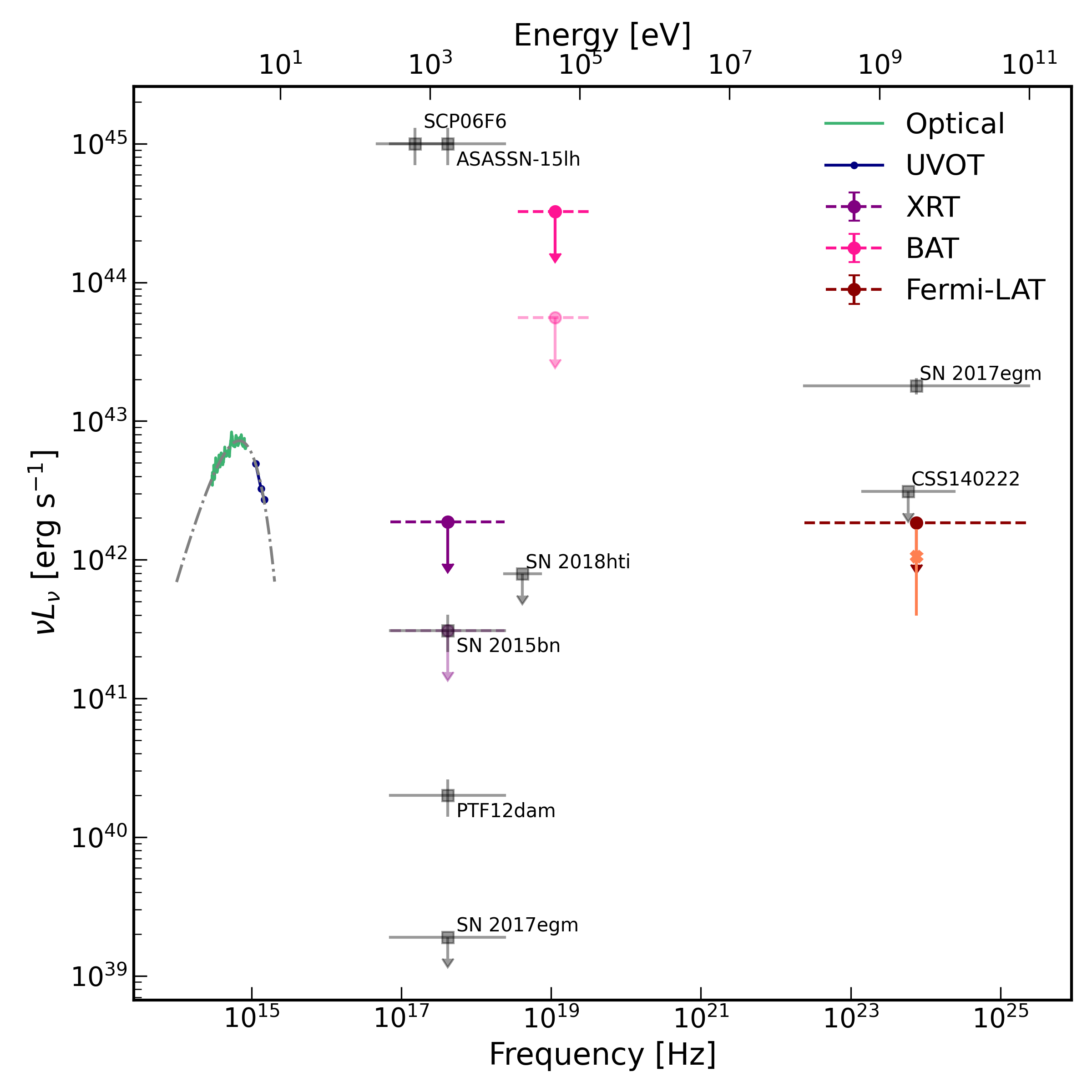}
    \caption{Luminosity SED of SN~2024jlc near maximum light, from optical to high-energy $\gamma$-rays. 
    A blackbody is fit to the optical+UV data. 
    For XRT and BAT, we plot the night-wise upper limit (solid line) and the integrated upper limit obtained by integrating over the entire observing period. 
    For \textit{Fermi}-LAT, we show the upper limit and the flux point as discussed in Sect.~\ref{sec:gamma}. 
    As a comparison, we plot the published fluxes or flux upper limits SCP06F6, ASASSN-15lh, SN~2018hti, PTF12dam, CSS140222 and SN~2017egm. 
    Reference data are adapted from \protect\citet{levan2013superluminous, margutti2017x, margutti2018results, renault2018search, andreoni2022hard, zhu2023sn, acero2026gamma}. 
    }
    \label{fig:sed}
    \end{center}
    \end{figure}

    Figure~\ref{fig:sed} shows the multiwavelength SED of SN~2024jlc at +47 days after explosion. 
    The SED includes data from optical to high-energy $\gamma$-rays, in the range $10^{15} - 10^{25}$ Hz. 
    We compare our XRT, BAT and \textit{Fermi}-LAT upper limits with published fluxes or flux upper limits of other SLSN-I events: SCP06F6, ASASSN-15lh, SN~2018hti, PTF12dam, CSS140222 and SN~2017egm \citep{levan2013superluminous, margutti2017x, margutti2018results, renault2018search, andreoni2022hard, zhu2023sn, acero2026gamma}.

    \subsection{X-rays: Constraints on the CSM density}

    Assuming that the IC is the only source of X-ray radiation at $t\sim t_{\text{max}}$, we can use the formulation from \citet{margutti2012inverse} and \citet{margutti2014relativistic} to constrain the density of the CSM using our XRT upper limits. 
    In this model, the IC luminosity depends on the density structure of the ejecta and the CSM, the relativistic electron distribution, the mass of the ejecta, the kinetic energy and the bolometric optical luminosity and, in particular, $L_{\text{IC}} \propto L_{\text{bol}}$
    We follow the same prescriptions as in \citet{margutti2018results}, assuming a wind-generated medium with $\rho_{\text{csm}}\sim R^{-2}$, a power-law electron distribution with $p=3$ and a fraction of post-shock energy into relativistic electrons of $\epsilon_e = 0.1$. 
    Combining the parameters estimated in this work with the XRT upper limit at +46 days, we constrain the mass-loss rate to wind velocity ratio of the progenitor star before the explosion to $\dot M / u_\text{w} \le 1.05\times10^{-2} \frac{M_\odot}{\text{yr}}\frac{\text{s}}{\text{km}}$ with $\kappa = 0.1 \,\text{cm}^2 \text{g}^{-1}$.  
    Similar results are obtained for each night of XRT observations. 
    Additionally, by defining $t_{\text{ref}}$ as the exposure weighted time of observation, equal to $\approx 61$ days, we can leverage the integral upper limit obtaining a final constrain of $\dot M / u_\text{w} \le 1.97\times10^{-3} \frac{M_\odot}{\text{yr}}\frac{\text{s}}{\text{km}}$ with $\kappa = 0.1 \,\text{cm}^2 \text{g}^{-1}$ and $\dot M / u_\text{w} \le 2.91\times10^{-3} \frac{M_\odot}{\text{yr}}\frac{\text{s}}{\text{km}}$ with $\kappa = 0.2\, \text{cm}^2 \text{g}^{-1}$.
    Unlike the case of SN~2015bn \citep{margutti2018results}, which has comparable explosion parameters and X-ray luminosity upper limits, our results do not rule out the possibility of a dense layer of CSM around the progenitor star, with the key difference being on the relative peak bolometric luminosity.  
    Notably, our upper limits constrain a region confined within $R_\text{s} \le 6\times10^{15}\,\text{cm}$ from the explosion site. 
    Assuming a steady wind with $u_\text{w} = 1000\,\text{km}\,\text{s}^{-1}$, this translates to material expelled $<$ 10 years before the explosion.

    \subsection{Gamma-rays: Hint of a signal, and implications for the powering model}\label{sec:gamma}

    To have a physically motivated time integration window, we followed the prescription of 
    \citet{crnogorvcevic2026gamma} to define the optimal search window for the GeV emission from SLSNe as [$t_{\rm min} = t_0 + 0.5 t_{\rm BH}(1+z)$, $t_{\rm max} = t_0 + 3 t_{\rm BH}(1+z)$], where 
    \begin{equation}
        t_{\rm BH} \sim 91 \left(\frac{M_{\rm ej}}{5 M_\odot}\right)^{1/2} \left(\frac{v_{\rm ej}}{6000 \,\rm km \, s^{-1}}\right)^{-1} \rm \, d 
    \end{equation}
    is the transparency time for pair-production absorption. 
    To define this time window, we adopted three sets of parameters for $M_{\rm ej}$ and $v_{\rm ej}$: $20.40\, M_\odot$ and $9800\,\rm km\,s^{-1}$, derived directly from model-independent analysis of the light curves and spectra (see Sect.~\ref{sec:diffusion} and Sect.~\ref{sec:Spectra}), $8.24\, M_\odot$ and $4815\,\rm km\,s^{-1}$, obtained from light-curve modeling under the assumption of a CSM+$^{56}$Ni powering mechanism (see Sect.~\ref{sec:model}), and $8.47\, M_\odot$ and $5380\,\rm km\,s^{-1}$ for the magnetar spin-down model.  
    In the first case, we obtained a time window of 58.5 -- 350.9 days after $t_0$ and a corresponding $\gamma$-ray $\rm TS = 4.9$ in the 100 MeV -- 100 GeV energy range (time window 1 in Fig.~\ref{fig:fermi-lat}). 
    For the model-dependent cases, we estimated a time window of 75.6 -- 453.8 and 68.6 -- 411.8 days after $t_0$ for the \texttt{csmni} and \texttt{slsnni} values, obtaining a $\gamma$-ray signal of $\rm TS = 12.9$ (time window 2 in Fig.~\ref{fig:fermi-lat}), and $\rm TS = 7.01$ (time window 3 in Fig.~\ref{fig:fermi-lat}), respectively. 
    Note that for one degree of freedom (power-law amplitude), these correspond to significance levels of $\sim 2.2\sigma$, $\sim 3.6\sigma$, and $\sim 2.6 \sigma$, respectively. 
    While no firm detection can be claimed, a significance of $\sim \, 3.6 \sigma$ represents an intriguing hint, also noted by \citet{crnogorvcevic2026gamma}\footnote{Notably, \citet{crnogorvcevic2026gamma} reported TS $\sim$ 7 in the range 50 -- 645 days. 
    However, as they also noted, this does not represent the optimal transparency window as it is not based on the definition of $t_{\rm BH}$.}.  
    For model comparison, we derive a 95\% confidence level integrated upper limit over the range 100 MeV -- 100 GeV of  $4.8 \times 10^{-13}$ erg cm$^{-2}$ s$^{-1}$ in the model dependent time window. 
    As a hint of signal is present in the data, we also report the flux level of this putative signal $2.7 ^{+1.2} _{-1.0} \times 10^{-13}$ erg cm$^{-2}$ s$^{-1}$. 

    At the distance of SN~2024jlc, this flux corresponds to a $\gamma$-ray luminosity of $L_\gamma = 1.06 ^{+0.50} _{-0.45} \times 10^{42}$ erg s$^{-1}$ (upper limit of $L_\gamma< 1.84 \times 10^{42}\,\rm erg \, s^{-1}$). 
    Following the prescription of \citet{crnogorvcevic2026gamma}, we define the GeV-to-optical efficiency $\eta$ as the ratio of $\gamma$-ray to bolometric optical luminosity at the time when the ejecta become transparent, $\eta \equiv L_\gamma/L_{\rm bol}(t=t_{\rm BH})$. 
    This yields $\eta=0.38$ (upper limit of $\eta<0.67$), providing important implications for the powering mechanism behind SN~2024jlc. 

    In the CSM-interaction scenario, the shock converts only a minor fraction of its kinetic energy into cosmic-rays, typically yielding $\eta\sim 10^{-2}-10^{-1}$, as observed in novae \citep{cheung2022fermi}. 
    Conversely, in a magnetar spin-down scenario, the efficiency can reach $\eta\sim1$ for weakly magnetized magnetar nebulae, where the spin-down power is converted equally between $\gamma$-rays and optical radiation \citep{vurm2021gamma}. 
    However, in a highly magnetized scenario, $\eta\ll1$ because a larger fraction of the energy is emitted via synchrotron radiation in the X-ray bands, which is subsequently reprocessed into the optical regime. 
    Among previous observations, only SN~2017egm yielded $\eta\sim 1$ \citep{acero2026gamma}, while \citet{crnogorvcevic2026gamma} constrained $\eta<1.3\times10^{-3}$ for a population of 223 SLSNe-I.  

    Our results indicate a comparable $\gamma$-ray and optical luminosity, consistent with the magnetar spin-down scenario. 
    This would also be the case even if the excess was not physical with the only caveat that our derived upper limit does not necessarily rule out either model.
    However, a difference of two orders of magnitude in efficiency is already a strong hint.  
    While these findings may suggest that the tentative $\gamma$-ray emission of SN~2024jlc is driven by magnetar spin-down, they do not exclude the coexistence of alternative mechanisms contributing to the optical and X-ray emission, as suggested by our spectrophotometric analysis.

    \section{Discussion}\label{sec:discussion}

\begin{figure*}[h!]
    \begin{center}
    \includegraphics[width=1\textwidth]{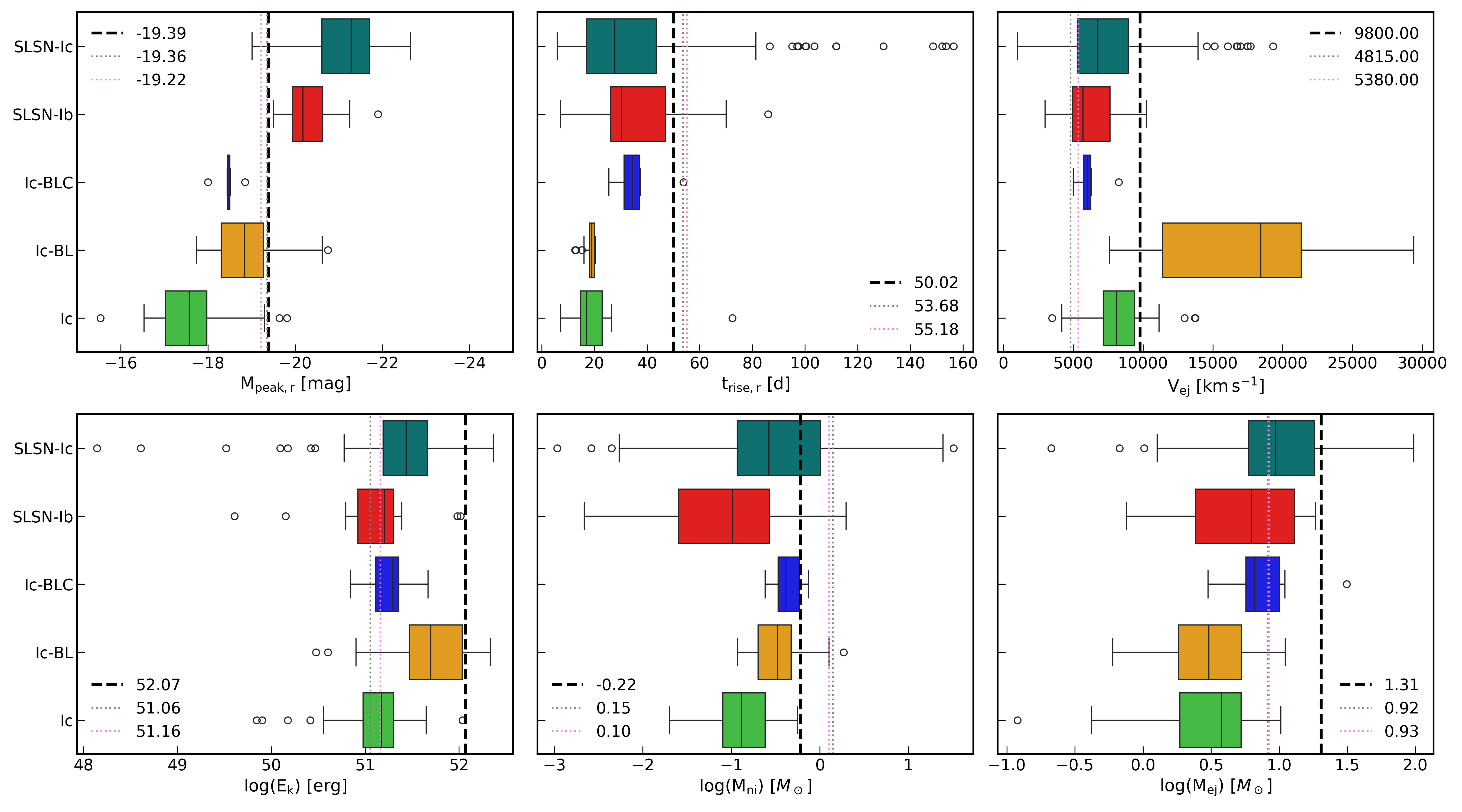}
    \caption{Comparison of the SN~2024jlc parameters with the distribution of the five samples, SLSN-Ic (teal), SLSN-Ib (red), Type Ic with broad light curves (blue), Type Ic-BL (orange), and Type Ic (green). 
    We used three sets of SN~2024jlc parameters: model-independent (dashed black vertical lines), and the results from \texttt{MOSFiT} for the \texttt{csmni} (dotted gray vertical lines) and the \texttt{slsnni} (dotted purple vertical lines) models. In the last two cases, the r-band peak magnitude and rise time are directly extracted from the fitted curve of Fig.~\ref{fig:mosfit} for ensure self-consistency. 
    The imperfect reproduction of the peak (especially in the \texttt{slsnni} case) may weight on the final results. 
    }
    \label{fig:parameters}
    \end{center}
    \end{figure*}

    We now summarize and discuss the observed properties of SN~2024jlc in a broader physical context. 
    SN~2024jlc shares many similarities with normal SLSNe, although its relatively low luminosity and the detection of helium, set it apart, rising the question whether SN~2024jlc is rather a SE-SN of some kind. 
    Quantitatively, we compare SN~2024jlc parameters with different samples \citep{taddia2015early, taddia2019analysis, Barbarino2021, gomez2022luminous, karamehmetoglu2023population, gomez2024type} and show the results in Fig.~\ref{fig:parameters} and Fig.~\ref{fig:total_comparison}.

    \subsection{Similarities with SLSNe}

    SN 2024jlc appears to be a good match to the SLSN-Ib class.  
    Its light curves are broad, characterized by a high diffusion time and large ejecta masses. 
    CSM+$^{56}$Ni and magnetar spin-down powering models both provide a good fit to its light curves. 
    The first scenario naturally explains the event's atypically low luminosity and is supported by the absence of the W-shaped features typically seen in other SLSNe, although the second scenario, with a slow magnetar spin-down, may as well explain the low luminosity.  
    Finally, the spectra evolve from a featureless blue-continuum dominated to metal-rich, with noticeable \ion{He}{i} features throughout the entire evolution.
    The nebular spectrum at 286 days post maximum exhibits strong [\ion{O}{i}] and [\ion{Ca}{ii}] lines, and we estimated $L_{7300/6300} = 0.48 \pm 0.02$. 
    This value is compatible with a previous estimation on a spectrum at 257 days performed by \citet{blanchard2025hydrogen}. 
    In their work, they collected the biggest sample of nebular spectra of SLSNe-I to date. 
    They found that the $L_{7300/6300}$ ratio is a good proxy for the level of ionization and that SLSNe-I have values spanning from $\approx 0.2$ to $\approx 6.2$ with a mean of $\approx 1.4$, with a decreasing trend over time that leads all events, apart from three outliers, to have a ratio below unity. 
    They found that the general trend of SLSNe-I differs from SE-SNe, which show ratios below unity at all times, and that the difference may be pointing to the presence of a magnetar engine. 
    Finally, \citet{blanchard2025hydrogen} argued that the origin of the outliers, which exhibit an increasing level of ionization with time, may be caused by late-time interaction with CSM. 

    All observed properties of SN~2024jlc are consistent with the physical picture proposed by \citet{yan2020helium} and \citet{chen2023hydrogenb}, which can naturally explain the light curve shape and the presence of helium in the ejecta.
    In this interpretation, the progenitor star shed its hydrogen envelope long before explosion, preventing hydrogen signatures from appearing in the spectra, as seen in other stripped-envelope SNe. 
    Helium, however, remained as a thin outer layer, and part of it was expelled shortly before core collapse, producing a denser He-rich CSM close to the star, surrounded by lower density regions.     
    This would also explain the absence of narrow spectral lines, as these would disappear fast or, alternatively, 
    these are suppressed by high temperatures that highly ionize the gas at early times.
    In this scenario, the excitation energy necessary to ionize helium would be reached by the classical transport of $^{56}$Ni, from the interior to the outer parts of ejecta. 
    
    \citet{chatzopoulos2013analytical} argued that instead of early-times narrow lines, the emission due to CSM interaction with a hydrogen-poor medium may appear as a late-time blue continuum or, as argued by \citet{blanchard2025hydrogen}, by a high $L_{7300/6300}$ ratio. 
    Given the absence of the two features, we exclude that SN~2024jlc light curves are powered by CSM interaction at late times. 

    \subsection{Similarities with SE-SNe}

    Among the (i)PTF sample of 220 SE-SN, \citet{karamehmetoglu2023population} estimated that a fraction of 13 per cent exhibits broad light curves, indicating large ejecta masses and high-mass progenitor stars. 
    They presented a sample of 8 Type Ibc SNe \footnote{Namely: PTF09dfk, PTF10inj, PTF11bov, PTF11mnb, PTF11rka, iPTF15dtg, iPTF16flq, iPTF16hgp.} with shared properties that are uncommon to the rest of the population. 
    These SNe exhibit rise times of the order of $\sim$ 40 days, higher than any other in the sample of Type Ibc events, and signs of undulations and/or multi-peak behavior. 
    They are all connected with $M_\text{ej}\gg 5 M_\odot$, while all previous Type Ibc samples had $1-5\,M_\odot$ \citep{taddia2019analysis, Barbarino2021}. 
    Similarly, they all show higher nickel masses, with an average of $M_{\text{Ni}} = 0.42 \pm 0.08 \,M_\odot$, more than double than the typical value of $0.2 \pm \, 0.14 \, M_\odot$. 
    Spectroscopically, they show typical \ion{Fe}{ii}, \ion{O}{i} and \ion{Ca}{ii} lines, with velocities of the order of $10^4\,\text{km}\,\text{s}^{-1}$. 
    The Type Ib SNe of the sample exhibit helium with high velocities, possibly blended with \ion{Na}{iD} at $\lambda$5876, whereas the Type Ic show \ion{Si}{ii} and no signs of helium features. 
    Their nebular spectra show strong [\ion{O}{i}] $\lambda\lambda$6300, 6364 and [\ion{Ca}{ii}] $\lambda\lambda$7319, 7324, with a relative luminosity ratio of $L_{L7774/6300} \le 0.81$, indicating high-mass progenitors. 
    Finally, they tend to prefer lower mass galaxies ($\le 10^{10}M_\odot$) with higher sSFR then normal SE-SNe, and more similar to the host galaxies of SLSNe-I and Type Ic-BL, and lower metallicity than 90 per cent of SE-SNe. 

    Given the strong similarities, it is natural to discuss whether SN~2024jlc should belong to this group of object. 
    The many similarities demonstrate a similar host environment and explosion mechanism and similar progenitor stars.
    However, a strong limiting factor is given by the difference in luminosity. 
    SN~2024jlc is $\sim$ 1 magnitude more luminous in every filter than the sample, which does not exhibit much scatter between single objects. 
    Classifying SN~2024jlc as a Type Ibc with broad light curves would make it the most luminous SN of the class, with a difference greater than the one between SN~2024jlc and the least luminous SLSN-I. 
    This may indicate that SLSNe-Ib are linked to broad SE-SNe, and the main difference could be found in the different powering mechanism: magnetar spin-down or CSM interaction with nickel decay in the first case and classical radioactive decay in the second. 

    Another possibility would be to link SN~2024jlc to the Type Ic-BL SNe, as they are the most luminous SE-SNe known, with some cases reaching even higher luminosities than SN~2024jlc \citep{taddia2019analysis}. 
    In this case, however, the key limiting factor would be the broadness of the light curves. 
    As also noted by \citet{chen2023hydrogen}, SNe Ic-BL exhibit narrower light curves, with a rise time lower than $\tau_{\text{rise},\, 10\%}\sim$ 25 days. 
    The brightest event from the sample of \citet{taddia2019analysis} shows a peak luminosity of $M_{\text{g}} \sim -19.4$ (after color correction by \citealt{chen2023hydrogen}), but a rise time of only 5 days. 
    In this case, SN~2024jlc would not only represent one of the brightest Type Ic-BL ever discovered, but also the one with the broadest light curves.

    \subsection{Statistical comparison}

    We performed random forest classification with \texttt{scikit-learn v1.9} \citep{Pendragosa2011} to quantify the probability of SN~2024jlc to belong to a certain class given the estimated value of its parameters. 
    As training samples, we used the data shown in Fig.~\ref{fig:parameters} and Fig.~\ref{fig:total_comparison}, focusing on the parameters that are common to every class: mass and velocity of the ejecta, kinetic energy, rise time, r-band peak magnitude, and mass of nickel. 
    In total, the samples are constituted of 254 SLSNe-Ic \citep{gomez2024type}, 44 Type Ic \citep{taddia2015early, Barbarino2021}, 30 Type Ic-BL \citep{taddia2015early, taddia2019analysis}, 16 SLSNe-Ib \citep{gomez2022luminous, gomez2024type, kumar2025detection, kumar2026sn, farah2026lense}, and 8 Type Ic with broad light curves (Ic-BLC; \citealt{Barbarino2021, karamehmetoglu2023population}). 
    From these, we removed the SNe with missing parameters, for a total of 5 Type Ic and 2 Type Ic-BLC. 
    To ensure a unbiased exploration towards the largely more populated class of SLSNe-Ic, we set \texttt{class\_weight='balanced'}, and we ran the classifier for a total of 5000 estimators. 
    We performed the classification for set of parameters of SN~2024jlc: first, using the model-independent values from Sect.~\ref{sec:diffusion}, and then using the results from the \texttt{csmni} and \texttt{slsnni} models reported in Table~\ref{tab:mosfit_csmni} and Table~\ref{table:mosfit_slsnni}. 

    In all cases, SLSN-Ic is the class with the highest probability ($\sim 40\%$), followed by SLSN-Ib ($\sim$ 20\%) and Ic-BLC ($\sim$ 25\%). 
    Type Ic  and Type Ic-BL ($\le 10\%$) are the least probable classes.
    When we excluded the population of SLSNe-Ic from the random forest exploration entirely, we obtained similar results, and the distance between SLSN-Ib and Ic-BLC from Types Ic and Ic-BL increased, as shown in Table~\ref{tab:classification}. 
    In particular, when model-dependent parameters were used, the probability associated with Type Ic-BLC increased to nearly $50\%$. 

    \begin{table*}[ht]
        \centering
        \caption{Results of the random forest classification.}
        \begin{tabular}{l|ccccc}
        \toprule 
        \toprule 
                        & SLSN-Ic & SLSN-Ib & Ic-BLC & Ic-BL & Ic \\ 
        \midrule
        Model independent (\%)      & 36.81 & 23.66  & 14.98 & 13.79 & 10.77  \\ 
        \texttt{csmni} (\%)   & 39.57 & 21.71  & 26.34 & 3.36  &  9.02 \\ 
        \texttt{slsnni} (\%)  & 34.25 & 19.70 & 34.20  & 3.09  & 8.77 \\ 
        \midrule 
        Model independent (\%) & -- & 35.69 & 30.97 & 21.68 & 11.67 \\ 
        \texttt{csmni} (\%)   & -- & 34.10 & 51.71 & 3.73  & 10.46 \\ 
        \texttt{slsnni} (\%)  & -- & 29.82 & 56.71 & 3.54  & 9.93 \\ 
        \bottomrule
        \end{tabular}
        \tablefoot{The probabilities were estimated for two sets of SN~2024jlc parameters: model independent refers to the estimations performed in Sect.~\ref{sec:diffusion}, while \texttt{csmni} and \texttt{slsnni} refer to the parameters estimated using \texttt{MOSFiT} shown in Table~\ref{tab:mosfit_csmni} and Table~\ref{table:mosfit_slsnni}. 
        The second block represents the case without the SLSN-Ic sample. }
        \label{tab:classification}
    \end{table*}

    The parameters that weight the most are the r-band peak magnitude ($\sim$ 35\%), and the rise time ($\sim 18$\%), which set the main difference between SE-SNe and SLSNe. 
    The velocity of the ejecta ($\sim$ 18\%) weights mostly in downgrading the Ic-BL class, while the mass of nickel ($\sim$ 10\%), kinetic energy ($\sim$ 9\%) and mass of the ejecta ($\sim$ 7.83\%) have only a relative weight. 

    The lack of statistically well populated samples prevent us from giving a definitive conclusion, although these results may already represent a strong indication that SN~2024jlc is bridging SLSNe and SE-SNe. 
    In all six realizations, the net separation between the SLSN and Ic-BLC samples from the normal SE-SNe ones represent a strong indication that SN~2024jlc is not a standard CCSN, while the boundary between SLSNe and Ic-BLC is less evident, if not for the very different magnitude peak.

\section{Conclusions}\label{sec:conclusions}

    We have presented a multiwavelength analysis of SN~2024jlc, the closest SLSN-I discovered in recent years, and one of the closest SLSN-I ever reported. 
    We included optical photometry and spectroscopy, covering the first year and a half after the explosion. 
    Additionally, we estimated upper limits on the X-ray flux and reported a possible hint of a $\gamma$-ray signal. 
    We highlight our major findings below. 
    \begin{enumerate}[i.]
    \item SN~2024jlc is one of the least luminous SLSNe discovered to date, with a peak $g$-band absolute magnitude of $-19.37\pm0.22$. 
    \item The evolution of the SN~2024jlc light curves is well fitted by the CSM+$^{56}$Ni and magnetar spin-down models.
    \item From a model-independent analysis, we found $M_{\text{ej}} \approx 20\, M_\odot$, $E_\text{k} \approx 10^{52}$ erg and $M_{\text{Ni}} \approx 0.6\, M_\odot$.
    \item The SN~2024jlc spectra exhibit a blue continuum until $\sim$ 12 days after maximum, followed by a steep decrease in temperature and the emergence of strong metal features. 
    \item The SN~2024jlc spectra exhibit features of \ion{He}{i} at 5876 \AA\,, 6678 \AA\,, 7065 \AA\,, and 7281 \AA. 
    The average expansion velocity at the peak luminosity is $9800\pm400$ km s$^{-1}$, in agreement with typical values for this class of objects. 
    \item From XRT and BAT observations, we found $L_{0.3-10 \,\text{keV}} \le  3.36 \times 10^{41}$ erg s$^{-1}$ and $L_{15-150 \,\text{keV}} \le  6.01 \times 10^{43}$ erg s$^{-1}$. 
    These upper limits were used to constrain the ratio of the mass loss to wind velocity to $<$ 10 years before the explosion to $\dot M / u_\text{w} \le 1.97\times10^{-3} \frac{M_\odot}{\text{yr}}\frac{\text{s}}{\text{km}}$. 
    \item From \textit{Fermi}-LAT observations, we found an intriguing hint of a signal at a TS = 12.9 ($\sim 3.6 \sigma$ level), in line with previous results from \citet{crnogorvcevic2026gamma}. 
    This confirms that SN~2024jlc is the SLSN with the highest TS ever found after SN~2017egm.
    \item From \textit{Fermi}-LAT observations, we found $L_{\gamma} = 1.06^{+0.50}_{-0.45}\times10^{42}\,\rm erg\, s^{-1}$ (upper limit of $L_{\gamma} < 1.84\times10^{42}\,\rm erg\, s^{-1}$), corresponding to a GeV-to-optical efficiency of $\eta = 0.38$ (upper limit of $\eta<0.67$). 
    \end{enumerate}

    Our spectrophotometric analysis indicates that SN~2024jlc belongs to the class of SLSNe-Ib, which are characterized by low peak luminosities, distinguishable helium features, and noticeable $^{56}$Ni contribution. 
    The light curves are well-fitted by CSM-interaction and magnetar spin-down models, although the spectra do not exhibit narrow features (which would support the first mechanism) or W-shaped features (which would support the second mechanism). 
    Similarly to other events of the same class, a possible explanation would be that of a dense He-rich CSM region close to the star, surrounded by low-density regions.    
    Our X-rays upper limits are consistent with this scenario and agree with previous results. 

    The strongest evidence in favor of a central-engine powering mechanism comes from our \textit{Fermi}-LAT analysis.  
    If the observed excess is real, it might indicate multiple powering mechanisms, similarly to the case of SN~2017egm, for which spectrophotometric analyses suggested CSM interaction \citep{zhu2023sn, 2023NatAs...7..779L}, while the $\gamma$-ray detection supports a magnetar spin-down acceleration \citep{acero2026gamma}. 

    Although it is statistically poorly populated so far, the SLSN-Ib class might represent a bridge between the classical SE-SNe events and SLSNe, with the key difference being the powering mechanism. 
    The observed similarities between the two populations suggest similar progenitors in similar environments. The physics behind the emission is left as a valuable candidate for explaining the intrinsic difference between the two classes. 
    It remains an open question whether these objects are simply very luminous Type Ic or Ic-BL events, and only future observations will shed light on this.
    The upcoming Vera C. Rubin Observatory’s Legacy Survey of Space and Time (LSST; \citealt{Ivezic2019}), with its unprecedented depth and coverage, will revolutionize this field. 
    Started in early 2026, LSST is expected to discover over 200 000 SLSNe during the nominal 10 years of operation, offering the statistical power necessary to better understand the full diversity of SLSNe and to populate the gap between Type Ic and SLSNe.  
    Objects such as SN~2024jlc will lie in the sweet spot of distance and luminosity to be precisely characterized by LSST \citep{simongini2025b}. 
    As the sample grows, events of this type will be key benchmarks for understanding the physical pathways that link ordinary core-collapse explosions to the most luminous stellar deaths. 

\subsubsection*{Data availability} 

Table~\ref{tab:photometry} is only available in electronic form at the CDS via anonymous ftp to cdsarc.u-strasbg.fr (130.79.128.5) or via http://cdsweb.u-strasbg.fr/cgi-bin/qcat?J/A+A/.
Data used for Fig.~\ref{fig:spectra} are available on \texttt{WISeREP} \citep{Yaron2012}.

\begin{acknowledgements}
We thank A. Segreto for his contribution in the BAT analysis and data reduction. 
Based on observations obtained with the Samuel Oschin Telescope 48-inch and the 60-inch Telescope at the Palomar Observatory as part of the Zwicky Transient Facility project. ZTF is supported by the National Science Foundation under Grant No. AST-2034437 and a collaboration including Caltech, IPAC, the Weizmann Institute of Science, the Oskar Klein Center at Stockholm University, the University of Maryland, Deutsches Elektronen-Synchrotron and Humboldt University, the TANGO Consortium of Taiwan, the University of Wisconsin at Milwaukee, Trinity College Dublin, Lawrence Livermore National Laboratories, and IN2P3, France. Operations are conducted by COO, IPAC, and UW.
SED Machine is based upon work supported by the National Science Foundation under Grant No. 1106171
The ZTF forced-photometry service was funded under the Heising-Simons Foundation grant $\#$12540303 (PI: Graham). 
Observations reported here were obtained at the MMT Observatory, a joint facility of the Smithsonian Institution and the University of Arizona.
The data presented here were obtained [in part] with ALFOSC, which is provided by the Instituto de Astrofisica de Andalucia (IAA) under a joint agreement with the University of Copenhagen and NOT.
We acknowledge the use of public data from the \textit{Swift} data archive (ID 16706, P.I. T. Moore). 
This research has made use of data and/or software provided by the High Energy Astrophysics Science Archive Research Center (HEASARC), which is a service of the Astrophysics Science Division at NASA/GSFC. 
The Pan-STARRS1 Surveys (PS1) and the PS1 public science archive have been made possible through contributions by the Institute for Astronomy, the University of Hawaii, the Pan-STARRS Project Office, the Max-Planck Society and its participating institutes, the Max Planck Institute for Astronomy, Heidelberg and the Max Planck Institute for Extraterrestrial Physics, Garching, The Johns Hopkins University, Durham University, the University of Edinburgh, the Queen's University Belfast, the Harvard-Smithsonian Center for Astrophysics, the Las Cumbres Observatory Global Telescope Network Incorporated, the National Central University of Taiwan, the Space Telescope Science Institute, the National Aeronautics and Space Administration under Grant No. NNX08AR22G issued through the Planetary Science Division of the NASA Science Mission Directorate, the National Science Foundation Grant No. AST-1238877, the University of Maryland, Eotvos Lorand University (ELTE), the Los Alamos National Laboratory, and the Gordon and Betty Moore Foundation.
N.R. is funded by a Northwestern University Presidential Fellowship. N.R.~is also partially supported by NSF grant \# AST-2421845. Zwicky Transient Facility access for N.R. was supported by Northwestern University and the Center for Interdisciplinary Exploration and Research in Astrophysics (CIERA).
This work made use of the Astro-COLIBRI platform \citep{Reichherzer2021}.  
\\ 
FA acknowledges financial support from the Centre national d'études spatiales (CNES), France (ROR: \url{https://ror.org/04h1h0y33}) within the framework of the Fermi mission.

\end{acknowledgements}
\bibliographystyle{aa}
\bibliography{Silmarillion}

@article{Simongini2024,
    author = {Simongini, Andrea and Ragosta, F and Piranomonte, S and Di Palma, I},
    title = "{Building spectral templates and reconstructing parameters for core-collapse supernovae with CASTOR}",
    journal = {\mnras},
    volume = {533},
    number = {3},
    pages = {3053-3067},
    year = {2024},
    month = {08},
    issn = {0035-8711},
    doi = {10.1093/mnras/stae1911},
    url = {https://doi.org/10.1093/mnras/stae1911},
    eprint = {https://academic.oup.com/mnras/article-pdf/533/3/3053/58985332/stae1911.pdf},
}

@ARTICLE{2020A&A...641A...6P,
       author = {{Planck Collaboration} and {Aghanim}, N. and {Akrami}, Y. and {Ashdown}, M. and {Aumont}, J. and {Baccigalupi}, C. and {Ballardini}, M. and {Banday}, A.~J. and {Barreiro}, R.~B. and {Bartolo}, N. and {Basak}, S. and {Battye}, R. and {Benabed}, K. and {Bernard}, J.-P. and {Bersanelli}, M. and {Bielewicz}, P. and {Bock}, J.~J. and {Bond}, J.~R. and {Borrill}, J. and {Bouchet}, F.~R. and {Boulanger}, F. and {Bucher}, M. and {Burigana}, C. and {Butler}, R.~C. and {Calabrese}, E. and {Cardoso}, J.-F. and {Carron}, J. and {Challinor}, A. and {Chiang}, H.~C. and {Chluba}, J. and {Colombo}, L.~P.~L. and {Combet}, C. and {Contreras}, D. and {Crill}, B.~P. and {Cuttaia}, F. and {de Bernardis}, P. and {de Zotti}, G. and {Delabrouille}, J. and {Delouis}, J.-M. and {Di Valentino}, E. and {Diego}, J.~M. and {Dor{\'e}}, O. and {Douspis}, M. and {Ducout}, A. and {Dupac}, X. and {Dusini}, S. and {Efstathiou}, G. and {Elsner}, F. and {En{\ss}lin}, T.~A. and {Eriksen}, H.~K. and {Fantaye}, Y. and {Farhang}, M. and {Fergusson}, J. and {Fernandez-Cobos}, R. and {Finelli}, F. and {Forastieri}, F. and {Frailis}, M. and {Fraisse}, A.~A. and {Franceschi}, E. and {Frolov}, A. and {Galeotta}, S. and {Galli}, S. and {Ganga}, K. and {G{\'e}nova-Santos}, R.~T. and {Gerbino}, M. and {Ghosh}, T. and {Gonz{\'a}lez-Nuevo}, J. and {G{\'o}rski}, K.~M. and {Gratton}, S. and {Gruppuso}, A. and {Gudmundsson}, J.~E. and {Hamann}, J. and {Handley}, W. and {Hansen}, F.~K. and {Herranz}, D. and {Hildebrandt}, S.~R. and {Hivon}, E. and {Huang}, Z. and {Jaffe}, A.~H. and {Jones}, W.~C. and {Karakci}, A. and {Keih{\"a}nen}, E. and {Keskitalo}, R. and {Kiiveri}, K. and {Kim}, J. and {Kisner}, T.~S. and {Knox}, L. and {Krachmalnicoff}, N. and {Kunz}, M. and {Kurki-Suonio}, H. and {Lagache}, G. and {Lamarre}, J.-M. and {Lasenby}, A. and {Lattanzi}, M. and {Lawrence}, C.~R. and {Le Jeune}, M. and {Lemos}, P. and {Lesgourgues}, J. and {Levrier}, F. and {Lewis}, A. and {Liguori}, M. and {Lilje}, P.~B. and {Lilley}, M. and {Lindholm}, V. and {L{\'o}pez-Caniego}, M. and {Lubin}, P.~M. and {Ma}, Y.-Z. and {Mac{\'\i}as-P{\'e}rez}, J.~F. and {Maggio}, G. and {Maino}, D. and {Mandolesi}, N. and {Mangilli}, A. and {Marcos-Caballero}, A. and {Maris}, M. and {Martin}, P.~G. and {Martinelli}, M. and {Mart{\'\i}nez-Gonz{\'a}lez}, E. and {Matarrese}, S. and {Mauri}, N. and {McEwen}, J.~D. and {Meinhold}, P.~R. and {Melchiorri}, A. and {Mennella}, A. and {Migliaccio}, M. and {Millea}, M. and {Mitra}, S. and {Miville-Desch{\^e}nes}, M.-A. and {Molinari}, D. and {Montier}, L. and {Morgante}, G. and {Moss}, A. and {Natoli}, P. and {N{\o}rgaard-Nielsen}, H.~U. and {Pagano}, L. and {Paoletti}, D. and {Partridge}, B. and {Patanchon}, G. and {Peiris}, H.~V. and {Perrotta}, F. and {Pettorino}, V. and {Piacentini}, F. and {Polastri}, L. and {Polenta}, G. and {Puget}, J.-L. and {Rachen}, J.~P. and {Reinecke}, M. and {Remazeilles}, M. and {Renzi}, A. and {Rocha}, G. and {Rosset}, C. and {Roudier}, G. and {Rubi{\~n}o-Mart{\'\i}n}, J.~A. and {Ruiz-Granados}, B. and {Salvati}, L. and {Sandri}, M. and {Savelainen}, M. and {Scott}, D. and {Shellard}, E.~P.~S. and {Sirignano}, C. and {Sirri}, G. and {Spencer}, L.~D. and {Sunyaev}, R. and {Suur-Uski}, A.-S. and {Tauber}, J.~A. and {Tavagnacco}, D. and {Tenti}, M. and {Toffolatti}, L. and {Tomasi}, M. and {Trombetti}, T. and {Valenziano}, L. and {Valiviita}, J. and {Van Tent}, B. and {Vibert}, L. and {Vielva}, P. and {Villa}, F. and {Vittorio}, N. and {Wandelt}, B.~D. and {Wehus}, I.~K. and {White}, M. and {White}, S.~D.~M. and {Zacchei}, A. and {Zonca}, A.},
        title = "{Planck 2018 results. VI. Cosmological parameters}",
      journal = {\aap},
         year = 2020,
        month = sep,
       volume = {641},
          eid = {A6},
        pages = {A6},
          doi = {10.1051/0004-6361/201833910},
archivePrefix = {arXiv},
       eprint = {1807.06209},
 primaryClass = {astro-ph.CO},
       adsurl = {https://ui.adsabs.harvard.edu/abs/2020A&A...641A...6P}
}

@inproceedings{Steele2004,
  title={The Liverpool Telescope: performance and first results},
  author={Steele, Iain A and Smith, Robert J and Rees, Paul C and Baker, Ian P and Bates, SD and Bode, Michael F and Bowman, Mark K and Carter, Dave and Etherton, Jason and Ford, Martyn J and others},
  booktitle={Ground-based Telescopes},
  volume={5489},
  pages={679--692},
  year={2004},
  organization={SPIE}
}

@ARTICLE{Sollerman2024,
       author = {{Sollerman}, J. and {Fremling}, C. and {Perley}, D. and {Laz}, T.~D.},
        title = "{ZTF Transient Discovery Report for 2024-05-28}",
      journal = {Transient Name Server Discovery Report},
         year = 2024,
        month = may,
       volume = {2024-1681},
        pages = {1},
       adsurl = {https://ui.adsabs.harvard.edu/abs/2024TNSTR1681....1S}
}

@ARTICLE{Perez2024,
       author = {{P{\'e}rez-Fournon}, I. and {Poidevin}, F. and {Delgado-Gonz{\'a}lez}, Z. and {Angel}, C.~J. and {Geier}, S. and {Marques-Chaves}, R. and {Shirley}, R.},
        title = "{SGLF Transient Classification Report for 2024-07-08}",
      journal = {Transient Name Server Classification Report},
         year = 2024,
        month = jul,
       volume = {2024-2320},
        pages = {1},
       adsurl = {https://ui.adsabs.harvard.edu/abs/2024TNSCR2320....1P}
}

@ARTICLE{Wise2024,
       author = {{Wise}, J. and {Hinds}, K. and {Perley}, D. and {Bochenek}, A. and {Rich}, R.~M. and {Huddleston}, A.},
        title = "{Shane Kast classification of 2 superluminous supernovae 2024-07-08}",
      journal = {Transient Name Server AstroNote},
         year = 2024,
        month = jul,
       volume = {193},
        pages = {1},
       adsurl = {https://ui.adsabs.harvard.edu/abs/2024TNSAN.193....1W}
}

@article{Nicholl2015,
  title={On the diversity of superluminous supernovae: ejected mass as the dominant factor},
  author={Nicholl, M and Smartt, Stephen J and Jerkstrand, Anders and Inserra, Cosimo and Sim, SA and Chen, T-W and Benetti, Stefano and Fraser, M and Gal-Yam, Avishay and Kankare, E and others},
  journal={\mnras},
  volume={452},
  number={4},
  pages={3869--3893},
  year={2015},
  publisher={The Royal Astronomical Society}
}

@article{Inserra2013,
  title={Super-luminous type Ic supernovae: catching a magnetar by the tail},
  author={Inserra, C and Smartt, SJ and Jerkstrand, Aea and Valenti, S and Fraser, M and Wright, Darryl and Smith, K and Chen, T-W and Kotak, R and Pastorello, A and others},
  journal={\apj},
  volume={770},
  number={2},
  pages={128},
  year={2013},
  publisher={IOP Publishing}
}

@ARTICLE{Yaron2012,
       author = {{Yaron}, Ofer and {Gal-Yam}, Avishay},
        title = "{WISeREP{\textemdash}An Interactive Supernova Data Repository}",
      journal = {\pasp},
         year = 2012,
        month = jul,
       volume = {124},
       number = {917},
        pages = {668},
          doi = {10.1086/666656},
archivePrefix = {arXiv},
       eprint = {1204.1891},
 primaryClass = {astro-ph.IM},
       adsurl = {https://ui.adsabs.harvard.edu/abs/2012PASP..124..668Y}
}

@ARTICLE{McCall2004,
       author = {{McCall}, Marshall L.},
        title = "{On Determining Extinction from Reddening}",
      journal = {\aj},
         year = 2004,
        month = nov,
       volume = {128},
       number = {5},
        pages = {2144-2169},
          doi = {10.1086/424933},
       adsurl = {https://ui.adsabs.harvard.edu/abs/2004AJ....128.2144M}
}

@article{zhu2023sn,
  title={SN 2017egm: A helium-rich superluminous supernova with multiple bumps in the light curves},
  author={Zhu, Jiazheng and Jiang, Ning and Dong, Subo and Filippenko, Alexei V and Rudy, Richard J and Pastorello, A and Ashall, Christopher and Bose, Subhash and Post, RS and Bersier, D and others},
  journal={\apj},
  volume={949},
  number={1},
  pages={23},
  year={2023},
  publisher={IOP Publishing}
}

@article{moriya2018superluminous,
  title={Superluminous supernovae},
  author={Moriya, Takashi J and Sorokina, Elena I and Chevalier, Roger A},
  journal={Space Science Reviews},
  volume={214},
  pages={1--37},
  year={2018},
  publisher={Springer}
}

@article{moriya2024superluminous,
  title={Superluminous supernovae},
  author={Moriya, Takashi J},
  journal={arXiv preprint arXiv:2407.12302},
  year={2024}
}

@article{farah2026lense,
  title={Lense--Thirring precessing magnetar engine drives a superluminous supernova},
  author={Farah, Joseph R and Prust, Logan J and Howell, D Andrew and Ni, Yuan Qi and McCully, Curtis and Andrews, Moira and Kumar, Harsh and Hiramatsu, Daichi and Gomez, Sebastian and Wynn, Kathryn and others},
  journal={Nature},
  volume={651},
  number={8105},
  pages={321--325},
  year={2026},
  publisher={Nature Publishing Group UK London}
}

@article{mukhopadhyay2026neutrino,
  title={Neutrino and electromagnetic signatures from Superluminous Supernovae: a case study for SN 2017egm},
  author={Mukhopadhyay, Mainak and Kimura, Shigeo S and Vurm, Indrek and Metzger, Brian D},
  journal={The Astrophysical Journal},
  volume={1006},
  number={1},
  pages={24},
  year={2026},
  publisher={The American Astronomical Society}
}

@article{inserra2019observational,
  title={Observational properties of extreme supernovae},
  author={Inserra, Cosimo},
  journal={Nature Astronomy},
  volume={3},
  number={8},
  pages={697--705},
  year={2019},
  publisher={Nature Publishing Group UK London}
}

@ARTICLE{Reichherzer2021,
       author = {{Reichherzer}, P. and {Sch{\"u}ssler}, F. and {Lefranc}, V. and {Yusafzai}, A. and {Alkan}, A.~K. and {Ashkar}, H. and {Becker Tjus}, J.},
        title = "{Astro-COLIBRI-The COincidence LIBrary for Real-time Inquiry for Multimessenger Astrophysics}",
      journal = {\apjs},
         year = 2021,
        month = sep,
       volume = {256},
       number = {1},
          eid = {5},
        pages = {5},
          doi = {10.3847/1538-4365/ac1517},
archivePrefix = {arXiv},
       eprint = {2109.01672},
 primaryClass = {astro-ph.IM},
       adsurl = {https://ui.adsabs.harvard.edu/abs/2021ApJS..256....5R}
}

@article{gehrels2004swift,
  title={The Swift gamma-ray burst mission},
  author={Gehrels, Neil and Chincarini, Guido and Giommi, PE and Mason, KO and Nousek, JA and Wells, AA and White, NE and Barthelmy, SD and Burrows, David N and Cominsky, Lynn R and others},
  journal={\apj},
  volume={611},
  number={2},
  pages={1005},
  year={2004},
  publisher={IOP Publishing}
}

@ARTICLE{bellm2018zwicky,
       author = {{Bellm}, Eric C. and {Kulkarni}, Shrinivas R. and {Graham}, Matthew J. and {Dekany}, Richard and {Smith}, Roger M. and {Riddle}, Reed and {Masci}, Frank J. and {Helou}, George and {Prince}, Thomas A. and {Adams}, Scott M. and {Barbarino}, C. and {Barlow}, Tom and {Bauer}, James and {Beck}, Ron and {Belicki}, Justin and {Biswas}, Rahul and {Blagorodnova}, Nadejda and {Bodewits}, Dennis and {Bolin}, Bryce and {Brinnel}, Valery and {Brooke}, Tim and {Bue}, Brian and {Bulla}, Mattia and {Burruss}, Rick and {Cenko}, S. Bradley and {Chang}, Chan-Kao and {Connolly}, Andrew and {Coughlin}, Michael and {Cromer}, John and {Cunningham}, Virginia and {De}, Kishalay and {Delacroix}, Alex and {Desai}, Vandana and {Duev}, Dmitry A. and {Eadie}, Gwendolyn and {Farnham}, Tony L. and {Feeney}, Michael and {Feindt}, Ulrich and {Flynn}, David and {Franckowiak}, Anna and {Frederick}, S. and {Fremling}, C. and {Gal-Yam}, Avishay and {Gezari}, Suvi and {Giomi}, Matteo and {Goldstein}, Daniel A. and {Golkhou}, V. Zach and {Goobar}, Ariel and {Groom}, Steven and {Hacopians}, Eugean and {Hale}, David and {Henning}, John and {Ho}, Anna Y.~Q. and {Hover}, David and {Howell}, Justin and {Hung}, Tiara and {Huppenkothen}, Daniela and {Imel}, David and {Ip}, Wing-Huen and {Ivezi{\'c}}, {\v{Z}}eljko and {Jackson}, Edward and {Jones}, Lynne and {Juric}, Mario and {Kasliwal}, Mansi M. and {Kaspi}, S. and {Kaye}, Stephen and {Kelley}, Michael S.~P. and {Kowalski}, Marek and {Kramer}, Emily and {Kupfer}, Thomas and {Landry}, Walter and {Laher}, Russ R. and {Lee}, Chien-De and {Lin}, Hsing Wen and {Lin}, Zhong-Yi and {Lunnan}, Ragnhild and {Giomi}, Matteo and {Mahabal}, Ashish and {Mao}, Peter and {Miller}, Adam A. and {Monkewitz}, Serge and {Murphy}, Patrick and {Ngeow}, Chow-Choong and {Nordin}, Jakob and {Nugent}, Peter and {Ofek}, Eran and {Patterson}, Maria T. and {Penprase}, Bryan and {Porter}, Michael and {Rauch}, Ludwig and {Rebbapragada}, Umaa and {Reiley}, Dan and {Rigault}, Mickael and {Rodriguez}, Hector and {van Roestel}, Jan and {Rusholme}, Ben and {van Santen}, Jakob and {Schulze}, S. and {Shupe}, David L. and {Singer}, Leo P. and {Soumagnac}, Maayane T. and {Stein}, Robert and {Surace}, Jason and {Sollerman}, Jesper and {Szkody}, Paula and {Taddia}, F. and {Terek}, Scott and {Van Sistine}, Angela and {van Velzen}, Sjoert and {Vestrand}, W. Thomas and {Walters}, Richard and {Ward}, Charlotte and {Ye}, Quan-Zhi and {Yu}, Po-Chieh and {Yan}, Lin and {Zolkower}, Jeffry},
        title = "{The Zwicky Transient Facility: System Overview, Performance, and First Results}",
      journal = {\pasp},
         year = 2019,
        month = jan,
       volume = {131},
       number = {995},
        pages = {018002},
          doi = {10.1088/1538-3873/aaecbe},
archivePrefix = {arXiv},
       eprint = {1902.01932},
 primaryClass = {astro-ph.IM},
       adsurl = {https://ui.adsabs.harvard.edu/abs/2019PASP..131a8002B}
}

@article{chen2023hydrogen,
  title={The Hydrogen-poor Superluminous Supernovae from the Zwicky Transient Facility Phase I Survey. I. Light Curves and Measurements},
  author={Chen, ZH and Yan, Lin and Kangas, T and Lunnan, Ragnhild and Schulze, Steve and Sollerman, Jesper and Perley, DA and Chen, T-W and Taggart, K and Hinds, KR and others},
  journal={\apj},
  volume={943},
  number={1},
  pages={41},
  year={2023},
  publisher={IOP Publishing}
}

@article{margutti2018results,
  title={Results from a systematic survey of X-ray emission from hydrogen-poor superluminous SNe},
  author={Margutti, Raffaella and Chornock, R and Metzger, BD and Coppejans, DL and Guidorzi, Cristiano and Migliori, G and Milisavljevic, D and Berger, E and Nicholl, M and Zauderer, BA and others},
  journal={\apj},
  volume={864},
  number={1},
  pages={45},
  year={2018},
  publisher={IOP Publishing}
}

@article{barthelmy2005burst,
  title={The burst alert telescope (BAT) on the SWIFT midex mission},
  author={Barthelmy, Scott D and Barbier, Louis M and Cummings, Jay R and Fenimore, Ed E and Gehrels, Neil and Hullinger, Derek and Krimm, Hans A and Markwardt, Craig B and Palmer, David M and Parsons, Ann and others},
  journal={Space Science Reviews},
  volume={120},
  pages={143--164},
  year={2005},
  publisher={Springer}
}

@article{segreto2010palermo,
  title={The Palermo Swift-BAT hard X-ray catalogue-I. Methodology},
  author={Segreto, A and Cusumano, G and Ferrigno, C and La Parola, V and Mangano, V and Mineo, T and Romano, P},
  journal={A$\&$A},
  volume={510},
  pages={A47},
  year={2010},
  publisher={EDP Sciences}
}

@techreport{MillerStone1994,
  title={Lick Obs},
  author={Miller, JS and Stone, RPS},
  year={1994},
  institution={Tech. Rep. 66. Santa Cruz: Lick Obs}
}

@article{roming2005swift,
  title={The Swift ultra-violet/optical telescope},
  author={Roming, Peter WA and Kennedy, Thomas E and Mason, Keith O and Nousek, John A and Ahr, Lindy and Bingham, Richard E and Broos, Patrick S and Carter, Mary J and Hancock, Barry K and Huckle, Howard E and others},
  journal={Space Science Reviews},
  volume={120},
  pages={95--142},
  year={2005},
  publisher={Springer}
}

@article{bekhti2016hi4pi,
  title={HI4PI: a full-sky H i survey based on EBHIS and GASS},
  author={Bekhti, N Ben and Fl{\"o}er, L and Keller, R and Kerp, J and Lenz, D and Winkel, B and Bailin, J and Calabretta, MR and Dedes, L and Ford, HA and others},
  journal={A$\&$A},
  volume={594},
  pages={A116},
  year={2016},
  publisher={EDP Sciences}
}

@article{burrows2005swift,
  title={The Swift X-ray telescope},
  author={Burrows, David N and Hill, JE and Nousek, JAea and Kennea, JA and Wells, A and Osborne, JP and Abbey, AF and Beardmore, A and Mukerjee, K and Short, ADT and others},
  journal={Space science reviews},
  volume={120},
  pages={165--195},
  year={2005},
  publisher={Springer}
}

@article{bhirombhakdi2018engine,
  title={Where is the engine hiding its missing energy? Constraints from a deep X-ray non-detection of the Superluminous SN 2015bn},
  author={Bhirombhakdi, Kornpob and Chornock, Ryan and Margutti, Raffaella and Nicholl, Matt and Metzger, Brian D and Berger, Edo and Margalit, Ben and Milisavljevic, Dan},
  journal={\apjl},
  volume={868},
  number={2},
  pages={L32},
  year={2018},
  publisher={IOP Publishing}
}

@article{andreoni2022hard,
  title={Hard X-Ray Observations of the Hydrogen-poor Superluminous Supernova SN 2018hti with NuSTAR},
  author={Andreoni, Igor and Lu, Wenbin and Grefenstette, Brian and Kasliwal, Mansi and Yan, Lin and Hare, Jeremy},
  journal={\apjl},
  volume={941},
  number={1},
  pages={L16},
  year={2022},
  publisher={IOP Publishing}
}

@article{Brose2022,
  title={Core-collapse supernovae in dense environments--particle acceleration and non-thermal emission},
  author={Brose, Robert and Sushch, Iurii and Mackey, Jonathan},
  journal={\mnras},
  volume={516},
  number={1},
  pages={492--505},
  year={2022},
  publisher={Oxford University Press}
}

@article{flewelling2020pan,
  title={The Pan-STARRS1 database and data products},
  author={Flewelling, HA ea and Magnier, EA and Chambers, KC and Heasley, JN and Holmberg, C and Huber, ME and Sweeney, W and Waters, CZ and Calamida, A and Casertano, S and others},
  journal={\apjs},
  volume={251},
  number={1},
  pages={7},
  year={2020},
  publisher={IOP Publishing}
}

@article{cheung2022fermi,
  title={Fermi LAT gamma-ray detection of the recurrent nova RS ophiuchi during its 2021 outburst},
  author={Cheung, CC and Johnson, TJ and Jean, P and Kerr, M and Page, KL and Osborne, JP and Beardmore, AP and Sokolovsky, KV and Teyssier, F and Ciprini, S and others},
  journal={\apj},
  volume={935},
  number={1},
  pages={44},
  year={2022},
  publisher={The American Astronomical Society}
}

@article{vurm2021gamma,
  title={Gamma-Ray Thermalization and Leakage from Millisecond Magnetar Nebulae: Toward a Self-consistent Model for Superluminous Supernovae},
  author={Vurm, Indrek and Metzger, Brian D},
  journal={\apj},
  volume={917},
  number={2},
  pages={77},
  year={2021},
  publisher={The American Astronomical Society}
}

@ARTICLE{2023NatAs...7..779L,
       author = {{Lin}, Weili and {Wang}, Xiaofeng and {Yan}, Lin and {Gal-Yam}, Avishay and {Mo}, Jun and {Brink}, Thomas G. and {Filippenko}, Alexei V. and {Xiang}, Danfeng and {Lunnan}, Ragnhild and {Zheng}, Weikang and {Brown}, Peter and {Kasliwal}, Mansi and {Fremling}, Christoffer and {Blagorodnova}, Nadejda and {Mirzaqulov}, Davron and {Ehgamberdiev}, Shuhrat A. and {Lin}, Han and {Zhang}, Kaicheng and {Zhang}, Jicheng and {Yan}, Shengyu and {Zhang}, Jujia and {Chen}, Zhihao and {Deng}, Licai and {Wang}, Kun and {Xiao}, Lin and {Wang}, Lingjun},
        title = "{A superluminous supernova lightened by collisions with pulsational pair-instability shells}",
      journal = {Nature Astronomy},
         year = 2023,
        month = jul,
       volume = {7},
        pages = {779-789},
          doi = {10.1038/s41550-023-01957-3},
archivePrefix = {arXiv},
       eprint = {2304.10416},
 primaryClass = {astro-ph.HE},
       adsurl = {https://ui.adsabs.harvard.edu/abs/2023NatAs...7..779L}
}

@article{gomez2024type,
  title={The Type I superluminous supernova catalogue I: light-curve properties, models, and catalogue description},
  author={Gomez, Sebastian and Nicholl, Matt and Berger, Edo and Blanchard, Peter K and Villar, V Ashley and Rest, Sofia and Hosseinzadeh, Griffin and Aamer, Aysha and Ajay, Yukta and Athukoralalage, Wasundara and others},
  journal={\mnras},
  volume={535},
  number={1},
  pages={471--515},
  year={2024},
  publisher={Oxford University Press}
}

@misc{extrabol_zenodo,
  author       = {Thornton, I.  and Villar, V. A. and Gomez, S. and Hosseinzadeh, G.},
  title        = {villrv/extrabol: RNAAS Release},
  month        = feb,
  year         = 2024,
  publisher    = {Zenodo},
  version      = {1.0.1},
  doi          = {10.5281/zenodo.10652250},
  url          = {https://doi.org/10.5281/zenodo.10652250},
}

@article{tinyanont2023supernova,
  title={Supernova 2020wnt: An Atypical Superluminous Supernova with a Hidden Central Engine},
  author={Tinyanont, Samaporn and Woosley, Stan E and Taggart, Kirsty and Foley, Ryan J and Yan, Lin and Lunnan, Ragnhild and Davis, Kyle W and Kilpatrick, Charles D and Siebert, Matthew R and Schulze, Steve and others},
  journal={\apj},
  volume={951},
  number={1},
  pages={34},
  year={2023},
  publisher={IOP Publishing}
}

@article{yan2020helium,
  title={Helium-rich superluminous supernovae from the Zwicky transient facility},
  author={Yan, Lin and Perley, DA and Schulze, S and Lunnan, Ragnhild and Sollerman, Jesper and De, K and Chen, ZH and Fremling, C and Gal-Yam, A and Taggart, K and others},
  journal={\apjl},
  volume={902},
  number={1},
  pages={L8},
  year={2020},
  publisher={IOP Publishing}
}

@article{quimby2018spectra,
  title={Spectra of hydrogen-poor superluminous supernovae from the palomar transient factory},
  author={Quimby, Robert M and De Cia, Annalisa and Gal-Yam, Avishay and Leloudas, Giorgos and Lunnan, Ragnhild and Perley, Daniel A and Vreeswijk, Paul M and Yan, Lin and Bloom, Joshua S and Cenko, S Bradley and others},
  journal={\apj},
  volume={855},
  number={1},
  pages={2},
  year={2018},
  publisher={IOP Publishing}
}

@article{kumar2025detection,
  title={A Detection of Helium in the Bright Superluminous Supernova SN 2024rmj},
  author={Kumar, Harsh and Berger, Edo and Blanchard, Peter K and Hiramatsu, Daichi and Gomez, Sebastian and Gagliano, Alex and Andrews, Moira and Bostroem, K Azalee and Farah, Joseph and Howell, D Andrew and others},
  journal={\apj},
  volume={992},
  number={1},
  pages={122},
  year={2025},
  publisher={IOP Publishing}
}

@article{kumar2025near,
  title={A Near-infrared Search for Helium in the Superluminous Supernova SN 2024ahr},
  author={Kumar, Harsh and Berger, Edo and Blanchard, Peter K and Gomez, Sebastian and Hiramatsu, Daichi and Andrews, Moira and Bostroem, K Azalee and Dong, Yize and Farah, Joseph and Padilla Gonzalez, Estefania and others},
  journal={\apj},
  volume={987},
  number={2},
  pages={127},
  year={2025},
  publisher={The American Astronomical Society}
}

@article{kumar2026sn,
  title={SN 2024afav: A Superluminous Supernova with Multiple Light-curve Bumps and Spectroscopic Signatures of Circumstellar Interaction},
  author={Kumar, Harsh and Blanchard, Peter K and Berger, Edo and Athukoralalage, Wasundara and Hiramatsu, Daichi and Gomez, Sebastian and Andrews, Moira and Bostroem, K Azalee and Farah, Joseph R and Howell, D Andrew and others},
  journal={\apjl},
  volume={998},
  number={1},
  pages={L3},
  year={2026},
  publisher={The American Astronomical Society}
}

@article{anderson2018nearby,
  title={A nearby super-luminous supernova with a long pre-maximum \& “plateau” and strong C II features},
  author={Anderson, Joseph P and Pessi, Priscila Jael and Dessart, L and Inserra, Cosimo and Hiramatsu, D and Taggart, K and Smartt, SJ and Leloudas, Giorgos and Chen, T-W and M{\"o}ller, A and others},
  journal={A$\&$A},
  volume={620},
  pages={A67},
  year={2018},
  publisher={EDP Sciences}
}

@ARTICLE{Dahiwale2019,
       author = {{Dahiwale}, A. and {Fremling}, C. and {Sharma}, Y.},
        title = "{ZTF Transient Classification Report for 2019-07-26}",
      journal = {Transient Name Server Classification Report},
         year = 2019,
        month = jul,
       volume = {2019-2837},
        pages = {1},
       adsurl = {https://ui.adsabs.harvard.edu/abs/2019TNSCR2837....1D}
}

@article{kasen2010supernova,
  title={Supernova light curves powered by young magnetars},
  author={Kasen, Daniel and Bildsten, Lars},
  journal={\apj},
  volume={717},
  number={1},
  pages={245},
  year={2010},
  publisher={IOP Publishing}
}

@article{gomez2022luminous,
  title={Luminous supernovae: unveiling a population between superluminous and normal core-collapse supernovae},
  author={Gomez, Sebastian and Berger, Edo and Nicholl, Matt and Blanchard, Peter K and Hosseinzadeh, Griffin},
  journal={\apj},
  volume={941},
  number={2},
  pages={107},
  year={2022},
  publisher={IOP Publishing}
}

@article{guillochon2018mosfit,
  title={MOSFiT: modular open source fitter for transients},
  author={Guillochon, James and Nicholl, Matt and Villar, V Ashley and Mockler, Brenna and Narayan, Gautham and Mandel, Kaisey S and Berger, Edo and Williams, Peter KG},
  journal={\apjs},
  volume={236},
  number={1},
  pages={6},
  year={2018},
  publisher={IOP Publishing}
}

@article{atwood2009large,
  title={The large area telescope on the Fermi gamma-ray space telescope mission},
  author={Atwood, WB and Abdo, Aous A and Ackermann, Markus and Althouse, W and Anderson, B and Axelsson, M and Baldini, Luca and Ballet, J and Band, DL and Barbiellini, Guido and others},
  journal={\apj},
  volume={697},
  number={2},
  pages={1071},
  year={2009},
  publisher={IOP Publishing}
}

@ARTICLE{2020ATel13970....1T,
       author = {{Terreran}, G. and {Blanchard}, P. and {DeMarchi}, L. and {Brethauer}, D. and {Margutti}, R. and {Miller}, A.~A. and {Berton}, M. and {Matthews}, D. and {Baldeschi}, A. and {Hajela}, A. and {Stroh}, M.~C. and {Alexander}, K.~D.},
        title = "{Spectroscopic classification of AT 2020oio and SN 2020qef with MMT and Binospec}",
      journal = {The Astronomer's Telegram},
         year = 2020,
        month = aug,
       volume = {13970},
        pages = {1},
       adsurl = {https://ui.adsabs.harvard.edu/abs/2020ATel13970....1T}
}

@article{ofek2013x,
  title={X-ray emission from supernovae in dense circumstellar matter environments: a search for collisionless shocks},
  author={Ofek, EO and Fox, D and Cenko, Stephen B and Sullivan, M and Gnat, O and Frail, DA and Horesh, Assaf and Corsi, A and Quimby, RM and Gehrels, N and others},
  journal={\apj},
  volume={763},
  number={1},
  pages={42},
  year={2013},
  publisher={IOP Publishing}
}

@article{inserra2017complexity,
  title={Complexity in the light curves and spectra of slow-evolving superluminous supernovae},
  author={Inserra, C and Nicholl, M and Chen, T-W and Jerkstrand, A and Smartt, SJ and Kr{\"u}hler, T and Anderson, JP and Baltay, C and Della Valle, Massimo and Fraser, M and others},
  journal={\mnras},
  volume={468},
  number={4},
  pages={4642--4662},
  year={2017},
  publisher={Oxford University Press}
}

@article{chen2023hydrogenb,
  title={The hydrogen-poor superluminous supernovae from the Zwicky Transient Facility phase I survey. II. Light-curve modeling and characterization of undulations},
  author={Chen, ZH and Yan, Lin and Kangas, Tuomas and Lunnan, Ragnhild and Sollerman, Jesper and Schulze, Steve and Perley, DA and Chen, T-W and Taggart, K and Hinds, KR and others},
  journal={\apj},
  volume={943},
  number={1},
  pages={42},
  year={2023},
  publisher={IOP Publishing}
}

@article{blanchard2025hydrogen,
  title={Hydrogen-poor superluminous supernovae in the nebular phase: spectral diversity due to ejecta ionization as a probe of the power source},
  author={Blanchard, Peter K and Berger, Edo and Gomez, Sebastian and Nicholl, Matt and Chornock, Ryan and Kumar, Harsh and Margutti, Raffaella and Hiramatsu, Daichi and Sears, Huei},
  journal={\apj},
  volume={999},
  number={1},
  pages={59},
  year={2026},
  publisher={The American Astronomical Society}
}

@article{levan2013superluminous,
  title={Superluminous X-rays from a superluminous supernova},
  author={Levan, Andrew J and Read, AM and Metzger, BD and Wheatley, PJ and Tanvir, NR},
  journal={\apj},
  volume={771},
  number={2},
  pages={136},
  year={2013},
  publisher={IOP Publishing}
}

@article{eftekhari2019radio,
  title={A radio source coincident with the superluminous supernova PTF10hgi: Evidence for a central engine and an analog of the repeating FRB 121102?},
  author={Eftekhari, T and Berger, E and Margalit, B and Blanchard, PK and Patton, L and Demorest, P and Williams, PKG and Chatterjee, S and Cordes, JM and Lunnan, Ragnhild and others},
  journal={\apjl},
  volume={876},
  number={1},
  pages={L10},
  year={2019},
  publisher={IOP Publishing}
}

@article{renault2018search,
  title={Search for $\gamma$-ray emission from superluminous supernovae with the Fermi-LAT},
  author={Renault-Tinacci, Nicolas and Kotera, Kumiko and Neronov, Andrii and Ando, Shin'ichiro},
  journal={A$\&$A},
  volume={611},
  pages={A45},
  year={2018},
  publisher={EDP Sciences}
}

@article{li2026evidence,
  title={Evidence for GeV Emission from the Superluminous Supernova SN 2017egm},
  author={Li, Shang and Liang, Yun-Feng and Liao, Neng-Hui and Lei, Lei and Fan, Yi-Zhong},
  journal={PhysRevLet},
  volume={136},
  number={11},
  pages={111402},
  year={2026},
  publisher={APS}
}

@article{acharyya2023veritas,
  title={VERITAS and Fermi-LAT constraints on the gamma-ray emission from superluminous supernovae SN2015bn and SN2017egm},
  author={Acharyya, A and Adams, CB and Bangale, P and Benbow, W and Buckley, JH and Capasso, M and Dwarkadas, VV and Errando, M and Falcone, A and Feng, Q and others},
  journal={\apj},
  volume={945},
  number={1},
  pages={30},
  year={2023},
  publisher={IOP Publishing}
}

@article{margutti2017x,
  title={X-Rays from the Location of the Double-humped Transient ASASSN-15lh},
  author={Margutti, Raffaella and Metzger, BD and Chornock, R and Milisavljevic, Danny and Berger, Edo and Blanchard, Peter Kelly and Guidorzi, Cristiano and Migliori, G and Kamble, Atish and Lunnan, R and others},
  journal={\apj},
  volume={836},
  number={1},
  pages={25},
  year={2017},
  publisher={IOP Publishing}
}

@article{margutti2023luminous,
  title={Luminous radio emission from the superluminous supernova 2017ens at 3.3 yr after explosion},
  author={Margutti, Raffaella and Bright, JS and Matthews, DJ and Coppejans, DL and Alexander, KD and Berger, E and Bietenholz, M and Chornock, R and DeMarchi, L and Drout, MR and others},
  journal={\apjl},
  volume={954},
  number={2},
  pages={L45},
  year={2023},
  publisher={IOP Publishing}
}

@article{de2018light,
  title={Light curves of hydrogen-poor Superluminous Supernovae from the Palomar Transient Factory},
  author={De Cia, Annalisa and Gal-Yam, A and Rubin, A and Leloudas, G and Vreeswijk, P and Perley, DA and Quimby, R and Yan, Lin and Sullivan, M and Fl{\"o}rs, A and others},
  journal={\apj},
  volume={860},
  number={2},
  pages={100},
  year={2018},
  publisher={IOP Publishing}
}

@article{gal2019most,
  title={The most luminous supernovae},
  author={Gal-Yam, Avishay},
  journal={\araa},
  volume={57},
  number={1},
  pages={305--333},
  year={2019},
  publisher={Annual Reviews}
}

@incollection{howell2017superluminous,
  title={Superluminous Supernovae},
  author={Howell, D Andrew},
  booktitle={Handbook of Supernovae},
  pages={431--458},
  year={2017},
  publisher={Springer}
}

@article{wheeler2017circumstellar,
  title={Circumstellar interaction models for the bolometric light curve of Type I superluminous SN 2017egm},
  author={Wheeler, J Craig and Chatzopoulos, Emmanouil and Vink{\'o}, Jozsef and Tuminello, Richard},
  journal={\apjl},
  volume={851},
  number={1},
  pages={L14},
  year={2017},
  publisher={IOP Publishing}
}

@article{chatzopoulos2013analytical,
  title={Analytical light curve models of superluminous supernovae: $\chi$2-minimization of parameter fits},
  author={Chatzopoulos, Emmanouil and Wheeler, J Craig and Vinko, Jozsef and Horvath, ZL and Nagy, A},
  journal={\apj},
  volume={773},
  number={1},
  pages={76},
  year={2013},
  publisher={IOP Publishing}
}

@article{margutti2012inverse,
  title={Inverse Compton X-ray emission from supernovae with compact progenitors: application to SN2011fe},
  author={Margutti, R and Soderberg, AM and Chomiuk, L and Chevalier, R and Hurley, K and Milisavljevic, D and Foley, RJ and Hughes, JP and Slane, P and Fransson, Claes and others},
  journal={\apj},
  volume={751},
  number={2},
  pages={134},
  year={2012},
  publisher={IOP Publishing}
}

@article{margutti2014relativistic,
  title={Relativistic supernovae have shorter-lived central engines or more extended progenitors: the case of SN 2012ap},
  author={Margutti, Raffaella and Milisavljevic, D and Soderberg, AM and Guidorzi, Cristiano and Morsony, BJ and Sanders, N and Chakraborti, S and Ray, A and Kamble, A and Drout, M and others},
  journal={\apj},
  volume={797},
  number={2},
  pages={107},
  year={2014},
  publisher={IOP Publishing}
}

@article{murase2011new,
  title={New class of high-energy transients from crashes of supernova ejecta with massive<? format?> circumstellar material shells},
  author={Murase, Kohta and Thompson, Todd A and Lacki, Brian C and Beacom, John F},
  journal={PhysRevD},
  volume={84},
  number={4},
  pages={043003},
  year={2011},
  publisher={APS}
}

@article{murase2015gamma,
  title={Gamma-ray and hard X-ray emission from pulsar-aided supernovae as a probe of particle acceleration in embryonic pulsar wind nebulae},
  author={Murase, Kohta and Kashiyama, Kazumi and Kiuchi, Kenta and Bartos, Imre},
  journal={\apj},
  volume={805},
  number={1},
  pages={82},
  year={2015},
  publisher={IOP Publishing}
}

@ARTICLE{Ivezic2019,
       author = {{Ivezi{\'c}}, {\v{Z}}eljko and {Kahn}, Steven M. and {Tyson}, J. Anthony and {Abel}, Bob and {Acosta}, Emily and {Allsman}, Robyn and {Alonso}, David and {AlSayyad}, Yusra and {Anderson}, Scott F. and {Andrew}, John and {Angel}, James Roger P. and {Angeli}, George Z. and {Ansari}, Reza and {Antilogus}, Pierre and {Araujo}, Constanza and {Armstrong}, Robert and {Arndt}, Kirk T. and {Astier}, Pierre and {Aubourg}, {\'E}ric and {Auza}, Nicole and {Axelrod}, Tim S. and {Bard}, Deborah J. and {Barr}, Jeff D. and {Barrau}, Aurelian and {Bartlett}, James G. and {Bauer}, Amanda E. and {Bauman}, Brian J. and {Baumont}, Sylvain and {Bechtol}, Ellen and {Bechtol}, Keith and {Becker}, Andrew C. and {Becla}, Jacek and {Beldica}, Cristina and {Bellavia}, Steve and {Bianco}, Federica B. and {Biswas}, Rahul and {Blanc}, Guillaume and {Blazek}, Jonathan and {Blandford}, Roger D. and {Bloom}, Josh S. and {Bogart}, Joanne and {Bond}, Tim W. and {Booth}, Michael T. and {Borgland}, Anders W. and {Borne}, Kirk and {Bosch}, James F. and {Boutigny}, Dominique and {Brackett}, Craig A. and {Bradshaw}, Andrew and {Brandt}, William Nielsen and {Brown}, Michael E. and {Bullock}, James S. and {Burchat}, Patricia and {Burke}, David L. and {Cagnoli}, Gianpietro and {Calabrese}, Daniel and {Callahan}, Shawn and {Callen}, Alice L. and {Carlin}, Jeffrey L. and {Carlson}, Erin L. and {Chandrasekharan}, Srinivasan and {Charles-Emerson}, Glenaver and {Chesley}, Steve and {Cheu}, Elliott C. and {Chiang}, Hsin-Fang and {Chiang}, James and {Chirino}, Carol and {Chow}, Derek and {Ciardi}, David R. and {Claver}, Charles F. and {Cohen-Tanugi}, Johann and {Cockrum}, Joseph J. and {Coles}, Rebecca and {Connolly}, Andrew J. and {Cook}, Kem H. and {Cooray}, Asantha and {Covey}, Kevin R. and {Cribbs}, Chris and {Cui}, Wei and {Cutri}, Roc and {Daly}, Philip N. and {Daniel}, Scott F. and {Daruich}, Felipe and {Daubard}, Guillaume and {Daues}, Greg and {Dawson}, William and {Delgado}, Francisco and {Dellapenna}, Alfred and {de Peyster}, Robert and {de Val-Borro}, Miguel and {Digel}, Seth W. and {Doherty}, Peter and {Dubois}, Richard and {Dubois-Felsmann}, Gregory P. and {Durech}, Josef and {Economou}, Frossie and {Eifler}, Tim and {Eracleous}, Michael and {Emmons}, Benjamin L. and {Fausti Neto}, Angelo and {Ferguson}, Henry and {Figueroa}, Enrique and {Fisher-Levine}, Merlin and {Focke}, Warren and {Foss}, Michael D. and {Frank}, James and {Freemon}, Michael D. and {Gangler}, Emmanuel and {Gawiser}, Eric and {Geary}, John C. and {Gee}, Perry and {Geha}, Marla and {Gessner}, Charles J.~B. and {Gibson}, Robert R. and {Gilmore}, D. Kirk and {Glanzman}, Thomas and {Glick}, William and {Goldina}, Tatiana and {Goldstein}, Daniel A. and {Goodenow}, Iain and {Graham}, Melissa L. and {Gressler}, William J. and {Gris}, Philippe and {Guy}, Leanne P. and {Guyonnet}, Augustin and {Haller}, Gunther and {Harris}, Ron and {Hascall}, Patrick A. and {Haupt}, Justine and {Hernandez}, Fabio and {Herrmann}, Sven and {Hileman}, Edward and {Hoblitt}, Joshua and {Hodgson}, John A. and {Hogan}, Craig and {Howard}, James D. and {Huang}, Dajun and {Huffer}, Michael E. and {Ingraham}, Patrick and {Innes}, Walter R. and {Jacoby}, Suzanne H. and {Jain}, Bhuvnesh and {Jammes}, Fabrice and {Jee}, M. James and {Jenness}, Tim and {Jernigan}, Garrett and {Jevremovi{\'c}}, Darko and {Johns}, Kenneth and {Johnson}, Anthony S. and {Johnson}, Margaret W.~G. and {Jones}, R. Lynne and {Juramy-Gilles}, Claire and {Juri{\'c}}, Mario and {Kalirai}, Jason S. and {Kallivayalil}, Nitya J. and {Kalmbach}, Bryce and {Kantor}, Jeffrey P. and {Karst}, Pierre and {Kasliwal}, Mansi M. and {Kelly}, Heather and {Kessler}, Richard and {Kinnison}, Veronica and {Kirkby}, David and {Knox}, Lloyd and {Kotov}, Ivan V. and {Krabbendam}, Victor L. and {Krughoff}, K. Simon and {Kub{\'a}nek}, Petr and {Kuczewski}, John and {Kulkarni}, Shri and {Ku}, John and {Kurita}, Nadine R. and {Lage}, Craig S. and {Lambert}, Ron and {Lange}, Travis and {Langton}, J. Brian and {Le Guillou}, Laurent and {Levine}, Deborah and {Liang}, Ming and {Lim}, Kian-Tat and {Lintott}, Chris J. and {Long}, Kevin E. and {Lopez}, Margaux and {Lotz}, Paul J. and {Lupton}, Robert H. and {Lust}, Nate B. and {MacArthur}, Lauren A. and {Mahabal}, Ashish and {Mandelbaum}, Rachel and {Markiewicz}, Thomas W. and {Marsh}, Darren S. and {Marshall}, Philip J. and {Marshall}, Stuart and {May}, Morgan and {McKercher}, Robert and {McQueen}, Michelle and {Meyers}, Joshua and {Migliore}, Myriam and {Miller}, Michelle and {Mills}, David J.},
        title = "{LSST: From Science Drivers to Reference Design and Anticipated Data Products}",
      journal = {\apj},
         year = 2019,
        month = mar,
       volume = {873},
       number = {2},
          eid = {111},
        pages = {111},
          doi = {10.3847/1538-4357/ab042c},
archivePrefix = {arXiv},
       eprint = {0805.2366},
 primaryClass = {astro-ph},
       adsurl = {https://ui.adsabs.harvard.edu/abs/2019ApJ...873..111I}
}

@software{2018ascl.soft12006W,
       author = {{Wood}, M. and {Caputo}, R. and {Charles}, E. and {Di Mauro}, M. and {Magill}, J. and {Perkins}, J.~S. and {Fermi-LAT Collaboration}},
        title = "{Fermipy: Fermi-LAT data analysis package}",
 howpublished = {Astrophysics Source Code Library, record ascl:1812.006},
         year = 2018,
        month = dec,
          eid = {ascl:1812.006},
       adsurl = {https://ui.adsabs.harvard.edu/abs/2018ascl.soft12006W}
}

@article{ballet2023fermi,
  title={Fermi Large Area Telescope fourth source catalog data release 4 (4FGL-DR4)},
  author={Ballet, J and Bruel, P and Burnett, TH and Lott, B and others},
  journal={arXiv preprint arXiv:2307.12546},
  year={2023}
}

@INPROCEEDINGS{uvot-breeveld2011,
       author = {{Breeveld}, A.~A. and {Landsman}, W. and {Holland}, S.~T. and {Roming}, P. and {Kuin}, N.~P.~M. and {Page}, M.~J.},
        title = "{An Updated Ultraviolet Calibration for the Swift/UVOT}",
    booktitle = {Gamma Ray Bursts 2010},
         year = 2011,
       editor = {{McEnery}, J.~E. and {Racusin}, J.~L. and {Gehrels}, N.},
       series = {American Institute of Physics Conference Series},
       volume = {1358},
        month = aug,
    publisher = {AIP},
        pages = {373-376},
          doi = {10.1063/1.3621807},
archivePrefix = {arXiv},
       eprint = {1102.4717},
 primaryClass = {astro-ph.IM},
       adsurl = {https://ui.adsabs.harvard.edu/abs/2011AIPC.1358..373B}
}

@article{fiore2026nearby,
  title={The nearby He-rich superluminous supernova SN 2021bnw during photospheric phases},
  author={Fiore, A and Kozyreva, A and Yan, L and Benetti, S and Anderson, JP and Baklanov, P and Cai, Y-Z and Cappellaro, E and Chen, T-W and Elias-Rosa, N and others},
  journal={\aap},
  volume={710},
  pages={A221},
  year={2026},
  publisher={EDP Sciences}
}

@ARTICLE{uvot-poole2008,
   author = {{Poole}, T.~S. and {Breeveld}, A.~A. and {Page}, M.~J. and {Landsman}, W. and 
	{Holland}, S.~T. and {Roming}, P. and {Kuin}, N.~P.~M. and {Brown}, P.~J. and 
	{Gronwall}, C. and {Hunsberger}, S. and {Koch}, S. and {Mason}, K.~O. and 
	{Schady}, P. and {vanden Berk}, D. and {Blustin}, A.~J. and 
	{Boyd}, P. and {Broos}, P. and {Carter}, M. and {Chester}, M.~M. and 
	{Cucchiara}, A. and {Hancock}, B. and {Huckle}, H. and {Immler}, S. and 
	{Ivanushkina}, M. and {Kennedy}, T. and {Marshall}, F. and {Morgan}, A. and 
	{Pandey}, S.~B. and {de Pasquale}, M. and {Smith}, P.~J. and 
	{Still}, M.},
    title = "{Photometric calibration of the Swift ultraviolet/optical telescope}",
  journal = {\mnras},
archivePrefix = "arXiv",
   eprint = {0708.2259},
     year = 2008,
    month = jan,
   volume = 383,
    pages = {627-645},
      doi = {10.1111/j.1365-2966.2007.12563.x},
   adsurl = {http://adsabs.harvard.edu/abs/2008MNRAS.383..627P}
}

@article{schlegel1998maps,
  title={Maps of dust infrared emission for use in estimation of reddening and cosmic microwave background radiation foregrounds},
  author={Schlegel, David J and Finkbeiner, Douglas P and Davis, Marc},
  journal={\apj},
  volume={500},
  number={2},
  pages={525},
  year={1998},
  publisher={IOP Publishing}
}

@article{schlafly2011measuring,
  title={Measuring reddening with Sloan Digital Sky Survey stellar spectra and recalibrating SFD},
  author={Schlafly, Edward F and Finkbeiner, Douglas P},
  journal={\apj},
  volume={737},
  number={2},
  pages={103},
  year={2011},
  publisher={IOP Publishing}
}

@article{tonry2018atlas,
  title={ATLAS: a high-cadence all-sky survey system},
  author={Tonry, JL and Denneau, L and Heinze, AN and Stalder, B and Smith, KW and Smartt, SJ and Stubbs, CW and Weiland, HJ and Rest, A},
  journal={\pasp},
  volume={130},
  number={988},
  pages={064505},
  year={2018},
  publisher={IOP Publishing}
}

@article{blagorodnova2018sed,
  title={The SED Machine: a robotic spectrograph for fast transient classification},
  author={Blagorodnova, Nadejda and Neill, James D and Walters, Richard and Kulkarni, Shrinivas R and Fremling, Christoffer and Ben-Ami, Sagi and Dekany, Richard G and Fucik, Jason R and Konidaris, Nick and Nash, Reston and others},
  journal={\pasp},
  volume={130},
  number={985},
  pages={035003},
  year={2018},
  publisher={IOP Publishing}
}

@article{oke1982efficient,
  title={An efficient low resolution and moderate resolution spectrograph for the Hale telescope},
  author={Oke, JB and Gunn, JE},
  journal={\pasp},
  volume={94},
  number={559},
  pages={586},
  year={1982},
  publisher={IOP Publishing}
}

@article{fabricant2019binospec,
  title={Binospec: a wide-field imaging spectrograph for the MMT},
  author={Fabricant, Daniel and Fata, Robert and Epps, Harland and Gauron, Thomas and Mueller, Mark and Zajac, Joseph and Amato, Stephen and Barberis, Jack and Bergner, Henry and Brennan, Patricia and others},
  journal={\pasp},
  volume={131},
  number={1001},
  pages={075004},
  year={2019},
  publisher={IOP Publishing}
}

@article{mazzali2016spectrum,
  title={Spectrum formation in superluminous supernovae (Type I)},
  author={Mazzali, PA and Sullivan, M and Pian, Elena and Greiner, J and Kann, DA},
  journal={\mnras},
  volume={458},
  number={4},
  pages={3455--3465},
  year={2016},
  publisher={Oxford University Press}
}

@incollection{djupvik2010nordic,
  title={The Nordic Optical Telescope},
  author={Djupvik, Anlaug Amanda and Andersen, Johannes},
  booktitle={Highlights of Spanish astrophysics V},
  pages={211--218},
  year={2010},
  publisher={Springer}
}

@misc{simongini2025a,
  author       = {Simongini, Andrea and
                  Ragosta, Fabio and
                  Piranomonte, Silvia and
                  Di Palma, Irene},
  title        = {CASTOR: v2.0 - 2025-03-05},
  month        = mar,
  year         = 2025,
  publisher    = {Zenodo},
  version      = {2.0},
  doi          = {10.5281/zenodo.14811842},
  url          = {https://doi.org/10.5281/zenodo.14811842},
}

@article{simongini2025b,
  title={Core-collapse supernova parameter estimation with the upcoming Vera C. Rubin Observatory},
  author={Simongini, Andrea and Ragosta, Fabio and Di Palma, Irene and Piranomonte, Silvia},
  journal={A$\&$A},
  volume={699},
  pages={A98},
  month = jul,
  year={2025},
  publisher={EDP Sciences}
}

@ARTICLE{Andrews2013a,
       author = {{Andrews}, Brett H. and {Martini}, Paul},
        title = "{The Mass-Metallicity Relation with the Direct Method on Stacked Spectra of SDSS Galaxies}",
      journal = {\apj},
         year = 2013,
        month = mar,
       volume = {765},
       number = {2},
          eid = {140},
        pages = {140},
          doi = {10.1088/0004-637X/765/2/140},
archivePrefix = {arXiv},
       eprint = {1211.3418},
 primaryClass = {astro-ph.CO},
       adsurl = {https://ui.adsabs.harvard.edu/abs/2013ApJ...765..140A}
}

@ARTICLE{Byler2017a,
       author = {{Byler}, Nell and {Dalcanton}, Julianne J. and {Conroy}, Charlie and
         {Johnson}, Benjamin D.},
        title = "{Nebular Continuum and Line Emission in Stellar Population Synthesis Models}",
      journal = {\apj},
         year = 2017,
        month = may,
       volume = {840},
       number = {1},
          eid = {44},
        pages = {44},
          doi = {10.3847/1538-4357/aa6c66},
archivePrefix = {arXiv},
       eprint = {1611.08305},
 primaryClass = {astro-ph.GA},
       adsurl = {https://ui.adsabs.harvard.edu/abs/2017ApJ...840...44B}
}

@ARTICLE{Curti2017a,
       author = {{Curti}, M. and {Cresci}, G. and {Mannucci}, F. and {Marconi}, A. and {Maiolino}, R. and {Esposito}, S.},
        title = "{New fully empirical calibrations of strong-line metallicity indicators in star-forming galaxies}",
      journal = {\mnras},
         year = 2017,
        month = feb,
       volume = {465},
       number = {2},
        pages = {1384-1400},
          doi = {10.1093/mnras/stw2766},
archivePrefix = {arXiv},
       eprint = {1610.06939},
 primaryClass = {astro-ph.GA},
       adsurl = {https://ui.adsabs.harvard.edu/abs/2017MNRAS.465.1384C}
}

@ARTICLE{Calzetti2000a,
   author = {{Calzetti}, D. and {Armus}, L. and {Bohlin}, R.~C. and {Kinney}, A.~L. and 
	{Koornneef}, J. and {Storchi-Bergmann}, T.},
    title = "{The Dust Content and Opacity of Actively Star-forming Galaxies}",
  journal = {\apj},
   eprint = {astro-ph/9911459},
     year = 2000,
    month = apr,
   volume = 533,
    pages = {682-695},
      doi = {10.1086/308692},
   adsurl = {http://adsabs.harvard.edu/abs/2000ApJ...533..682C}
}

@ARTICLE{Chabrier2003a,
   author = {{Chabrier}, G.},
    title = "{Galactic Stellar and Substellar Initial Mass Function}",
  journal = {\pasp},
   eprint = {astro-ph/0304382},
     year = 2003,
    month = jul,
   volume = 115,
    pages = {763},
      doi = {10.1086/376392},
   adsurl = {http://adsabs.harvard.edu/abs/2003PASP..115..763C}
}

@ARTICLE{Chambers2016a,
       author = {{Chambers}, K.~C. and {Magnier}, E.~A. and {Metcalfe}, N. and {Flewelling}, H.~A. and {Huber}, M.~E. and {Waters}, C.~Z. and {Denneau}, L. and {Draper}, P.~W. and {Farrow}, D. and {Finkbeiner}, D.~P. and {Holmberg}, C. and {Koppenhoefer}, J. and {Price}, P.~A. and {Rest}, A. and {Saglia}, R.~P. and {Schlafly}, E.~F. and {Smartt}, S.~J. and {Sweeney}, W. and {Wainscoat}, R.~J. and {Burgett}, W.~S. and {Chastel}, S. and {Grav}, T. and {Heasley}, J.~N. and {Hodapp}, K.~W. and {Jedicke}, R. and {Kaiser}, N. and {Kudritzki}, R. -P. and {Luppino}, G.~A. and {Lupton}, R.~H. and {Monet}, D.~G. and {Morgan}, J.~S. and {Onaka}, P.~M. and {Shiao}, B. and {Stubbs}, C.~W. and {Tonry}, J.~L. and {White}, R. and {Ba{\~n}ados}, E. and {Bell}, E.~F. and {Bender}, R. and {Bernard}, E.~J. and {Boegner}, M. and {Boffi}, F. and {Botticella}, M.~T. and {Calamida}, A. and {Casertano}, S. and {Chen}, W. -P. and {Chen}, X. and {Cole}, S. and {Deacon}, N. and {Frenk}, C. and {Fitzsimmons}, A. and {Gezari}, S. and {Gibbs}, V. and {Goessl}, C. and {Goggia}, T. and {Gourgue}, R. and {Goldman}, B. and {Grant}, P. and {Grebel}, E.~K. and {Hambly}, N.~C. and {Hasinger}, G. and {Heavens}, A.~F. and {Heckman}, T.~M. and {Henderson}, R. and {Henning}, T. and {Holman}, M. and {Hopp}, U. and {Ip}, W. -H. and {Isani}, S. and {Jackson}, M. and {Keyes}, C.~D. and {Koekemoer}, A.~M. and {Kotak}, R. and {Le}, D. and {Liska}, D. and {Long}, K.~S. and {Lucey}, J.~R. and {Liu}, M. and {Martin}, N.~F. and {Masci}, G. and {McLean}, B. and {Mindel}, E. and {Misra}, P. and {Morganson}, E. and {Murphy}, D.~N.~A. and {Obaika}, A. and {Narayan}, G. and {Nieto-Santisteban}, M.~A. and {Norberg}, P. and {Peacock}, J.~A. and {Pier}, E.~A. and {Postman}, M. and {Primak}, N. and {Rae}, C. and {Rai}, A. and {Riess}, A. and {Riffeser}, A. and {Rix}, H.~W. and {R{\"o}ser}, S. and {Russel}, R. and {Rutz}, L. and {Schilbach}, E. and {Schultz}, A.~S.~B. and {Scolnic}, D. and {Strolger}, L. and {Szalay}, A. and {Seitz}, S. and {Small}, E. and {Smith}, K.~W. and {Soderblom}, D.~R. and {Taylor}, P. and {Thomson}, R. and {Taylor}, A.~N. and {Thakar}, A.~R. and {Thiel}, J. and {Thilker}, D. and {Unger}, D. and {Urata}, Y. and {Valenti}, J. and {Wagner}, J. and {Walder}, T. and {Walter}, F. and {Watters}, S.~P. and {Werner}, S. and {Wood-Vasey}, W.~M. and {Wyse}, R.},
        title = "{The Pan-STARRS1 Surveys}",
      journal = {arXiv e-prints},
         year = 2016,
        month = dec,
          eid = {arXiv:1612.05560},
        pages = {arXiv:1612.05560},
archivePrefix = {arXiv},
       eprint = {1612.05560},
 primaryClass = {astro-ph.IM},
       adsurl = {https://ui.adsabs.harvard.edu/abs/2016arXiv161205560C}
}

@ARTICLE{Conroy2009a,
       author = {{Conroy}, Charlie and {Gunn}, James E. and {White}, Martin},
        title = "{The Propagation of Uncertainties in Stellar Population Synthesis Modeling. I. The Relevance of Uncertain Aspects of Stellar Evolution and the Initial Mass Function to the Derived Physical Properties of Galaxies}",
      journal = {\apj},
         year = 2009,
        month = jul,
       volume = {699},
       number = {1},
        pages = {486-506},
          doi = {10.1088/0004-637X/699/1/486},
archivePrefix = {arXiv},
       eprint = {0809.4261},
 primaryClass = {astro-ph},
       adsurl = {https://ui.adsabs.harvard.edu/abs/2009ApJ...699..486C}
}

@MISC{ForemanMackey2014a,
       author = {{Foreman-Mackey}, Dan and {Sick}, Jonathan and {Johnson}, Ben},
        title = "{Python-Fsps: Python Bindings To Fsps (V0.1.1)}",
         year = 2014,
        month = oct,
          eid = {10.5281/zenodo.12157},
          doi = {10.5281/zenodo.12157},
      version = {v0.1.1},
    publisher = {Zenodo},
       adsurl = {https://ui.adsabs.harvard.edu/abs/2014zndo.....12157F}
}

@ARTICLE{Johnson2021a,
       author = {{Johnson}, Benjamin D. and {Leja}, Joel and {Conroy}, Charlie and {Speagle}, Joshua S.},
        title = "{Stellar Population Inference with Prospector}",
      journal = {\apjs},
         year = 2021,
        month = jun,
       volume = {254},
       number = {2},
          eid = {22},
        pages = {22},
          doi = {10.3847/1538-4365/abef67},
archivePrefix = {arXiv},
       eprint = {2012.01426},
 primaryClass = {astro-ph.GA},
       adsurl = {https://ui.adsabs.harvard.edu/abs/2021ApJS..254...22J}
}

@ARTICLE{Kennicutt1998a,
       author = {{Kennicutt}, Robert C., Jr.},
        title = "{Star Formation in Galaxies Along the Hubble Sequence}",
      journal = {\araa},
         year = 1998,
        month = jan,
       volume = {36},
        pages = {189-232},
          doi = {10.1146/annurev.astro.36.1.189},
archivePrefix = {arXiv},
       eprint = {astro-ph/9807187},
 primaryClass = {astro-ph},
       adsurl = {https://ui.adsabs.harvard.edu/abs/1998ARA&A..36..189K}
}

@ARTICLE{Lang2014a,
       author = {{Lang}, D.},
        title = "{unWISE: Unblurred Coadds of the WISE Imaging}",
      journal = {\aj},
         year = "2014",
        month = "May",
       volume = {147},
          eid = {108},
        pages = {108},
          doi = {10.1088/0004-6256/147/5/108},
archivePrefix = {arXiv},
       eprint = {1405.0308},
 primaryClass = {astro-ph.IM},
       adsurl = {https://ui.adsabs.harvard.edu/\#abs/2014AJ....147..108L}
}

@ARTICLE{Madau2014a,
       author = {{Madau}, Piero and {Dickinson}, Mark},
        title = "{Cosmic Star-Formation History}",
      journal = {\araa},
         year = 2014,
        month = aug,
       volume = {52},
        pages = {415-486},
          doi = {10.1146/annurev-astro-081811-125615},
archivePrefix = {arXiv},
       eprint = {1403.0007},
 primaryClass = {astro-ph.CO},
       adsurl = {https://ui.adsabs.harvard.edu/abs/2014ARA&A..52..415M}
}

@BOOK{Osterbrock2006a,
       author = {{Osterbrock}, Donald E. and {Ferland}, Gary J.},
        title = "{Astrophysics of gaseous nebulae and active galactic nuclei}",
         year = 2006,
    publisher = "{University Science Books}",
       adsurl = {https://ui.adsabs.harvard.edu/abs/2006agna.book.....O}
}

@ARTICLE{Pettini2004a,
       author = {{Pettini}, Max and {Pagel}, Bernard E.~J.},
        title = "{[OIII]/[NII] as an abundance indicator at high redshift}",
      journal = {\mnras},
         year = 2004,
        month = mar,
       volume = {348},
       number = {3},
        pages = {L59-L63},
          doi = {10.1111/j.1365-2966.2004.07591.x},
archivePrefix = {arXiv},
       eprint = {astro-ph/0401128},
 primaryClass = {astro-ph},
       adsurl = {https://ui.adsabs.harvard.edu/abs/2004MNRAS.348L..59P}
}

@ARTICLE{Schulze2021a,
       author = {{Schulze}, Steve and {Yaron}, Ofer and {Sollerman}, Jesper and {Leloudas}, Giorgos and {Gal}, Amit and {Wright}, Angus H. and {Lunnan}, Ragnhild and {Gal-Yam}, Avishay and {Ofek}, Eran O. and {Perley}, Daniel A. and {Filippenko}, Alexei V. and {Kasliwal}, Mansi M. and {Kulkarni}, Shrinivas R. and {Neill}, James D. and {Nugent}, Peter E. and {Quimby}, Robert M. and {Sullivan}, Mark and {Strotjohann}, Nora Linn and {Arcavi}, Iair and {Ben-Ami}, Sagi and {Bianco}, Federica and {Bloom}, Joshua S. and {De}, Kishalay and {Fraser}, Morgan and {Fremling}, Christoffer U. and {Horesh}, Assaf and {Johansson}, Joel and {Kelly}, Patrick L. and {Kne{\v{z}}evi{\'c}}, Nikola and {Kne{\v{z}}evi{\'c}}, Sladjana and {Maguire}, Kate and {Nyholm}, Anders and {Papadogiannakis}, Sem{\'e}li and {Petrushevska}, Tanja and {Rubin}, Adam and {Yan}, Lin and {Yang}, Yi and {Adams}, Scott M. and {Bufano}, Filomena and {Clubb}, Kelsey I. and {Foley}, Ryan J. and {Green}, Yoav and {Harmanen}, Jussi and {Ho}, Anna Y.~Q. and {Hook}, Isobel M. and {Hosseinzadeh}, Griffin and {Howell}, D. Andrew and {Kong}, Albert K.~H. and {Kotak}, Rubina and {Matheson}, Thomas and {McCully}, Curtis and {Milisavljevic}, Dan and {Pan}, Yen-Chen and {Poznanski}, Dovi and {Shivvers}, Isaac and {van Velzen}, Sjoert and {Verbeek}, Kars K.},
        title = "{The Palomar Transient Factory Core-collapse Supernova Host-galaxy Sample. I. Host-galaxy Distribution Functions and Environment Dependence of Core-collapse Supernovae}",
      journal = {\apjs},
         year = 2021,
        month = aug,
       volume = {255},
       number = {2},
          eid = {29},
        pages = {29},
          doi = {10.3847/1538-4365/abff5e},
archivePrefix = {arXiv},
       eprint = {2008.05988},
 primaryClass = {astro-ph.GA},
       adsurl = {https://ui.adsabs.harvard.edu/abs/2021ApJS..255...29S}
}

@ARTICLE{Speagle2020a,
       author = {{Speagle}, Joshua S.},
        title = "{DYNESTY: a dynamic nested sampling package for estimating Bayesian posteriors and evidences}",
      journal = {\mnras},
         year = 2020,
        month = apr,
       volume = {493},
       number = {3},
        pages = {3132-3158},
          doi = {10.1093/mnras/staa278},
archivePrefix = {arXiv},
       eprint = {1904.02180},
 primaryClass = {astro-ph.IM},
       adsurl = {https://ui.adsabs.harvard.edu/abs/2020MNRAS.493.3132S}
}

@ARTICLE{Wright2010a,
       author = {{Wright}, Edward L. and {Eisenhardt}, Peter R.~M. and {Mainzer}, Amy K. and {Ressler}, Michael E. and {Cutri}, Roc M. and {Jarrett}, Thomas and {Kirkpatrick}, J. Davy and {Padgett}, Deborah and {McMillan}, Robert S. and {Skrutskie}, Michael and {Stanford}, S.~A. and {Cohen}, Martin and {Walker}, Russell G. and {Mather}, John C. and {Leisawitz}, David and {Gautier}, Thomas N., III and {McLean}, Ian and {Benford}, Dominic and {Lonsdale}, Carol J. and {Blain}, Andrew and {Mendez}, Bryan and {Irace}, William R. and {Duval}, Valerie and {Liu}, Fengchuan and {Royer}, Don and {Heinrichsen}, Ingolf and {Howard}, Joan and {Shannon}, Mark and {Kendall}, Martha and {Walsh}, Amy L. and {Larsen}, Mark and {Cardon}, Joel G. and {Schick}, Scott and {Schwalm}, Mark and {Abid}, Mohamed and {Fabinsky}, Beth and {Naes}, Larry and {Tsai}, Chao-Wei},
        title = "{The Wide-field Infrared Survey Explorer (WISE): Mission Description and Initial On-orbit Performance}",
      journal = {\aj},
         year = 2010,
        month = dec,
       volume = {140},
       number = {6},
        pages = {1868-1881},
          doi = {10.1088/0004-6256/140/6/1868},
archivePrefix = {arXiv},
       eprint = {1008.0031},
 primaryClass = {astro-ph.IM},
       adsurl = {https://ui.adsabs.harvard.edu/abs/2010AJ....140.1868W}
}

@ARTICLE{Wright2016a,
   author = {{Wright}, A.~H. and {Robotham}, A.~S.~G. and {Bourne}, N. and 
	{Driver}, S.~P. and {Dunne}, L. and {Maddox}, S.~J. and {Alpaslan}, M. and 
	{Andrews}, S.~K. and {Bauer}, A.~E. and {Bland-Hawthorn}, J. and 
	{Brough}, S. and {Brown}, M.~J.~I. and {Clarke}, C. and {Cluver}, M. and 
	{Davies}, L.~J.~M. and {Grootes}, M.~W. and {Holwerda}, B.~W. and 
	{Hopkins}, A.~M. and {Jarrett}, T.~H. and {Kafle}, P.~R. and 
	{Lange}, R. and {Liske}, J. and {Loveday}, J. and {Moffett}, A.~J. and 
	{Norberg}, P. and {Popescu}, C.~C. and {Smith}, M. and {Taylor}, E.~N. and 
	{Tuffs}, R.~J. and {Wang}, L. and {Wilkins}, S.~M.},
    title = "{Galaxy And Mass Assembly: accurate panchromatic photometry from optical priors using LAMBDAR}",
  journal = {\mnras},
archivePrefix = "arXiv",
   eprint = {1604.01923},
     year = 2016,
    month = jul,
   volume = 460,
    pages = {765},
      doi = {10.1093/mnras/stw832},
   adsurl = {http://adsabs.harvard.edu/abs/2016MNRAS.460..765W}
}

@article{mainzer2014initial,
  title={Initial performance of the NEOWISE reactivation mission},
  author={Mainzer, Amy and Bauer, J and Cutri, RM and Grav, T and Masiero, J and Beck, R and Clarkson, P and Conrow, T and Dailey, J and Eisenhardt, P and others},
  journal={\apj},
  volume={792},
  number={1},
  pages={30},
  year={2014},
  publisher={IOP Publishing}
}

@article{meisner2017full,
  title={Full-depth Coadds of the WISE and First-year NEOWISE-reactivation Images},
  author={Meisner, Aaron M and Lang, Dustin and Schlegel, David J},
  journal={\aj},
  volume={153},
  number={1},
  pages={38},
  year={2017},
  publisher={IOP Publishing}
}

@article{karamehmetoglu2023population,
  title={A population of Type Ibc supernovae with massive progenitors-Broad lightcurves not uncommon in (i) PTF},
  author={Karamehmetoglu, Emir and Sollerman, Jesper and Taddia, Francesco and Barbarino, Cristina and Feindt, Ulrich and Fremling, C and Gal-Yam, Avishay and Kasliwal, MM and Petrushevska, Tanja and Schulze, Steve and others},
  journal={A$\&$A},
  volume={678},
  pages={A87},
  year={2023},
  publisher={EDP Sciences}
}

@article{taddia2019analysis,
  title={Analysis of broad-lined Type Ic supernovae from the (intermediate) Palomar Transient Factory},
  author={Taddia, Francesco and Sollerman, Jesper and Fremling, C and Barbarino, Cristina and Karamehmetoglu, Emir and Arcavi, I and Cenko, SB and Filippenko, AV and Gal-Yam, A and Hiramatsu, D and others},
  journal={A$\&$A},
  volume={621},
  pages={A71},
  year={2019},
  publisher={EDP Sciences}
}

@ARTICLE{arnett1982,
       author = {{Arnett}, W.~D.},
        title = "{Type I supernovae. I - Analytic solutions for the early part of the light curve}",
      journal = {\apj},
         year = 1982,
        month = feb,
       volume = {253},
        pages = {785-797},
          doi = {10.1086/159681},
       adsurl = {https://ui.adsabs.harvard.edu/abs/1982ApJ...253..785A}
}

@article{schweyer2025sn,
  title={SN 2019odp--A massive oxygen-rich Type Ib supernova},
  author={Schweyer, Tassilo and Sollerman, Jesper and Jerkstrand, Anders and Ergon, Mattias and Chen, T-W and Omand, CMB and Schulze, Steve and Coughlin, MW and Andreoni, I and Fremling, C and others},
  journal={A$\&$A},
  volume={693},
  pages={A13},
  year={2025},
  publisher={EDP Sciences}
}

@article{nicholl2019nebular,
  title={Nebular-phase spectra of superluminous supernovae: physical insights from observational and statistical properties},
  author={Nicholl, Matt and Berger, Edo and Blanchard, Peter K and Gomez, Sebastian and Chornock, Ryan},
  journal={\apj},
  volume={871},
  number={1},
  pages={102},
  year={2019},
  publisher={IOP Publishing}
}

@article{sarin2024redback,
  title={REDBACK: a Bayesian inference software package for electromagnetic transients},
  author={Sarin, Nikhil and H{\"u}bner, Moritz and Omand, Conor MB and Setzer, Christian N and Schulze, Steve and Adhikari, Naresh and Sagu{\'e}s-Carracedo, Ana and Galaudage, Shanika and Wallace, Wendy F and Lamb, Gavin P and others},
  journal={\mnras},
  volume={531},
  number={1},
  pages={1203--1227},
  year={2024},
  publisher={Oxford University Press}
}

@article{graham2019zwicky,
  title={The zwicky transient facility: science objectives},
  author={Graham, Matthew J and Kulkarni, SR and Bellm, Eric C and Adams, Scott M and Barbarino, Cristina and Blagorodnova, Nadejda and Bodewits, Dennis and Bolin, Bryce and Brady, Patrick R and Cenko, S Bradley and others},
  journal={\pasp},
  volume={131},
  number={1001},
  pages={078001},
  year={2019},
  publisher={IOP Publishing}
}

@article{masci2018zwicky,
  title={The zwicky transient facility: Data processing, products, and archive},
  author={Masci, Frank J and Laher, Russ R and Rusholme, Ben and Shupe, David L and Groom, Steven and Surace, Jason and Jackson, Edward and Monkewitz, Serge and Beck, Ron and Flynn, David and others},
  journal={\pasp},
  volume={131},
  number={995},
  pages={018003},
  year={2018},
  publisher={IOP Publishing}
}

@article{dekany2020zwicky,
  title={The zwicky transient facility: Observing system},
  author={Dekany, Richard and Smith, Roger M and Riddle, Reed and Feeney, Michael and Porter, Michael and Hale, David and Zolkower, Jeffry and Belicki, Justin and Kaye, Stephen and Henning, John and others},
  journal={\pasp},
  volume={132},
  number={1009},
  pages={038001},
  year={2020},
  publisher={IOP Publishing}
}

@article{van2019skyportal,
  title={SkyPortal: an astronomical data platform},
  author={van der Walt, St{\'e}fan J and Crellin-Quick, Arien and Bloom, Joshua S},
  journal={Journal of Open Source Software},
  volume={4},
  number={37},
  pages={1247},
  year={2019}
}

@article{coughlin2023data,
  title={A data science platform to enable time-domain astronomy},
  author={Coughlin, Michael W and Bloom, Joshua S and Nir, Guy and Antier, Sarah and Du Laz, Theophile Jegou and Van Der Walt, St{\'e}fan and Crellin-Quick, Arien and Culino, Thomas and Duev, Dmitry A and Goldstein, Daniel A and others},
  journal={\apjs},
  volume={267},
  number={2},
  pages={31},
  year={2023},
  publisher={IOP Publishing}
}

@article{rigault2019fully,
  title={Fully automated integral field spectrograph pipeline for the SEDMachine: pysedm},
  author={Rigault, M and Neill, JD and Blagorodnova, N and Dugas, A and Feeney, M and Walters, R and Brinnel, V and Copin, Y and Fremling, C and Nordin, J and others},
  journal={A$\&$A},
  volume={627},
  pages={A115},
  year={2019},
  publisher={EDP Sciences}
}

@article{kim2022new,
  title={New modules for the SEDMachine to remove contaminations from cosmic rays and non-target light: BYECR and CONTSEP},
  author={Kim, Y-L and Rigault, M and Neill, JD and Briday, M and Copin, Y and Lezmy, J and Nicolas, N and Riddle, R and Sharma, Y and Smith, M and others},
  journal={\pasp},
  volume={134},
  number={1032},
  pages={024505},
  year={2022},
  publisher={IOP Publishing}
}

@article{smith2020design,
  title={Design and operation of the ATLAS transient science server},
  author={Smith, KW and Smartt, SJ and Young, DR and Tonry, JL and Denneau, L and Flewelling, H and Heinze, AN and Weiland, HJ and Stalder, B and Rest, A and others},
  journal={\pasp},
  volume={132},
  number={1014},
  pages={085002},
  year={2020},
  publisher={IOP Publishing}
}

@ARTICLE{Shingles2021,
       author = {{Shingles}, L. and {Smith}, K.~W. and {Young}, D.~R. and {Smartt}, S.~J. and {Tonry}, J. and {Denneau}, L. and {Heinze}, A. and {Weiland}, H. and {Flewelling}, H. and {Stalder}, B. and {Clocchiatti}, A. and {F{\"o}rster}, F. and {Pignata}, G. and {Rest}, A. and {Anderson}, J. and {Stubbs}, C. and {Erasmus}, N.},
        title = "{Release of the ATLAS Forced Photometry server for public use}",
      journal = {Transient Name Server AstroNote},
         year = 2021,
        month = jan,
       volume = {7},
        pages = {1-7},
       adsurl = {https://ui.adsabs.harvard.edu/abs/2021TNSAN...7....1S}
}

@ARTICLE{Barbarino2021,
       author = {{Barbarino}, C. and {Sollerman}, J. and {Taddia}, F. and {Fremling}, C. and {Karamehmetoglu}, E. and {Arcavi}, I. and {Gal-Yam}, A. and {Laher}, R. and {Schulze}, S. and {Wozniak}, P. and {Yan}, Lin},
        title = "{Type Ic supernovae from the (intermediate) Palomar Transient Factory}",
      journal = {\aap},
         year = 2021,
        month = jul,
       volume = {651},
          eid = {A81},
        pages = {A81},
          doi = {10.1051/0004-6361/202038890},
archivePrefix = {arXiv},
       eprint = {2010.08392},
 primaryClass = {astro-ph.SR},
       adsurl = {https://ui.adsabs.harvard.edu/abs/2021A&A...651A..81B}
}

@article{prochaska2020pypeit,
  title={PypeIt: The Python Spectroscopic Data Reduction Pipeline},
  author={Prochaska, J and Hennawi, Joseph and Westfall, Kyle and Cooke, Ryan and Wang, Feige and Hsyu, Tiffany and Davies, Frederick and Farina, Emanuele and Pelliccia, Debora},
  journal={The Journal of Open Source Software},
  volume={5},
  number={56},
  pages={2308},
  year={2020}
}

@inproceedings{piascik2014sprat,
  title={SPRAT: spectrograph for the rapid acquisition of transients},
  author={Piascik, AS and Steele, Iain A and Bates, Stuart D and Mottram, Christopher J and Smith, RJ and Barnsley, RM and Bolton, B},
  booktitle={Ground-based and Airborne Instrumentation for Astronomy V},
  volume={9147},
  pages={2703--2718},
  year={2014},
  organization={SPIE}
}

@article{albareti201713th,
  title={The 13th data release of the Sloan Digital Sky Survey: First spectroscopic data from the SDSS-IV survey mapping nearby galaxies at Apache Point Observatory},
  author={Albareti, Franco D and Prieto, Carlos Allende and Almeida, Andres and Anders, Friedrich and Anderson, Scott and Andrews, Brett H and Arag{\'o}n-Salamanca, Alfonso and Argudo-Fern{\'a}ndez, Maria and Armengaud, Eric and Aubourg, Eric and others},
  journal={\apj Supplement Series},
  volume={233},
  number={2},
  pages={25},
  year={2017},
  publisher={IOP Publishing}
}

@article{fremling2016ptf12os,
  title={PTF12os and iPTF13bvn-Two stripped-envelope supernovae from low-mass progenitors in NGC 5806},
  author={Fremling, Christoffer and Sollerman, Jesper and Taddia, Francesco and Ergon, Mattias and Fraser, M and Karamehmetoglu, Emir and Valenti, S and Jerkstrand, A and Arcavi, I and Bufano, FILOMENA and others},
  journal={\aap},
  volume={593},
  pages={A68},
  year={2016},
  publisher={EDP Sciences}
}

@ARTICLE{Siebert20,
       author = {{Siebert}, Matthew R. and {Dimitriadis}, Georgios and {Polin}, Abigail and {Foley}, Ryan J.},
        title = "{Strong Calcium Emission Indicates that the Ultraviolet-flashing SN Ia 2019yvq Was the Result of a Sub-Chandrasekar-mass Double-detonation Explosion}",
      journal = {\apjl},
         year = 2020,
        month = sep,
       volume = {900},
       number = {2},
          eid = {L27},
        pages = {L27},
          doi = {10.3847/2041-8213/abae6e},
archivePrefix = {arXiv},
       eprint = {2007.13793},
 primaryClass = {astro-ph.HE},
       adsurl = {https://ui.adsabs.harvard.edu/abs/2020ApJ...900L..27S}
}

@ARTICLE{Foley03,
       author = {{Foley}, Ryan J. and {Papenkova}, Marina S. and {Swift}, Brandon J. and {Filippenko}, Alexei V. and {Li}, Weidong and {Mazzali}, Paolo A. and {Chornock}, Ryan and {Leonard}, Douglas C. and {Van Dyk}, Schuyler D.},
        title = "{Optical Photometry and Spectroscopy of the SN 1998bw-like Type Ic Supernova 2002ap}",
      journal = {\pasp},
         year = 2003,
        month = oct,
       volume = {115},
       number = {812},
        pages = {1220-1235},
          doi = {10.1086/378242},
archivePrefix = {arXiv},
       eprint = {astro-ph/0307136},
 primaryClass = {astro-ph},
       adsurl = {https://ui.adsabs.harvard.edu/abs/2003PASP..115.1220F}
}

@ARTICLE{Silverman2012,
       author = {{Silverman}, Jeffrey M. and {Foley}, Ryan J. and {Filippenko}, Alexei V. and {Ganeshalingam}, Mohan and {Barth}, Aaron J. and {Chornock}, Ryan and {Griffith}, Christopher V. and {Kong}, Jason J. and {Lee}, Nicholas and {Leonard}, Douglas C. and {Matheson}, Thomas and {Miller}, Emily G. and {Steele}, Thea N. and {Barris}, Brian J. and {Bloom}, Joshua S. and {Cobb}, Bethany E. and {Coil}, Alison L. and {Desroches}, Louis-Benoit and {Gates}, Elinor L. and {Ho}, Luis C. and {Jha}, Saurabh W. and {Kandrashoff}, Michael T. and {Li}, Weidong and {Mandel}, Kaisey S. and {Modjaz}, Maryam and {Moore}, Matthew R. and {Mostardi}, Robin E. and {Papenkova}, Marina S. and {Park}, Sung and {Perley}, Daniel A. and {Poznanski}, Dovi and {Reuter}, Cassie A. and {Scala}, James and {Serduke}, Franklin J.~D. and {Shields}, Joseph C. and {Swift}, Brandon J. and {Tonry}, John L. and {Van Dyk}, Schuyler D. and {Wang}, Xiaofeng and {Wong}, Diane S.},
        title = "{Berkeley Supernova Ia Program - I. Observations, data reduction and spectroscopic sample of 582 low-redshift Type Ia supernovae}",
      journal = {\mnras},
         year = 2012,
        month = sep,
       volume = {425},
       number = {3},
        pages = {1789-1818},
          doi = {10.1111/j.1365-2966.2012.21270.x},
archivePrefix = {arXiv},
       eprint = {1202.2128},
 primaryClass = {astro-ph.CO},
       adsurl = {https://ui.adsabs.harvard.edu/abs/2012MNRAS.425.1789S}
}

@ARTICLE{Horne86,
       author = {{Horne}, K.},
        title = "{An optimal extraction algorithm for CCD spectroscopy.}",
      journal = {\pasp},
         year = 1986,
        month = jun,
       volume = {98},
        pages = {609-617},
          doi = {10.1086/131801},
       adsurl = {https://ui.adsabs.harvard.edu/abs/1986PASP...98..609H}
}

@article{astropy2022astropy,
  title={The Astropy Project: sustaining and growing a community-oriented open-source project and the latest major release (v5. 0) of the core package},
  author={Price-Whelan, Adrian M and Lim, Pey Lian and Earl, Nicholas and Starkman, Nathaniel and Bradley, Larry and Shupe, David L and Patil, Aarya A and Corrales, Lia and Brasseur, CE and others},
  journal={\apj},
  volume={935},
  number={2},
  pages={167},
  year={2022},
  publisher={The American Astronomical Society}
}

@article{van2011numpy,
  title={The NumPy array: a structure for efficient numerical computation},
  author={Van Der Walt, Stefan and Colbert, S Chris and Varoquaux, Gael},
  journal={Computing in science \& engineering},
  volume={13},
  number={2},
  pages={22--30},
  year={2011},
  publisher={IEEE}
}

@article{virtanen2020scipy,
  title={SciPy 1.0: fundamental algorithms for scientific computing in Python},
  author={Virtanen, Pauli and Gommers, Ralf and Oliphant, Travis E and Haberland, Matt and Reddy, Tyler and Cournapeau, David and Burovski, Evgeni and Peterson, Pearu and Weckesser, Warren and Bright, Jonathan and others},
  journal={Nature methods},
  volume={17},
  number={3},
  pages={261--272},
  year={2020},
  publisher={Nature Publishing Group US New York}
}

@article{van2001cosmic,
  title={Cosmic-ray rejection by laplacian edge detection},
  author={Van Dokkum, Pieter G},
  journal={\pasp},
  volume={113},
  number={789},
  pages={1420--1427},
  year={2001},
  publisher={The University of Chicago Press}
}

@article{acero2026gamma,
  title={Gamma-ray signature of superluminous supernovae: Fermi-LAT GeV detection of SN 2017egm and evidence of a central engine},
  author={Acero, F and Acharyya, A and Adelfio, A and Ajello, M and Aviano, E and Baldini, L and Ballet, J and Bartolini, C and Bastieri, D and Becerra Gonzalez, J and others},
  journal={\aap},
  volume={709},
  pages={A229},
  year={2026},
  publisher={EDP Sciences}
}

@article{crnogorvcevic2026gamma,
  title={On the Gamma-ray Efficiency of Superluminous Supernovae: Potential Detections and Population-Level Constraints},
  author={Crnogor{\v{c}}evi{\'c}, Milena and Linden, Tim and Goobar, Ariel and Metzger, Brian D},
  journal={arXiv preprint arXiv:2604.16595},
  year={2026}
}

@article{taddia2015early,
  title={Early-time light curves of Type Ib/c supernovae from the SDSS-II Supernova Survey},
  author={Taddia, Francesco and Sollerman, Jesper and Leloudas, G and Stritzinger, MD and Valenti, S and Galbany, Lluis and Kessler, R and Schneider, DP and Wheeler, JC},
  journal={\aap},
  volume={574},
  pages={A60},
  year={2015},
  publisher={EDP Sciences}
}

@article{Pendragosa2011,
  title={Scikit-learn: Machine Learning in {P}ython},
  author={Pedregosa, F. and Varoquaux, G. and Gramfort, A. and Michel, V.
          and Thirion, B. and Grisel, O. and Blondel, M. and Prettenhofer, P.
          and Weiss, R. and Dubourg, V. and Vanderplas, J. and Passos, A. and
          Cournapeau, D. and Brucher, M. and Perrot, M. and Duchesnay, E.},
  journal={Journal of Machine Learning Research},
  volume={12},
  pages={2825--2830},
  year={2011}
}

\begin{appendix}
\onecolumn 
\section{Tables and Data}

    \begin{figure*}[h!]
    \sidecaption
    \includegraphics[width=12cm]{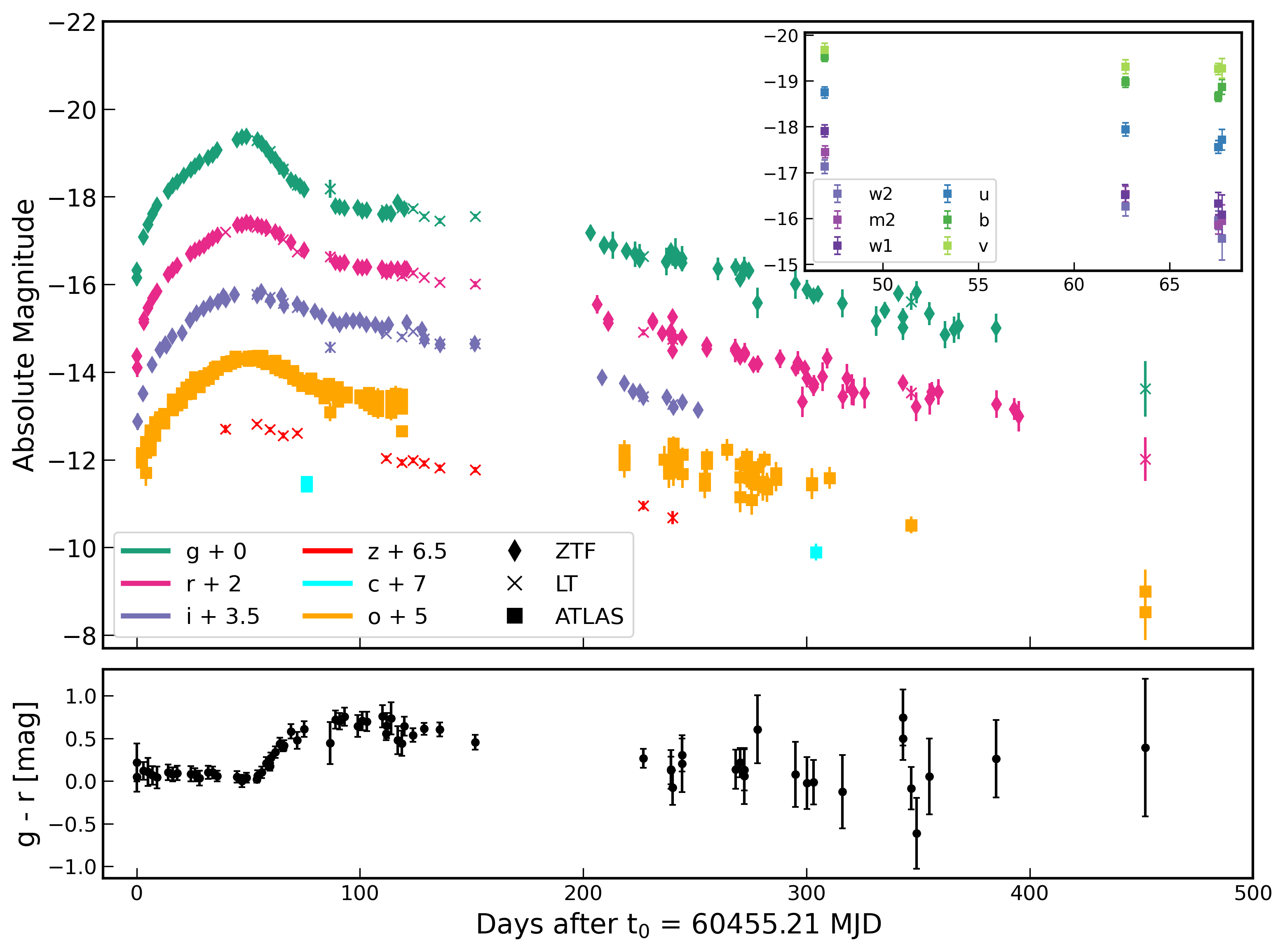}
    \caption{Photometry of SN~2024jlc.
    All bands are in the AB system and presented since the time of explosion $t_0$. 
    The main panel shows the optical evolution of the $griz$ and $c$ and $o$ bands over the first 450 days. 
    The insert panel shows the \textit{Swift}-UVOT filters across the three observation epochs.
    The lower panel shows the $g-r$ color evolution over the same time range.  
    Data are reported in Table~\ref{tab:photometry}.  
    }
    \label{fig:lcs}

    \end{figure*}

    \begin{figure*}[h!]
    \sidecaption
    \includegraphics[width=12cm]{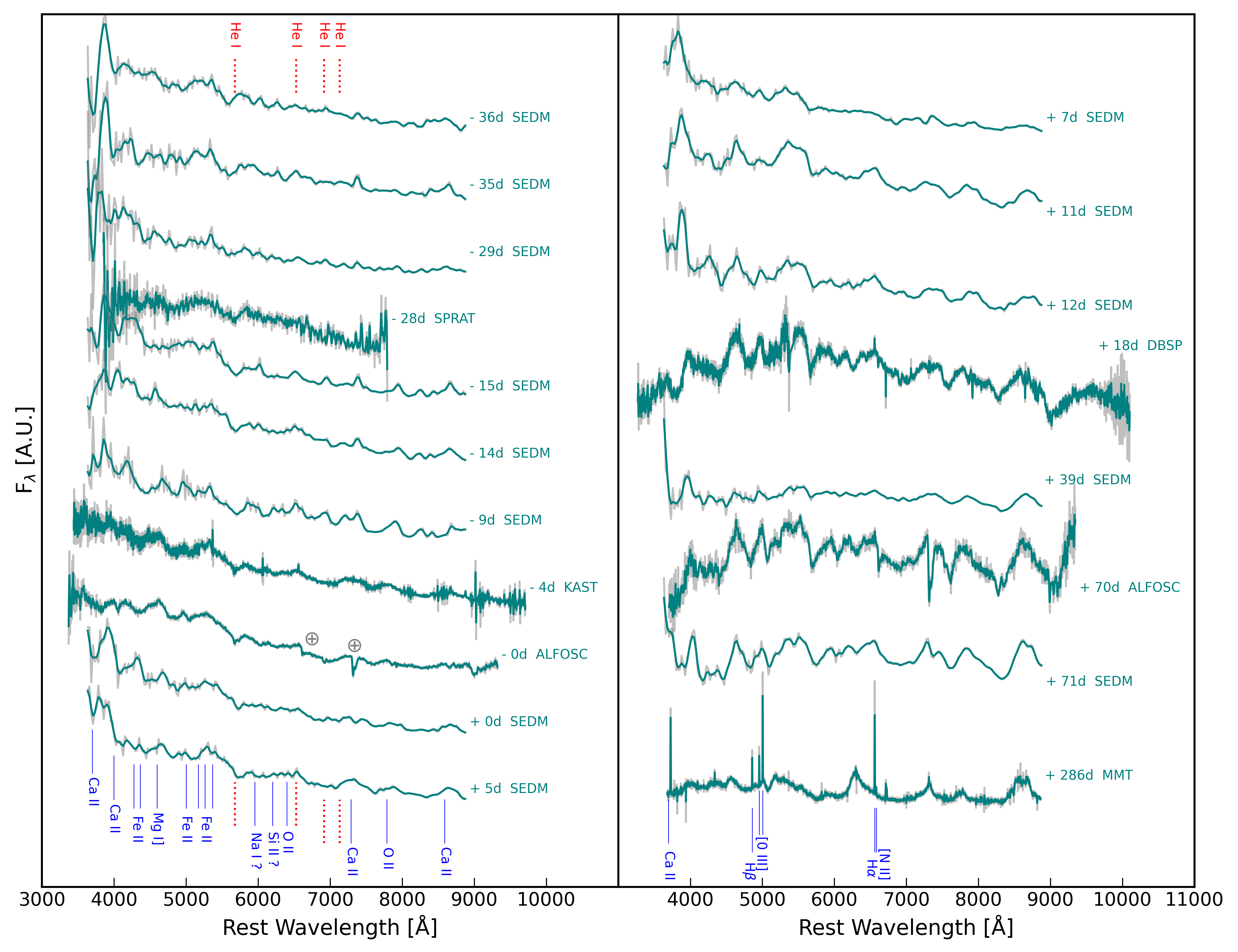}
    \caption{Spectral data of SN~2024jlc.
    Epochs are given with respect to the $g$-band time of maximum $t_{\rm max}$. 
    Fluxes are normalized and offset for better visualization. 
    Original spectra are plotted in light gray in the background, while the smoothed interpolated spectra are shown in teal. 
    Wavelengths are in rest-frame.
    We highlight some spectral features. 
    The spectra at $t_0 + 20.81\,\rm d$, $t_0 + 35.75\,\rm d$, and $t_0 + 213.34\,\rm d$ are not shown due to low signal to noise ratio. 
    Data are available on \texttt{WISeREP}. 
    } 
    \label{fig:spectra} 

    \end{figure*}

\begin{table}[h!]
\centering
\caption{Measurements of host-galaxy brightness.}
\label{tab:host}
\begin{tabular}{ccc} 
\toprule 
\toprule 
Source  &Band & Brightness \\ 
        &     &(mag, AB)   \\ 
\midrule 
PanSTARRS &g  &17.78 $\pm$ 0.03 \\ 
PanSTARRS &i  &17.27 $\pm$ 0.07 \\ 
PanSTARRS &r  &17.36 $\pm$ 0.03 \\ 
PanSTARRS &z  &17.16 $\pm$ 0.07 \\
unWISE    &W1 &17.88 $\pm$ 0.09 \\ 
unWISE    &W2 &18.57 $\pm$ 0.13 \\ 
\bottomrule
\end{tabular}
\tablefoot{Not corrected for MW extinction. }
\end{table}

\begin{table}[h!]
\centering
\caption{Photometric data of SN~2024jlc (extract) } 
\label{tab:photometry} 
\begin{tabular}{lccccc}
\toprule 
\toprule 
Date & Phase & Mag & Err & Band & Instrument \\ 
MJD  & day   & AB  & AB  &      &             \\ 
\midrule
60455.21&0.0&20.16&0.1&g&ZTF\\
60455.27&0.06&20.32&0.17&g&ZTF\\
60455.29&0.08&20.03&0.14&r&ZTF\\
60455.39&0.18&20.3&0.21&r&ZTF\\
60455.59&0.39&19.98&0.19&i&ZTF\\
60457.41&2.21&19.35&0.27&o&ATLAS\\
60457.42&2.21&19.43&0.26&o&ATLAS\\
60457.42&2.22&19.22&0.22&o&ATLAS\\
:        &  :     & :     &  :   & : & : \\ 
:        &  :     & :     &  :   & : & : \\ 
\bottomrule
\end{tabular}
\tablefoot{Phases are expressed relative to $t_0$ = 60455.21 MJD. 
Magnitudes are in the AB system.
The full table is available at the CDS. }
\end{table}

    \begin{table*}[h]
    \centering
    \caption{Log of spectral observations.}
    \label{tab:spectra}
    \begin{tabular}{l|ccccccc} 
    \toprule 
    \toprule  
    Date       & Phase & $\lambda_\text{min}$ & $\lambda_{\text{max}}$ & Exposure time & Instrument & Telescope \\ 
    MJD        & (day)   & ($\AA$)       & ($\AA$)                  & (s)        &            &             \\ 
    \midrule 
    60468.328 & 13.12 & 3776 & 9223 & 2700.0 & SEDM & P60 \\
    60469.328 & 14.12 & 3776 & 9223 & 2700.0 & SEDM & P60 \\
    60475.415 & 20.21 & 3776 & 9223 & 2700.0 & SEDM & P60 \\
    60476.02 & 20.81 & 4002 & 8098 & 1700.0 & SPRAT & LT \\
    60489.328 & 34.12 & 3776 & 9223 & 2160.0 & SEDM & P60 \\
    60490.181 & 34.98 & 3776 & 9223 & 2160.0 & SEDM & P60 \\
    60490.952 & 35.75 & 4002 & 8097 & 2000.0 & SPRAT & LT \\
    60495.275 & 40.07 & 3776 & 9223 & 2160.0 & SEDM & P60 \\
    60500.202 & 45.0 & 3574 & 10091 & 600 & KAST & Lick-3m \\
    60503.918 & 48.71 & 3501 & 9688 & 1200.0 & ALFOSC & NOT \\
    60505.219 & 50.01 & 3776 & 9223 & 2160.0 & SEDM & P60 \\
    60510.283 & 55.08 & 3776 & 9223 & 2160.0 & SEDM & P60 \\
    60512.255 & 57.05 & 3776 & 9223 & 2160.0 & SEDM & P60 \\
    60516.173 & 60.97 & 3776 & 9223 & 2160.0 & SEDM & P60 \\
    60517.285 & 62.08 & 3776 & 9223 & 2160.0 & SEDM & P60 \\
    60523.264 & 68.06 & 3400 & 10498 & 600 & DBSP & P200 \\
    60544.232 & 89.03 & 3776 & 9223 & 2160.0 & SEDM & P60 \\
    60574.91 & 119.7 & 3850 & 9712 & 900.0 & ALFOSC & NOT \\
    60576.12 & 120.91 & 3776 & 9223 & 2160.0 & SEDM & P60 \\
    60668.546 & 213.34 & 3776 & 9223 & 414.0 & SEDM & P60 \\
    60791.389 & 336.18 & 3824 & 9212 & 1200 & BINOSPEC & MMT \\
    \midrule 
    60791.39 & 336.18 & 3824 & 9212 & 1200 & BINOSPEC & MMT \\
    \bottomrule 
    \end{tabular}     
    \tablefoot{Phases are expressed with respect to the time of explosion $t_0$ in the observed frame. }
    \end{table*}

    \begin{table}[h]
    \centering
    \caption{\textit{Swift}-XRT observations.}
    \label{tab:xrt}
    \begin{tabular}{l|cccc} 
    \toprule 
    \toprule 
    Date    & Counts        &  Flux                    & Luminosity     & Exposure   \\
            &  $10^{-3} $   & $10^{-13} $              & $10^{41} $     &            \\
    MJD     & (cts s$^{-1}$)  &  (erg cm$^{-2}$ s$^{-1}$)  & (erg s$^{-1}$)   & (s)       \\ 
    \midrule  
        60502  & 14.7      &  5.076    &   20.33    &  770    \\ 
        60517  & 4.53      &  1.564    &   6.27     &  1910   \\
        60522  & 5.10      &  1.761    &   7.05     &  2200   \\ 
    \midrule   
        ---    & $\le$ 2.43     &  $\le$ 0.839     &  $\le$ 3.36     & 4880 \\ 
    \bottomrule
    \end{tabular}
    \tablefoot{Count rates, fluxes and luminosities are $3\sigma$ upper limits. We refer to flux as the absorbed flux, while the luminosity is obtained from the unabsorbed flux. The last line is obtained by integrating over the entire time period.}
    \end{table}

    \begin{table}[h!]
    \centering
    \caption{Input and output parameters for the \texttt{csmni} model from \texttt{MOSFiT}}
    \label{tab:mosfit_csmni}
    \begin{tabular}{l|ccc}
    \toprule
    \toprule
    Parameter & Prior range & Prior type & Posterior \\
    \midrule
    $t_{\rm exp}\,{\rm (d)}$            & [-10, 0]  & Uniform & $-3.66_{\scriptscriptstyle-0.22}^{\scriptscriptstyle+0.19}$ \\
    $M_\text{ej} (M_\odot)$ & [0.1, 30]               &   Uniform         &   $8.24^{\scriptscriptstyle+0.83}_{\scriptscriptstyle-0.56}$ \\ 
    $M_\text{cm} (M_\odot)$ & [0.1, 50]               &   Uniform         &   $1.81^{\scriptscriptstyle+0.08}_{\scriptscriptstyle-0.21}$ \\\ 
    $f_{\rm Ni}$   & [$10^{-3}$, 1] & Log-uniform     & $0.17_{\scriptscriptstyle-0.01}^{\scriptscriptstyle+0.01}$ \\
    $T_{\min}\,{\rm (10^3\,K)}$ &  [6, 20]    & Uniform & $6.73_{\scriptscriptstyle-0.06}^{\scriptscriptstyle+0.13}$ \\
    $E_{\rm k}\,(10^{51}\,{\rm erg})$ & [1, 10]  & Log-uniform & $1.14_{\scriptscriptstyle-0.01}^{\scriptscriptstyle+0.01}$ \\
    $\sigma$ &  [$10^{-5}$, 10]                  & Log-uniform & $0.17_{\scriptscriptstyle-0.01}^{\scriptscriptstyle+0.01}$ \\
    $\rho\ (10^{-12} \rm g\,cm^{-3})$&  [$10^{-2}$, 10]  & Log-uniform & $4.49_{\scriptscriptstyle-1.50}^{\scriptscriptstyle+1.50}$ \\
    $r_0$ & [1, 10] & Uniform &  $9.8_{-0.95}^{+0.98}$ \\ 
    $n_\text{H, host} (10^{20}\,\rm cm^{-3})$   & [$10^{-2}$, 10] &   Log-uniform         &   $9.72^{\scriptscriptstyle+0.32}_{\scriptscriptstyle-0.23}$ \\
    $\kappa_\gamma\,(10^2\,{\rm cm}^{2}\,{\rm g}^{-1})$ & [0.1, $10^4$]  & Log-uniform & $5.8_{\scriptscriptstyle-0.5}^{\scriptscriptstyle+1.5}$ \\
    \bottomrule
    \end{tabular}
    \tablefoot{The fit was executed giving as input redshift, Galactic extinction, luminosity distance, and number of iterations.}
    \end{table}

    \begin{table}[h!]
    \centering
    \caption{Input and output parameters for the \texttt{slsnni} model from \texttt{MOSFiT}}
    \begin{tabular}{l|ccc}
    \toprule
    \toprule
    Parameter & Prior range & Prior type & Posterior \\
    \midrule

    $t_{\rm exp}\,{\rm (d)}$    & [-10, 0]  & Uniform & $-5.16_{\scriptscriptstyle-0.37}^{\scriptscriptstyle+0.78}$ \\
    $M_{\rm ej}\,(M_\odot)$     & [1, 30] & Uniform & $8.47_{\scriptscriptstyle-1.14}^{\scriptscriptstyle+1.39}$ \\
    $P_{\rm spin}\,{\rm (ms)}$  & [1, 30] & Uniform & $5.22_{\scriptscriptstyle-0.69}^{\scriptscriptstyle+0.38}$ \\
    $B\,(10^{14}\,{\rm G})$     & [1, 30] & Uniform & $2.09_{\scriptscriptstyle-0.37}^{\scriptscriptstyle+0.50}$ \\
    $f_{\rm Ni}$                & [$10^{-3}$, 1] & Log-uniform & $0.15_{\scriptscriptstyle-0.04}^{\scriptscriptstyle+0.03}$ \\
    $T_{\min}\,{\rm (10^3\,K)}$ &  [5, 15]  & Uniform & $7.3_{\scriptscriptstyle-0.20}^{\scriptscriptstyle+0.29}$ \\
    $\sigma$                    & [$10^{-3}$, 100]  & Log-uniform & $0.16_{\scriptscriptstyle-0.01}^{\scriptscriptstyle+0.01}$ \\
    $M_{\rm NS}\,(M_\odot)$     & [1, 2] & Uniform & $1.02_{\scriptscriptstyle-0.01}^{\scriptscriptstyle+0.03}$ \\
    $\alpha$ & [0, 5] & Uniform & $0.17_{\scriptscriptstyle-0.13}^{\scriptscriptstyle+1.61}$ \\
    $\kappa\,({\rm cm}^{2}\,{\rm g}^{-1})$ & [0.01, 0.2]   & Uniform & $0.11_{\scriptscriptstyle-0.02}^{\scriptscriptstyle+0.01}$ \\
    $\kappa_\gamma\,({10^3\,\rm cm}^{2}\,{\rm g}^{-1})$ & [0.1, $10^4$] & Log-uniform & $4.49_{\scriptscriptstyle-1.78}^{\scriptscriptstyle+2.34}$ \\
    $v_{\rm ej}\,({\rm 10^3\,km\,s}^{-1})$ & [5, 13]  & Uniform & $5.38_{\scriptscriptstyle-0.23}^{\scriptscriptstyle+0.21}$ \\
    $\theta_{\rm PB}$ & [0, 1.57] & Uniform & $1.12_{\scriptscriptstyle-0.27}^{\scriptscriptstyle+0.32}$ \\
    $\lambda_{\rm cutoff}\,(10^3)$      & [2, 8]      & Uniform & $3.57_{\scriptscriptstyle-1.29}^{\scriptscriptstyle+4.19}$ \\
    $n_\text{H, host} (10^{20}\,\rm cm^{-3})$   & [$10^{-2}$, 10] &   Log-uniform         &   $9.49_{\scriptscriptstyle-1.28}^{\scriptscriptstyle+0.38}$ \\
    \bottomrule
    \end{tabular}
    \tablefoot{The fit was executed giving as input redshift, Galactic extinction, luminosity distance, and number of iterations.
    }
    \label{table:mosfit_slsnni}
    \end{table}

\clearpage

    \begin{figure*}[h!]
    \begin{center}
    \includegraphics[width=1\textwidth]{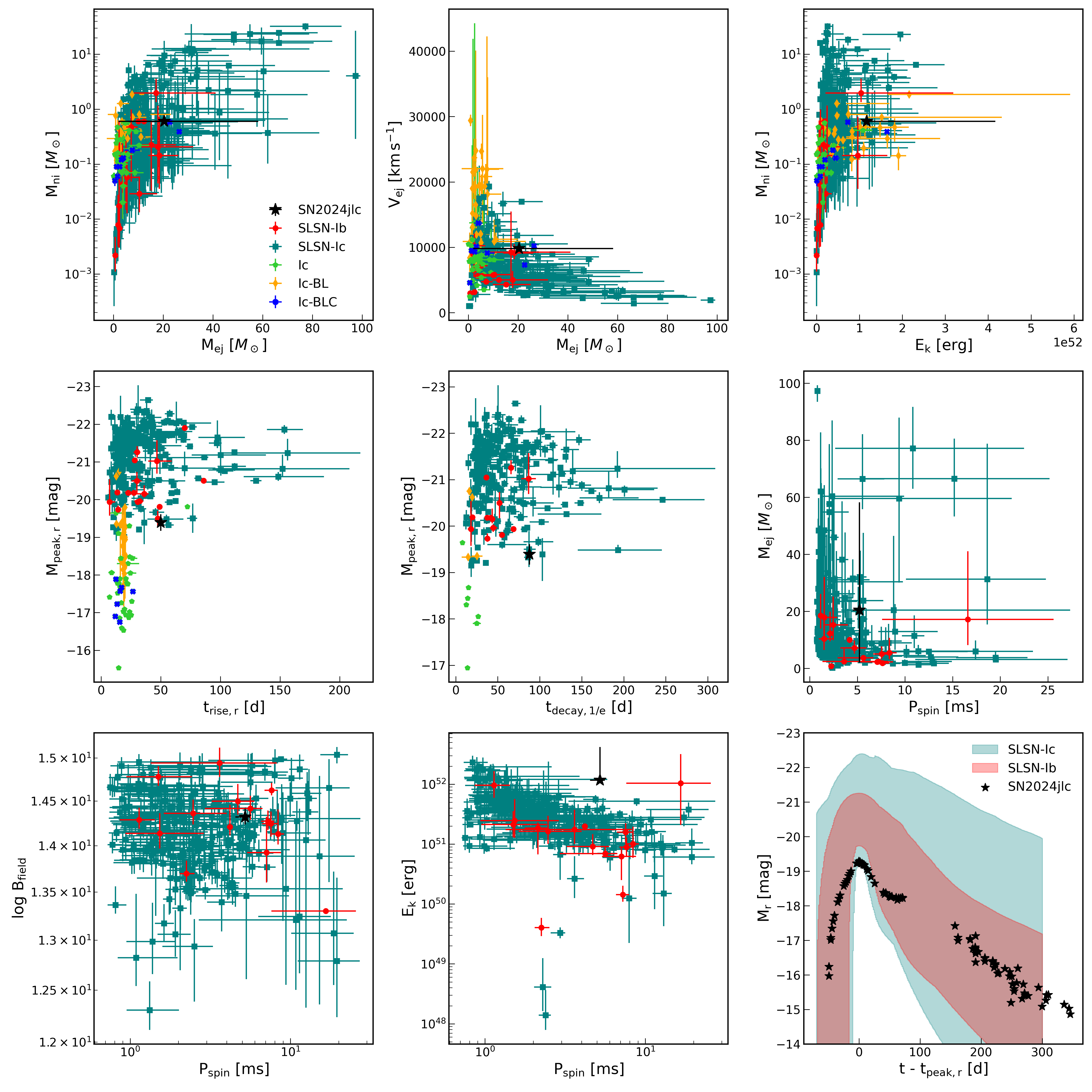}
    \caption{Comparison between SN~2024jlc parameters (black star) and SLSN-Ic (teal), SLSN-Ib (red), Ic (green), Ic-BL (orange), and Ic with broad light curves (Ic-BLC, blue) samples. 
    SLSNe data are taken from the \texttt{slsne} library \protect\citep{gomez2024type}. 
    Additional data for the SLSN-Ib sample are extracted from \protect\citet{gomez2022luminous}, \protect\citet{kumar2025detection}, \protect\citet{kumar2026sn}, \protect\citet{farah2026lense}. 
    Type Ic-BL data are taken from \protect\citet{taddia2015early, taddia2019analysis}. 
    Type Ic data are taken from \protect\citet{taddia2015early, Barbarino2021}. 
    Type Ic-BLC are taken from \protect\citet{Barbarino2021, karamehmetoglu2023population}. 
    Additional notes: i. the rise time ($t_{\rm rise, r}$) and the peak magnitude ($M_{\rm peak, r}$) are defined relative to the r-band; ii. the velocity of ejecta $v_{\rm ej}$ is taken as the expansion velocity relative to the maximum light; in some cases, this is a direct result of \texttt{MOSFiT}; iii. when missing, the kinetic energy ($E_{\rm k}$) or the velocity of the ejecta are derived from Arnett's formula \protect\citep{arnett1982}; iv. as in \protect\citep{chen2023hydrogen}, iPTF12gty \protect\citep{Barbarino2021} and iPTF15eov \protect\citep{taddia2019analysis} are included in the SLSN-Ic class. 
    }
    \label{fig:total_comparison}
    \end{center}
    \end{figure*}

\end{appendix}
\end{document}